\documentclass[12pt]{article}

\usepackage[utf8]{inputenc}
\usepackage{amsmath, amssymb, amsfonts, amscd}
\usepackage{mathtools}
\usepackage{dsfont}
\usepackage{graphicx}
\usepackage{float}
\usepackage{caption}
\usepackage{subcaption}
\usepackage{enumitem}
\usepackage{booktabs}
\usepackage{multirow}
\usepackage{threeparttable}
\usepackage{wrapfig}
\usepackage{adjustbox}
\usepackage{geometry}
\usepackage{longtable}
\usepackage{setspace}
\usepackage{fancyhdr}
\usepackage{placeins}
\fancypagestyle{plain}{%
  \fancyhf{}
  \fancyfoot[C]{\thepage}

}
\usepackage{tikz}
\colorlet{teal}{black}
\usepackage{yhmath}
\usepackage{wasysym}
\usepackage{floatpag}
\floatpagestyle{plain}

\usepackage[breaklinks=true]{hyperref}
\hypersetup{
    colorlinks=true,
    linkcolor=blue,
    filecolor=magenta,
    urlcolor=cyan,
    pdftitle={Socio-Spatial Patterns of Suicide Mortality in the United States},
}
\usepackage{xurl}
\usepackage{newunicodechar}
\newunicodechar{ }{\,}

\title{\vspace{-40pt}\LARGE{Socio-Spatial Patterns of Suicide Mortality in the United States}}
\author{\vspace{-5pt}
\normalsize{Kushagra Tiwari$^{1}$, M. Amin Rahimian$^{1\ast}$, Marie-Laure Charpignon$^{2\ast}$,} \\
\normalsize{Philippe J. Giabbanelli$^{3}$, Praveen Kumar$^{4}$} \\
\scriptsize{$^1$Department of Industrial Engineering, University of Pittsburgh, Pittsburgh, PA} \\[-0.5em]
\scriptsize{$^2$Division of Research, Kaiser Permanente Northern California, Pleasanton, CA;} \\[-0.5em]
\scriptsize{Departments of Biostatistics and Epidemiology, School of Public Health, University of California Berkeley, Berkeley, CA} \\[-0.5em]
\scriptsize{Computational Health Informatics Program, Boston Children's Hospital, Boston, MA;} \\[-0.5em]
\scriptsize{Institut Pierre Louis d’Epidémiologie et de Santé Publique, Sorbonne Université, INSERM, Paris, France} \\[-0.5em]
\scriptsize{$^3$Virginia Modeling, Analysis, and Simulation Center (VMASC), Old Dominion University, Norfolk, VA} \\[-0.5em]
\scriptsize{$^4$Department of Health Policy and Management, School of Public Health, University of Pittsburgh, Pittsburgh, PA;} \\[-0.5em]
\scriptsize{Public Health Dynamics Laboratory, School of Public Health, University of Pittsburgh, Pittsburgh, PA} \\[-0.5em]
\scriptsize{$^\ast$~Correspondence: \texttt{rahimian@pitt.edu} and \texttt{mariecharpignon@berkeley.edu}}
\vspace{-20pt}
}
\date{}

\begin{document}

\maketitle

\begin{abstract}

\noindent
Suicides cause more than 49,000 deaths annually in the United States, including 55\% involving firearms. Across states and counties in the US, suicide mortality exhibits substantial geographical and sociodemographic heterogeneity. \textcolor{teal}{Since suicide risk and attitude toward firearm access can spread through social ties, large-scale social networks may shape this variation, yet their role remains underexplored}. \textcolor{teal}{To assess whether suicide mortality and exposure to firearm-restriction policies are associated across inter-county social ties, we integrate data on county-level suicide mortality (2010–2022) and the Facebook Social Connectedness Index (SCI), which measures the strength of inter-county social ties.} First, using two-way fixed effects regression models with sociodemographic, economic, and spatial controls\textcolor{teal}{, as well as COVID-19, drug-overdose, and alcohol-related mortality, poor mental health days, and state-level firearm policies}, we find that a one-standard-deviation increase in the SCI-weighted average suicide mortality rate of connected counties is associated with an increase of  \textcolor{teal}{2.47 suicide deaths per 100,000 people in the focal county (95\% CI: 0.97-4.00)}. Second, we examine Extreme Risk Protection Orders (ERPOs), state-level policies that allow temporary restriction of firearm access for individuals at risk of self-harm. \textcolor{teal}{We find that counties more strongly connected to ERPO-implementing states had lower suicide mortality, including counties whose own states had not enacted an ERPO.} 
A one-standard-deviation increase in ERPO social exposure is associated with \textcolor{teal}{0.266 fewer suicide deaths per 100,000 people in the focal county (95\% CI: 0.128-0.403)}. Such a protective association persists when adjusting for geographical proximity and including state-by-year fixed effects that capture time-varying state-level factors. Our findings suggest that social networks can facilitate the diffusion of both harmful exposures and protective interventions. This socio-spatial structuring of suicide mortality underscores the need for prevention strategies that incorporate social network topology, alongside more traditional approaches based on geographical targeting.
\end{abstract}

\section*{Introduction}\label{sec:intro}

Suicide mortality remains a major public health concern in the United States (US), with approximately 49{,}000 deaths per year~\cite{CDC2025}. Its burden varies substantially across age groups, disproportionately affecting adolescents and working-age adults: suicide is the second and fourth leading cause of death among individuals aged 10--34 and 35--44, respectively~\cite{CDC2025}. The social ecological framework provides a valuable lens {for understanding this heterogeneity by situating suicide} risk factors and preventive measures at the individual, community, or societal level~\cite{Prades2025}. At these levels, established risk factors include mental health and substance use disorders~\cite{NIMH2024, CCSA2024}, social isolation~\cite{Cacioppo2018}, bullying and cyberbullying~\cite{Hinduja2018}, and access to lethal means---particularly firearms~\cite{Miller2022}. In addition, knowing someone who died by suicide or being exposed to suicide-related information through social media {has been associated with increased suicide risk}~\cite{Swanson2013,pirkis2024public}.

Here, we \textcolor{teal}{examine how social networks, from close interpersonal ties to broader community-level connections, shape suicide risk.} Social networks {may influence suicide risk} through multiple mechanisms. First, network structure can increase exposure to risk factors such as social isolation or {maladaptive peer influences}. Second, exposure to suicide death within social networks can {contribute to} suicidal ideation and {suicidal behavior}. For instance, exposure to suicide by specific means (e.g., firearm, poison) can precipitate one's decision to end their life in the same manner~\cite{Calvo2024}---a phenomenon termed \textit{suicide contagion}~\cite{Gould2003, niederkrotenthaler2020tv,shaman2024quantifying}. \textcolor{teal}{A related but distinct theoretical tradition originating with Durkheim~\cite{jones1987emile} emphasizes the role of social integration, proposing that lower levels of social cohesion are associated with higher suicide mortality rates at the population level.} Building on these complementary perspectives, empirical studies have demonstrated that the risk of suicide mortality increases following exposure to suicidal acts within interpersonal or community networks~\cite{gould2013contagion,phillips1974effects,mueller2015suicidal,shaman2024quantifying}. \textcolor{teal}{In sum, both theoretical and empirical studies suggest that suicide-related outcomes may be structured, in part, by social ties.} {Such social exposures are} especially prominent among adolescents and young adults who frequently encounter suicide-related content via digital media~\cite{arendt2019effects,niederkrotenthaler2020role}. 

Importantly, social networks may also confer protection: greater social connectedness within families, schools, workplaces, and communities has been associated with lower rates of suicidal ideation and attempts~\cite{janiri2020risk,verlenden2024mental,marraccini2017school,arango2024social,kleiman2013social}. Moreover, interventions aimed at fostering social connections (e.g., caring contacts) have been shown to reduce suicide risk ~\cite{motto2001randomized}. \textcolor{teal}{Together, these studies motivate adopting a network science perspective to examine suicide mortality across interconnected populations}~\cite{Pescosolido2024}.


Until recently, efforts to study population-level associations between social networks and suicide mortality were limited by data constraints. Researchers relied on small samples or qualitative studies because high-resolution social network data were lacking. The 2018 release of the social connectedness index (SCI) by Meta has enabled the quantification of social ties between geographical regions using aggregated Facebook friendship data~\cite{bailey2018social}. The SCI measures the relative probability that a Facebook user in county $i$ is a friend of a Facebook user in county $j$, accounting for the number of active users in each county. This metric allows comparisons of social connectedness between pairs of counties, with higher SCI values indicating stronger interpersonal ties. {Derived} from billions of anonymized friendship links and periodically updated, the SCI provides a scalable proxy for real-world social connectedness, enabling studies of how network structure relates to spatial patterns in behavioral and health outcomes. Bailey et al.~\cite{bailey2018social} demonstrate the empirical relevance of the SCI, showing that it correlates with inter-county migration patterns, trade volumes, job search behavior, public health outcomes, and patent citations.
The SCI has also enabled studies examining how behavioral and health outcomes propagate across {socially connected} regions. For instance, Charoenwong et al.~\cite{charoenwong2020social} found that counties socially connected to those first affected by COVID-19 outbreaks had earlier reductions in mobility. {Using the SCI,} Holtz et al.~\cite{holtz2020interdependence} showed that compliance with public health mandates during the COVID-19 pandemic was strongly correlated with {social proximity} to counties with more stringent policies. Similarly, Tiwari et al.~\cite{tiwari2024measuring} demonstrated that opioid overdose death rates in socially connected counties were positively associated with rates in the focal county. \textcolor{teal}{Together, these studies illustrate the utility of the SCI for examining how behavioral and health outcomes are patterned across socially connected regions.}


However, most studies of suicide in social networks have been restricted to small samples or specific settings (e.g., school-based cohorts, military units, clinically high-risk youth) and have not evaluated whether spatial variation in suicide mortality or the diffusion of preventive policies correlates with inter-county friendship networks across the US. Thus, the broader socio-spatial dynamics of suicide risk--including the indirect transmission of protective interventions through social ties--remain insufficiently understood. One prominent example is the use of Extreme Risk Protection Orders (ERPOs), also known as ``red flag laws.'' ERPOs are civil court orders that allow temporary restriction of firearm access for individuals deemed to pose a risk to themselves or others~\cite{miller2024updated}. ERPO implementation relies on social contacts who identify warning signs of suicide risk~\cite{swanson2017implementation}, enabling not only law enforcement but also family members and friends to file petitions\textcolor{teal}{; petitions are filed and orders enforced within the enacting state, with jurisdictional reach varying across statutes}. \textcolor{teal}{Although evidence remains limited, studies from several states suggest that ERPOs may reduce suicide mortality.} For example, \textcolor{teal}{one suicide is estimated to have been averted for every 10 to 20 orders issued} in Connecticut~\cite{swanson2017implementation}. We hypothesize that social ties serve as channels through which awareness, norms, and information about interventions to reduce firearm-related suicide risk diffuse across communities\textcolor{teal}{, rather than through the extension of orders across jurisdictional boundaries}. 
ERPOs therefore provide a useful case study to assess whether exposure to ERPO implementation through social connectedness is associated with lower suicide mortality rates.

The contributions of our study, which integrates data on county-level suicide mortality and social connectedness, are three-fold:
\begin{enumerate}
    \item We define and quantify county-level social exposure and spatial exposure to suicide mortality. 
    \item We estimate associations between county-level suicide mortality and social exposure, with and without adjustment for spatial exposure.
    \item We evaluate whether suicide mortality rates are reduced in non-ERPO-implementing counties that are more socially connected to counties located in ERPO-implementing states.
\end{enumerate}

We test two hypotheses to evaluate whether and how socially-mediated exposures are associated with county-level suicide mortality in the US:
\begin{itemize}
    \item[\textbf{H1}] a one-standard-deviation increase in the SCI-weighted average suicide death rate in socially connected counties is positively associated with the focal county’s suicide death rate, controlling for spatial exposure.
    \item[\textbf{H2}] a one-standard-deviation increase in social exposure to firearm restriction policy is negatively associated with the focal county’s suicide death rate.
\end{itemize}

We estimate county–year panel models with county and state–year fixed effects, controlling for time-varying sociodemographic covariates \textcolor{teal}{, county-level COVID-19, drug-overdose, and alcohol-related mortality, poor mental health days, and state firearm policies}. Although the associations identified in our study do not establish causality, they highlight how social networks can function as pathways for both detrimental socio-spatial influence and beneficial dissemination of preventive policies. In conclusion, our findings underscore the role of social networks in suicide mortality and support the integration of network‑based approaches into the design of public health interventions.


\section*{Results}
\subsection*{Socio-spatial patterns of suicide mortality in the US (H1)}   

We analyzed county-level suicide death data from the US National Vital Statistics System (NVSS) Multiple Cause of Death files for the 2010--2022 period, encompassing 40,794 county-year observations nationwide. Our primary outcome was the overall suicide mortality rate per county (expressed as the number of deaths per 100,000 people), across all age groups. Our primary objective was to assess whether a higher rate of suicide mortality in socially connected counties was positively associated with a higher rate of suicide mortality in the focal county (\textbf{H1}). To test whether the association between suicide mortality and inter-county social connectedness was confounded by geographical distance, we compared two approaches, with and without controlling for spatial proximity.



To estimate these associations, we used two-way fixed effects regression models with county and year indicators, effectively controlling for time-invariant county characteristics and for annual patterns common to all counties. \textcolor{teal}{All models adjust for time-varying county-level characteristics: population density, age distribution, racial and ethnic composition, median household income, limited English proficiency, unemployment, educational attainment, COVID-19 mortality, drug-overdose mortality, alcohol-related mortality, poor mental health days, and state firearm policies. Standard errors are clustered at the state level.} For individuals living in county $i$, exposure to suicide mortality outcomes in other counties than $i$ (denoted by $-i$) was captured through two metrics. The first, denoted by $s_{-it}$, corresponds to ``deaths in social proximity.'' It is defined as the weighted average of suicide mortality rates in year $t$ in counties other than the focal county $i$, where the weight of a given county is equal to its SCI with the focal county. The second, denoted by $d_{-it}$, corresponds to ``deaths in spatial proximity''. It is defined as the weighted average of suicide mortality rates in year $t$ in counties other than the focal county $i$, where the weight of a given county is equal to the inverse of the geographical distance between its centroid and the focal county's centroid. Detailed mathematical formulae appear in equations~\ref{eq:social_spaital_proximity} and~\ref{eq:weights} in the \hyperref[sec:meth]{\it Methods} section.

\begin{figure}[htbp]
    \centering
    \includegraphics[width=0.8\textwidth]{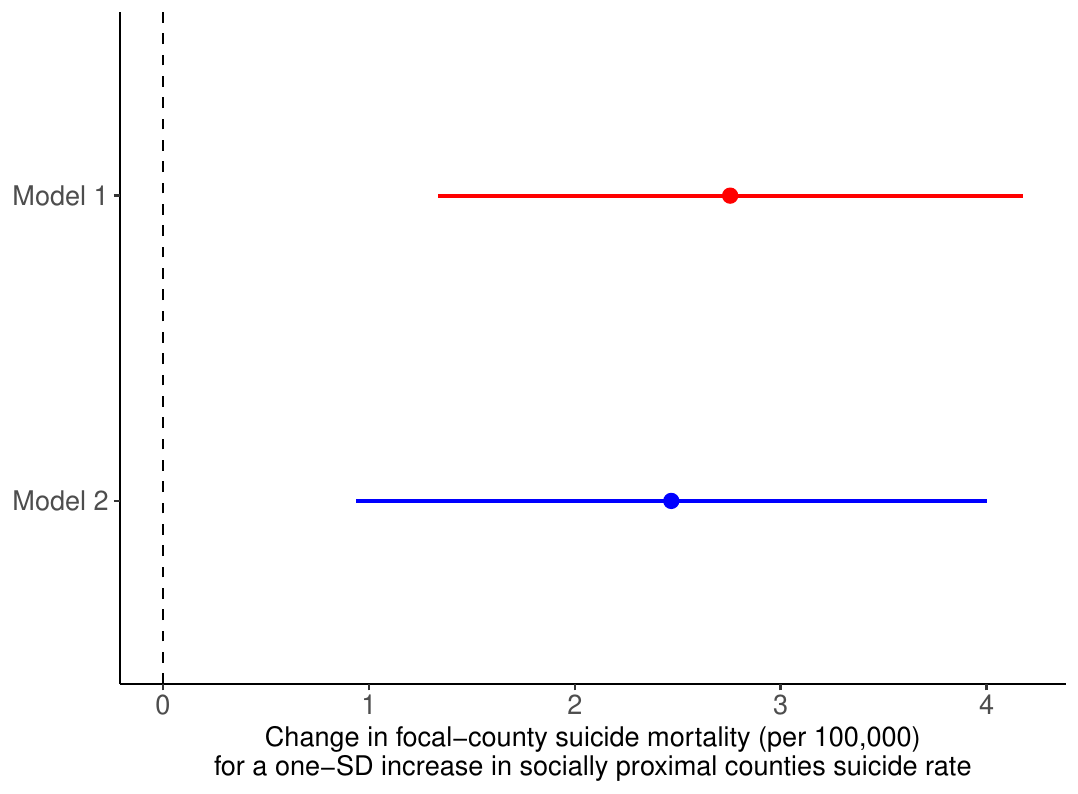}
    \caption{{\bf Role of social ties in county-level suicide mortality.}
    Estimated regression coefficients ($\hat\zeta_1$) for suicide mortality rates in socially connected counties ($s_{-it}$) in two models. Model 1 (red): without adjustment for deaths in spatial proximity ($d_{-it}$). Model 2 (blue): with adjustment for deaths in spatial proximity ($d_{-it}$). Horizontal lines denote 95\% confidence intervals (CI). The vertical dashed line indicates the null hypothesis ($\zeta_1 = 0$). Point estimate in Model 1: \textcolor{teal}{2.75} (cluster-robust 95\% CI: [\textcolor{teal}{1.33}, \textcolor{teal}{4.18}]); point estimate in Model 2: \textcolor{teal}{2.47} (cluster-robust 95\% CI: [\textcolor{teal}{0.97}, \textcolor{teal}{4.00}]). Both models include county and year fixed effects, sociodemographic control variables\textcolor{teal}{, COVID-19 mortality, drug-overdose mortality, alcohol-related mortality, poor mental health days, and state firearm policies} (see Table \ref{tab:socio_spatial_model}).
    }
    \label{fig:ci_peer_effects}
\end{figure}

We first estimated the relation between suicide mortality in the focal county and ``deaths in social proximity'' without controlling for spatial proximity (Model~1). A one-standard-deviation (1-SD) increase in ``deaths in social proximity'' was associated with an increase of \textcolor{teal}{2.75} suicides per 100{,}000 people in the focal county (cluster-robust 95\% CI: [\textcolor{teal}{1.33}, \textcolor{teal}{4.18}], $p<0.01$; Fig.~\ref{fig:ci_peer_effects}, in red). Since the strength of social connectedness is often correlated with geographical distance, we next evaluated whether this observed population-level association was confounded by spatial proximity.

To that end, we built a model that included both variables, i.e., ``deaths in social proximity'' and ``deaths in spatial proximity'', to disentangle their independent associations with suicide mortality (Model~2). The association between the suicide mortality rate in the focal county and ``deaths in social proximity'' remained statistically significant and substantial (\textcolor{teal}{2.47} suicides per 100{,}000 people, cluster-robust 95\% CI: [\textcolor{teal}{0.97}, \textcolor{teal}{4.00}], $p<0.01$; Fig.~\ref{fig:ci_peer_effects}, in blue). \textcolor{teal}{The coefficient on ``deaths in spatial proximity'' was positive but not statistically distinguishable from zero. The association with deaths in social proximity therefore persisted after adjustment for spatial proximity.}

\textcolor{teal}{Controlling for $d_{-it}$ adjusts for the mortality of nearby counties, but does not establish that the connectedness measure itself is distinct from geographical proximity. We therefore assessed whether $s_{-it}$ captures social structure beyond distance, and whether the association is robust to the set of counties over which the SCI-weighted average is computed. Across all $4{,}921{,}953$ county pairs, the SCI and inverse centroid distance are only moderately associated (Pearson $r = 0.53$ on the logarithmic scale), so that geographical distance accounts for approximately $28\%$ of the variation in social connectedness. We then re-estimated Model~2 under two restrictions on the SCI weights. Under a distance-based restriction, in which the SCI-weighted average was computed only over counties located more than $100$~km from the focal county, which excludes all contiguous counties, the association was $2.23$ deaths per $100{,}000$ ($95\%$ CI: $[1.13, 3.34]$). Under a second restriction, in which the average was computed only over the one percent of counties having the highest SCI with the focal county, the association was $1.57$ ($[0.40, 2.73]$; see Supplementary Table~\ref{tab:sci_robustness} and Supplementary Fig.~\ref{fig:sci_distance}). 
The association therefore persists when geographically proximate counties are excluded and when the social neighborhood is restricted to the most strongly connected counties. These findings indicate that the association is not readily explained by geographical proximity alone.
}


Full regression results, including the effects of population density, age structure, racial/ethnic composition, median household income, unemployment, educational attainment, English proficiency\textcolor{teal}{, COVID-19 mortality, drug-overdose mortality, alcohol-related mortality, poor mental health days, and state firearm policies}, appear in Table~\ref{tab:socio_spatial_model}. All continuous variables (i.e., population density, median household income, deaths in social proximity, and deaths in spatial proximity) were standardized to have mean zero and unit variance, thus facilitating the interpretation of estimated coefficients. Consistent with both national- and county-level analyses, higher population density, higher median household income, and higher proportions of Asian and Hispanic individuals were associated with lower suicide mortality rates. \textcolor{teal}{Alcohol-related mortality was positively and significantly associated with suicide mortality. The coefficients on drug-overdose mortality, COVID-19 mortality, and the three state firearm policy indicators were not statistically distinguishable from zero in Model~2, with the exception of waiting periods, which carried a positive coefficient.} The associations for percent Asian, percent Black, percent Hispanic, and median household income remained statistically significant in Model~2, consistent with prior national- and county-level analyses ~\cite{cammack2024vital,steelesmith2019contextual,garnett2023suicide}.

To test the robustness of these associations to outcome definition, we fitted the two models again, using \textit{age-adjusted} rather than \textit{crude} suicide mortality rates as the outcome variable to account for differences in age structure among counties. These sensitivity analyses (see Supplementary Table S1) also yielded a positive association between the suicide mortality rate in the focal county and the suicide mortality rate in socially connected counties, although the estimated magnitudes were approximately 60\% smaller than when using \textit{crude} suicide mortality rates as the outcome variable. Such an attenuation was expected, as adjustment for a county's age structure removes demographic variability in the baseline risk of suicide of its population. Nevertheless, the consistency of the positive association, irrespective of outcome definition (i.e., with or without age adjustment), underscores the robustness of our findings. \textcolor{teal}{We also examined how this association varies with age and sex, and whether it is present for both firearm and non-firearm suicides, by re-estimating Model~2 within cohorts defined by these characteristics (see Supplementary Table~\ref{tab:subgroup} and Figure~\ref{fig:subgroup}). The association was positive and significant among ages $10$--$24$ and $25$--$64$ but statistically indistinguishable from zero among adults aged $65$ and older, consistent with the age-graded susceptibility to social influence reported in the contagion literature~\cite{mueller2015suicidal}. On a scale standardized by each cohort's own mortality dispersion, the two non-elderly cohorts were of comparable magnitude ($0.075$ and $0.100$ within-cohort standard deviations per one-standard-deviation increase in social exposure, respectively). The association was significant in both sexes and, importantly, for both firearm and non-firearm suicides (standardized $0.081$ and $0.089$). That the social channel operates irrespective of manner of death, controlling for spatial proximity and county-level covariates, indicates that it does not simply reflect the geographical concentration of lethal means.}

\textcolor{teal}{Finally, we tested whether these associations are sensitive to residual spatial autocorrelation. Although both models include $d_{-it}$ as a regressor and cluster standard errors by state, neither approach models the correlation structure of the error term. If errors are correlated between neighboring counties, the reported standard errors are too small. We therefore re-estimated Model~2 as a two-way fixed effects spatial-error model, in which each county's error is permitted to depend on the errors of its five nearest neighbors in the same year, through a spatial-error parameter $\rho$ estimated jointly with the regression coefficients by maximum likelihood~\cite{anselin1988spatial,baltagi2003testing,millo2012splm}. The estimated $\rho$ is small but statistically significant ($-0.057$, $p<0.001$), confirming that spatial dependence is present; once it is absorbed into the error covariance, the association with deaths in social proximity remains positive and highly significant, and it is stable across alternative neighbor counts (see Supplementary Tables~\ref{tab:sem_fe} and~\ref{tab:sem_fe_knn}, and Supplementary Fig.~\ref{fig:sem_fe_knn}). Correcting for spatial autocorrelation therefore does not attenuate the social channel, as would be expected were the association an artifact of residual spatial structure.}

The subsequent section examines whether exposure to firearm restriction policies similarly propagates through socio-spatial channels (\textbf{H2}).



\begin{table}[!htbp] \centering 
\scriptsize
\tiny
  \caption{{\bf Estimates of socio-spatial correlates of county-level suicide mortality obtained via two-way fixed effects regressions.} In Model~(1), county-level suicide mortality rates are regressed on standardized deaths in social proximity ($s_{-it}$). Model~(2) additionally controls for standardized deaths in spatial proximity ($d_{-it}$) to disentangle the role of social ties from that of geographical proximity. Both models include county and year fixed effects and adjust for time-varying county-level characteristics: population density, age distribution (percent aged 0--17, 18--44 and 45--64), racial composition (percent Asian, Black, and Other racial subgroups), ethnic composition (percent Hispanic), median household income, percent with limited English proficiency, percent unemployed, and percent with less than high school education\textcolor{teal}{, as well as COVID-19 mortality, drug-overdose mortality, alcohol-related mortality, poor mental health days, and state-level firearm policies (waiting periods, child-access prevention laws, and minimum purchase age of 21)}. Standard errors are clustered at the state level.} 
  \label{tab:socio_spatial_model} 
\begin{tabular}{@{\extracolsep{1pt}}lcc} 
\\[-1.8ex]\hline 
\hline \\[-1.8ex] 
 & \multicolumn{2}{c}{Outcome variable: county-level crude suicide mortality rate} \\ 
\cline{2-3} 
 & Model 1 & Model 2 \\ 
\hline \\[-1.8ex] 
 Deaths in social proximity $s_{-it}$ & \textcolor{teal}{2.755$^{***}$} & \textcolor{teal}{2.469$^{***}$} \\ 
  & \textcolor{teal}{(0.707)} & \textcolor{teal}{(0.763)} \\ 
  Deaths in spatial proximity $d_{-it}$ &  & \textcolor{teal}{0.409} \\ 
  &  & \textcolor{teal}{(0.324)} \\ 
  Population density & \textcolor{teal}{$-$0.633$^{*}$} & \textcolor{teal}{$-$0.551} \\ 
  & \textcolor{teal}{(0.378)} & \textcolor{teal}{(0.363)} \\ 
  Percent aged 0-17 & \textcolor{teal}{0.018} & \textcolor{teal}{0.024} \\ 
  & \textcolor{teal}{(0.369)} & \textcolor{teal}{(0.366)} \\ 
  Percent aged 18-44 & \textcolor{teal}{0.027} & \textcolor{teal}{$-$0.070} \\ 
  & \textcolor{teal}{(0.493)} & \textcolor{teal}{(0.502)} \\ 
  Percent aged 45-64 & \textcolor{teal}{$-$0.426} & \textcolor{teal}{$-$0.441} \\ 
  & \textcolor{teal}{(0.360)} & \textcolor{teal}{(0.360)} \\ 
  Percent Asian & \textcolor{teal}{$-$0.651$^{***}$} & \textcolor{teal}{$-$0.641$^{***}$} \\ 
  & \textcolor{teal}{(0.234)} & \textcolor{teal}{(0.233)} \\ 
  Percent Black & \textcolor{teal}{$-$1.935$^{**}$} & \textcolor{teal}{$-$1.948$^{**}$} \\ 
  & \textcolor{teal}{(0.875)} & \textcolor{teal}{(0.882)} \\ 
  Percent Other & \textcolor{teal}{0.395$^{*}$} & \textcolor{teal}{0.365$^{*}$} \\ 
  & \textcolor{teal}{(0.205)} & \textcolor{teal}{(0.191)} \\ 
  Percent Hispanic & \textcolor{teal}{$-$3.012$^{***}$} & \textcolor{teal}{$-$2.961$^{***}$} \\ 
  & \textcolor{teal}{(0.832)} & \textcolor{teal}{(0.855)} \\ 
  Median household income  & \textcolor{teal}{$-$0.477$^{***}$} & \textcolor{teal}{$-$0.477$^{**}$} \\ 
  & \textcolor{teal}{(0.178)} & \textcolor{teal}{(0.181)} \\ 
  Percent with limited English proficiency & \textcolor{teal}{$-$0.021} & \textcolor{teal}{$-$0.006} \\ 
  & \textcolor{teal}{(0.072)} & \textcolor{teal}{(0.068)} \\ 
  Percent unemployed & \textcolor{teal}{0.043} & \textcolor{teal}{0.048} \\ 
  & \textcolor{teal}{(0.159)} & \textcolor{teal}{(0.157)} \\ 
  Percent with less than high school education & \textcolor{teal}{0.003} & \textcolor{teal}{$-$0.001} \\ 
  & \textcolor{teal}{(0.122)} & \textcolor{teal}{(0.121)} \\ 
  \textcolor{teal}{COVID-19 deaths per 100k} & \textcolor{teal}{0.070} & \textcolor{teal}{0.068} \\ 
  & \textcolor{teal}{(0.066)} & \textcolor{teal}{(0.065)} \\ 
  \textcolor{teal}{Drug-overdose deaths per 100k} & \textcolor{teal}{$-$0.942} & \textcolor{teal}{$-$0.917} \\ 
  & \textcolor{teal}{(1.010)} & \textcolor{teal}{(1.027)} \\ 
  \textcolor{teal}{Alcohol-related deaths per 100k} & \textcolor{teal}{9.534$^{***}$} & \textcolor{teal}{9.498$^{***}$} \\ 
  & \textcolor{teal}{(3.248)} & \textcolor{teal}{(3.273)} \\ 
  \textcolor{teal}{Poor mental health days} & \textcolor{teal}{0.126$^{*}$} & \textcolor{teal}{0.132$^{*}$} \\ 
  & \textcolor{teal}{(0.073)} & \textcolor{teal}{(0.073)} \\ 
  \textcolor{teal}{Firearm: waiting period} & \textcolor{teal}{0.335$^{*}$} & \textcolor{teal}{0.376$^{**}$} \\ 
  & \textcolor{teal}{(0.174)} & \textcolor{teal}{(0.157)} \\ 
  \textcolor{teal}{Firearm: child-access prevention} & \textcolor{teal}{$-$0.067} & \textcolor{teal}{$-$0.018} \\ 
  & \textcolor{teal}{(0.157)} & \textcolor{teal}{(0.160)} \\ 
  \textcolor{teal}{Firearm: minimum purchase age 21} & \textcolor{teal}{$-$0.296} & \textcolor{teal}{$-$0.282} \\ 
  & \textcolor{teal}{(0.192)} & \textcolor{teal}{(0.189)} \\ 
\hline \\[-1.8ex] 
Observations & 40,794 & 40,794 \\ 
R$^{2}$ & \textcolor{teal}{0.949} & \textcolor{teal}{0.949} \\ 
Adjusted R$^{2}$ & \textcolor{teal}{0.945} & \textcolor{teal}{0.945} \\ 
\hline \\[-1.8ex] 
\multicolumn{3}{l}{\footnotesize Robust standard errors in parentheses. * p$<$0.1; ** p$<$0.05; *** p$<$0.01}
\end{tabular} 
\end{table}

\subsection*{Direct and indirect exposures to ERPO policies through social networks (H2)}


\textbf{Using staggered implementation of ERPOs to assess the role of social networks in suicide mortality.} In this section, we investigate whether indirect exposure to firearm access restrictions, through social network connections, is associated with reductions in suicide mortality. A county is deemed \textit{directly exposed} if the corresponding state has enacted an ERPO; a county is deemed \textit{indirectly exposed} if it is socially connected to counties in states that have enacted ERPOs, while its own state has not. We hypothesize that counties with greater social connectedness to counties located in ERPO-implementing states have lower suicide mortality rates, even in the absence of local ERPO implementation (\textbf{H2}). 

To test hypothesis (\textbf{H2}), we first constructed a standardized metric of ERPO social exposure using the SCI. We subsequently estimated the association between ERPO social exposure and suicide mortality using a two-way fixed effects regression, accounting for county- and time-specific patterns. Finally, to disentangle the role of ERPO social exposure from that of geographical distance, we implemented two distinct regressions, with and without controlling for spatial proximity. 

In what follows, we report the empirical effectiveness of ERPOs, provide estimates of their direct and indirect effects on suicide mortality, and comment on the robustness of our results to confounding by spatial proximity. In the Supplementary Materials, we provide more details about ERPOs and refer to recent literature relevant to the topic~\cite{kivisto2018effects,swanson2024suicide,zeoli2021use}.

\textbf{Direct effects of ERPOs on suicide mortality.}
We found a statistically significant association between ERPO implementation and suicide mortality rates. On average, counties located in ERPO-implementing states had lower suicide mortality rates. ERPO implementation was associated with a decrease of $\widehat{\psi}=\textcolor{teal}{-0.481}$ suicide deaths per 100,000 people (cluster-robust 95\% CI: [\textcolor{teal}{-0.917}, \textcolor{teal}{-0.0453}], $p<\textcolor{teal}{0.05}$). Full results of the direct effect model with county and year fixed effects appear in Table~\ref{policy_exposure}. Additional information about model specification appears in \hyperref[sec:meth]{\it Methods} (see Equation~\ref{eq:direct_effect}).


\textbf{Indirect effects of ERPOs on suicide mortality.} Beyond the direct effect of ERPO implementation on suicide mortality, we investigated its indirect effect, through social networks. To that end, we defined the $ERPO\ Social\ Exposure_{it}$ metric. It quantifies the share of social ties in the focal county $i$ directed toward counties located in ERPO-implementing states in year $t$ (see Equation~\ref{eq:social_exposure} in \hyperref[sec:meth]{\it Methods}). To test hypothesis \textbf{(H2)} that ERPO social exposure is negatively associated with the suicide mortality rate of the focal county, we implemented a two-way fixed-effects regression (see Equation~\ref{eq:indirect_exposure} in \hyperref[sec:meth]{\it Methods}). 

\begin{figure}[!htbp]
  \centering
  \includegraphics[width=\linewidth]{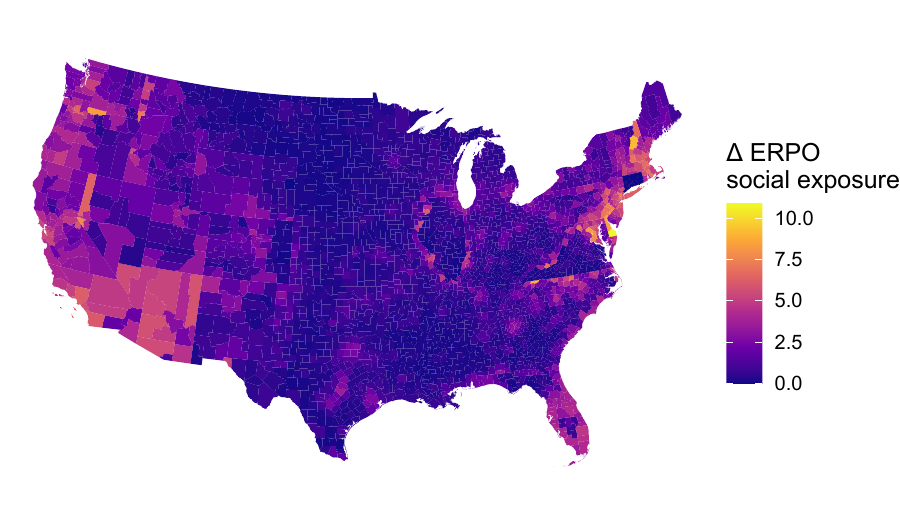}
\caption{\textbf{County-level change in ERPO social exposure, 2010--2022.}
\textcolor{teal}{Colors show ($\Delta_i=E_{i,2022}-E_{i,2010}$), the change in standardized ERPO social exposure between 2010 and 2022. Higher values (yellow) indicate increased exposure
through out-of-state social ties to ERPO-adopting states; lower values (dark purple) indicate smaller increases or no change in exposure.} Although the analysis covers all US counties, the map shows only the 48 contiguous states and the District of Columbia; Alaska, Hawaii, and US territories are not shown.}
  \label{fig:delta_erpo_exposure_contigUS}
\end{figure}

{\color{teal}
Figure~\ref{fig:delta_erpo_exposure_contigUS} summarizes changes in ERPO social exposure between 2010 and 2022. For county \(i\), we define $\Delta_i = E_{i,2022} - E_{i,2010}$, where \(E_{it}\) denotes the standardized ERPO social exposure defined in Equation~\ref{eq:social_exposure} in the \hyperref[sec:meth]{\it Methods}. Because this measure includes only ties to counties in other states, \(\Delta_i\) is not mechanically determined by whether the focal county's own state adopted an ERPO. Counties with large positive values of \(\Delta_i\) are therefore found in both ERPO-adopting and non-adopting states. For example, some counties in California, Florida, Massachusetts, New York, and Washington experienced large increases because a relatively large share of their out-of-state social ties connected them to states that adopted ERPOs over the study period. Large increases also occurred in some counties in non-adopting states, including Pennsylvania, Utah, and Arizona, when those counties were strongly connected to adopting states. Whereas Figure~\ref{fig:delta_erpo_exposure_contigUS} displays the endpoint change \(\Delta_i\), the panel regression uses the full annual series \(E_{it}\). The coefficient on ERPO social exposure is identified from annual county-level variation in \(E_{it}\) that remains after county fixed effects absorb time-invariant differences across counties and state-by-year fixed effects absorb factors shared by all counties in the same state and year.}

\begin{figure}[htbp]
    \centering
    \includegraphics[width=\linewidth]{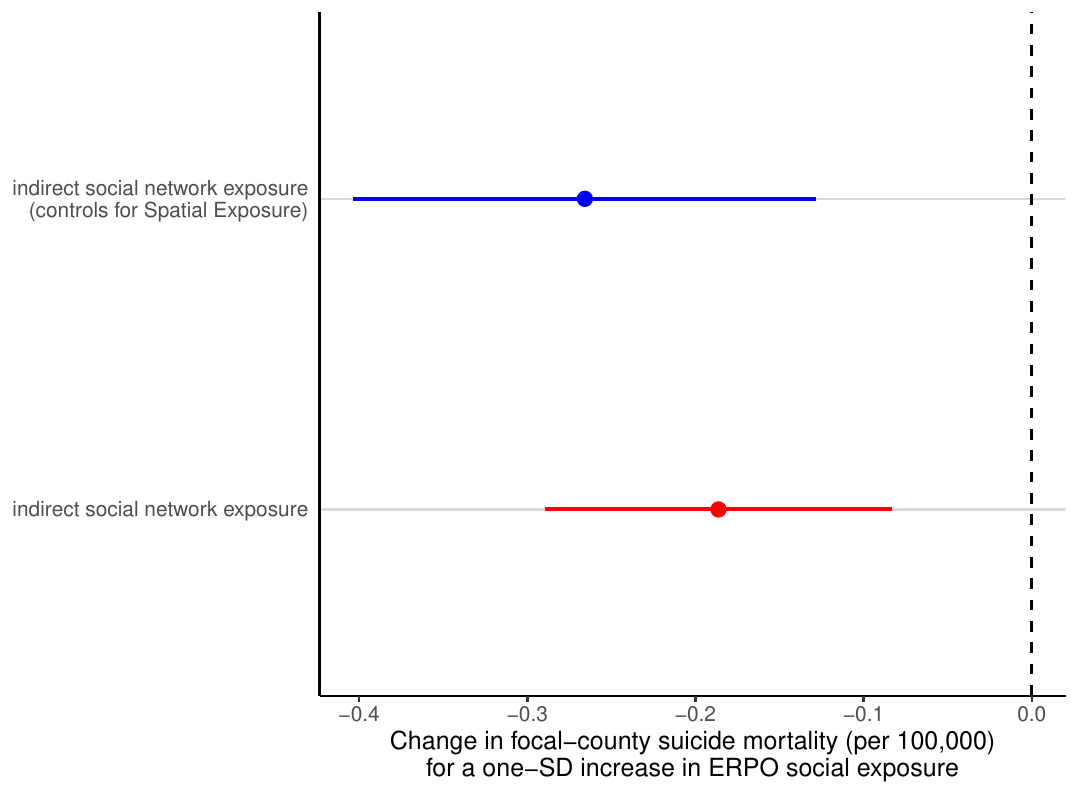}
    \caption{\textbf{Estimated coefficients ($\hat{\delta}_1$, $\hat{\theta}_1$) for ERPO social exposure in two specifications.} 
    Red point indicates estimate from the baseline model without spatial exposure ($\hat{\delta}_1 = \textcolor{teal}{-0.186}$, cluster-robust 95\% CI: [$\textcolor{teal}{-0.29}$, $\textcolor{teal}{-0.0831}$]); 
    blue point indicates estimate from the specification controlling for $\textit{ERPO Spatial Exposure}_{it}$ ($\hat{\theta}_1 = \textcolor{teal}{-0.266}$, cluster-robust 95\% CI: [$\textcolor{teal}{-0.403}$, $\textcolor{teal}{-0.128}$]). 
    Horizontal lines denote 95\% confidence intervals; vertical dashed line denotes the null hypothesis ($\delta_1 = 0$). 
    Both models include county and state--year fixed effects ($\phi_i$, $\gamma_{st}$) and sociodemographic controls ($\overline{X}_{it}$)\textcolor{teal}{, as well as COVID-19 mortality, drug-overdose mortality, alcohol-related mortality, and poor mental health days (see Table~\ref{policy_exposure})}. Consistent negative and statistically significant estimates indicate the association between suicide mortality and indirect social exposure to ERPO policies is robust to spatial confounding.}
    \label{fig:erpo_social_robustness}
\end{figure}

We found a statistically significant negative association between social connectedness to counties located in ERPO-implementing states and suicide mortality, accounting for state-by-year fixed effects (Figure~\ref{fig:erpo_social_robustness}). Specifically, a one-standard-deviation increase in ERPO exposure through out-of-state social ties with counties located in ERPO-implementing states was associated with a decrease of $\widehat{\delta}_1=\textcolor{teal}{-0.186}$ suicide deaths per \(100{,}000\) people (cluster-robust 95\% CI: [\textcolor{teal}{-0.29}, \textcolor{teal}{-0.0831}], $p<\textcolor{teal}{0.001}$). \textcolor{teal}{Local ERPO implementation and the three state firearm policies (waiting periods, child-access prevention laws, and a minimum purchase age of 21) are constant within each state-year, so they are perfectly collinear with the state-by-year fixed effects. They therefore cannot be estimated in Columns~2 and~3 of Table~\ref{policy_exposure}, and appear only in the direct-effect model of Column~1, which uses county and year fixed effects.} Full regression results are in Table~\ref{policy_exposure}.



\textbf{Robustness to confounding by geographical proximity.} To assess the robustness of this indirect effect of ERPO implementation on suicide mortality and disentangle the role of social networks from that of geographical proximity, we introduced a second variable: ERPO spatial exposure. $ERPO\ Spatial\ Exposure_{it}$ quantifies the share of geographical proximity of the focal county $i$ accounted for by counties located in ERPO-implementing states in year $t$, where proximity is measured by the inverse of the distance between county centroids (see Equation~\ref{eq:spatial_exposure} in \hyperref[sec:meth]{\it Methods}). Even when accounting for spatial proximity (see Equation \ref{eq:social_spatial_exposure} in \hyperref[sec:meth]{\it Methods}), our prior finding of a statistically significant negative association between indirect exposure to ERPOs through social networks and suicide mortality persisted (Figure \ref{fig:erpo_social_robustness}). A one-standard-deviation increase in ERPO social exposure was associated with a decrease of $\widehat{\theta}_1=\textcolor{teal}{-0.266}$ suicide deaths per 100,000 people (cluster-robust 95\% CI: [\textcolor{teal}{-0.403}, \textcolor{teal}{-0.128}], $p<\textcolor{teal}{0.001}$). In contrast, the coefficient on ERPO spatial exposure was positive but not statistically distinguishable from zero ($\widehat{\theta}_2=\textcolor{teal}{0.478}$ cluster-robust 95\% CI: [\textcolor{teal}{-0.161}, \textcolor{teal}{1.12}], $p<0.10$). Conditional on social exposure and state--year fixed effects, the spatial ERPO exposure term carries little independent signal, and we place emphasis on the robust negative association for ERPO social exposure. Full regression results appear in Table~\ref{policy_exposure}.


\textbf{Robustness to confounding by age structure.} To evaluate the robustness of our findings to outcome definition, we estimated the indirect effect of ERPO implementation using \textit{age-adjusted} rather than \textit{crude} suicide mortality as the outcome variable. In both models, with and without adjustment for spatial proximity, we again found a negative association between indirect exposure to ERPOs through social networks and suicide mortality (see Supplementary Figure S2 and Table S2). 

\textbf{Robustness to residual confounding from suicide deaths in social proximity.} To assess robustness to potential residual confounding from correlated suicide mortality patterns across the social network, we controlled for suicide deaths in social proximity. Whether using crude or age-adjusted suicide mortality rates as outcomes, the association between indirect exposure to ERPOs and suicide mortality through social networks persisted (see Supplementary Figure S3 and Tables S3 and S4). 

Through these analyses, we show the existence of a robust negative association between social ties and suicide-related events. This investigation validates our hypothesis that ERPO social exposure is negatively associated with suicide mortality. We posit that these indirect protective effects stem from the reduced diffusion of information on suicidal ideation and death events within social networks.

\begin{table}[p]
\thisfloatpagestyle{empty}
\centering 
\tiny
\caption{\textbf{Estimated effects of ERPO policy exposure on county-level crude suicide mortality rates (expressed in terms of the number of deaths per 100,000 people).} Column~(1) reports direct effects of local ERPO implementation with county and year fixed effects. Column~(2) reports indirect effects of ERPO social exposure, measured through inter-county social ties, estimated with county and state--year fixed effects. Column~(3) reports results from the indirect social exposure model, with an additional control for ERPO \emph{spatial} exposure as well as county and state--year fixed effects. All models adjust for population density, age distribution (percent aged 0--17, 18--44 and 45--64), racial composition (percent Asian, Black, and Other racial subgroups), ethnic composition (percent Hispanic), median household income, percent with limited English proficiency, percent unemployed, percent with less than high school education, political affiliation\textcolor{teal}{, COVID-19 mortality, drug-overdose mortality, alcohol-related mortality, and poor mental health days. Local ERPO implementation and the state firearm policies (waiting periods, child-access prevention laws, and a minimum purchase age of 21) are constant within each state--year and therefore perfectly collinear with the state--year fixed effects; they can be estimated only in Column~(1)}. Standard errors are clustered at the state level.}
\label{policy_exposure} 
\setlength{\tabcolsep}{3pt}
\renewcommand{\arraystretch}{0.85}
\begin{tabular}{@{\extracolsep{1pt}}lccc} 
\hline 
 & \multicolumn{3}{c}{Outcome variable: county-level crude suicide mortality rate} \\ 
\cline{2-4} 
 & (1) & (2) & (3) \\ 
\hline 
ERPO & \textcolor{teal}{$-$0.481$^{**}$} &  &  \\ 
  & \textcolor{teal}{(0.217)} &  &  \\ 
ERPO social exposure &  & \textcolor{teal}{$-$0.186$^{***}$} & \textcolor{teal}{$-$0.266$^{***}$} \\ 
  &  & \textcolor{teal}{(0.051)} & \textcolor{teal}{(0.069)} \\ 
ERPO spatial exposure &  &  & \textcolor{teal}{0.478} \\ 
  &  &  & \textcolor{teal}{(0.318)} \\ 
Population density & \textcolor{teal}{$-$0.897$^{**}$} & \textcolor{teal}{$-$0.304} & \textcolor{teal}{$-$0.349} \\ 
  & \textcolor{teal}{(0.360)} & \textcolor{teal}{(0.563)} & \textcolor{teal}{(0.536)} \\ 
Percent aged below 18 & \textcolor{teal}{0.089} & \textcolor{teal}{$-$0.257} & \textcolor{teal}{$-$0.235} \\ 
  & \textcolor{teal}{(0.378)} & \textcolor{teal}{(0.346)} & \textcolor{teal}{(0.346)} \\ 
Percent aged 18-44 & \textcolor{teal}{0.162} & \textcolor{teal}{$-$0.476} & \textcolor{teal}{$-$0.393} \\ 
  & \textcolor{teal}{(0.509)} & \textcolor{teal}{(0.497)} & \textcolor{teal}{(0.485)} \\ 
Percent aged 45-64 & \textcolor{teal}{$-$0.518} & \textcolor{teal}{$-$0.538} & \textcolor{teal}{$-$0.477} \\ 
  & \textcolor{teal}{(0.384)} & \textcolor{teal}{(0.365)} & \textcolor{teal}{(0.374)} \\ 
Percent Asian & \textcolor{teal}{$-$0.780$^{***}$} & \textcolor{teal}{$-$0.492$^{*}$} & \textcolor{teal}{$-$0.476$^{**}$} \\ 
  & \textcolor{teal}{(0.263)} & \textcolor{teal}{(0.246)} & \textcolor{teal}{(0.230)} \\ 
Percent Black & \textcolor{teal}{$-$1.904$^{**}$} & \textcolor{teal}{$-$2.841$^{***}$} & \textcolor{teal}{$-$2.766$^{***}$} \\ 
  & \textcolor{teal}{(0.847)} & \textcolor{teal}{(0.692)} & \textcolor{teal}{(0.696)} \\ 
Percent Other & \textcolor{teal}{0.475$^{**}$} & \textcolor{teal}{0.077} & \textcolor{teal}{0.068} \\ 
  & \textcolor{teal}{(0.218)} & \textcolor{teal}{(0.205)} & \textcolor{teal}{(0.207)} \\ 
Percent Hispanic & \textcolor{teal}{$-$3.315$^{***}$} & \textcolor{teal}{$-$2.151$^{***}$} & \textcolor{teal}{$-$2.271$^{***}$} \\ 
  & \textcolor{teal}{(0.928)} & \textcolor{teal}{(0.689)} & \textcolor{teal}{(0.713)} \\ 
Median household income  & \textcolor{teal}{$-$0.440$^{**}$} & \textcolor{teal}{$-$0.401$^{**}$} & \textcolor{teal}{$-$0.374$^{*}$} \\ 
  & \textcolor{teal}{(0.183)} & \textcolor{teal}{(0.183)} & \textcolor{teal}{(0.187)} \\ 
Percent with limited English proficiency & \textcolor{teal}{$-$0.045} & \textcolor{teal}{$-$0.050} & \textcolor{teal}{$-$0.066} \\ 
  & \textcolor{teal}{(0.072)} & \textcolor{teal}{(0.116)} & \textcolor{teal}{(0.114)} \\ 
Percent unemployed & \textcolor{teal}{0.056} & \textcolor{teal}{$-$0.052} & \textcolor{teal}{$-$0.039} \\ 
  & \textcolor{teal}{(0.165)} & \textcolor{teal}{(0.129)} & \textcolor{teal}{(0.130)} \\ 
Percent with less than high school education & \textcolor{teal}{0.011} & \textcolor{teal}{$-$0.083} & \textcolor{teal}{$-$0.088} \\ 
  & \textcolor{teal}{(0.139)} & \textcolor{teal}{(0.110)} & \textcolor{teal}{(0.112)} \\ 
Political affiliation & \textcolor{teal}{$-$0.057} & \textcolor{teal}{0.160} & \textcolor{teal}{0.140} \\ 
  & \textcolor{teal}{(0.168)} & \textcolor{teal}{(0.142)} & \textcolor{teal}{(0.147)} \\ 
\textcolor{teal}{COVID-19 deaths per 100k} & \textcolor{teal}{0.088} & \textcolor{teal}{0.098} & \textcolor{teal}{0.083} \\ 
  & \textcolor{teal}{(0.068)} & \textcolor{teal}{(0.096)} & \textcolor{teal}{(0.099)} \\ 
\textcolor{teal}{Drug-overdose deaths per 100k} & \textcolor{teal}{$-$1.020} & \textcolor{teal}{$-$0.926} & \textcolor{teal}{$-$0.923} \\ 
  & \textcolor{teal}{(0.983)} & \textcolor{teal}{(1.123)} & \textcolor{teal}{(1.125)} \\ 
\textcolor{teal}{Alcohol-related deaths per 100k} & \textcolor{teal}{9.671$^{***}$} & \textcolor{teal}{9.671$^{***}$} & \textcolor{teal}{9.668$^{***}$} \\ 
  & \textcolor{teal}{(3.243)} & \textcolor{teal}{(3.442)} & \textcolor{teal}{(3.440)} \\ 
\textcolor{teal}{Poor mental health days} & \textcolor{teal}{0.104} & \textcolor{teal}{0.093} & \textcolor{teal}{0.083} \\ 
  & \textcolor{teal}{(0.078)} & \textcolor{teal}{(0.090)} & \textcolor{teal}{(0.091)} \\ 
\textcolor{teal}{Firearm: waiting period} & \textcolor{teal}{0.399$^{*}$} &  &  \\ 
  & \textcolor{teal}{(0.208)} &  &  \\ 
\textcolor{teal}{Firearm: child-access prevention} & \textcolor{teal}{0.101} &  &  \\ 
  & \textcolor{teal}{(0.196)} &  &  \\ 
\textcolor{teal}{Firearm: minimum purchase age 21} & \textcolor{teal}{$-$0.016} &  &  \\ 
  & \textcolor{teal}{(0.184)} &  &  \\ 
 \hline \\[-1.8ex] 
Observations & 40,794 & 40,794 & 40,794 \\ 
R$^{2}$ & \textcolor{teal}{0.949} & \textcolor{teal}{0.950} & \textcolor{teal}{0.950} \\ 
Adjusted R$^{2}$ & \textcolor{teal}{0.945} & \textcolor{teal}{0.945} & \textcolor{teal}{0.945} \\ 
\hline 
\multicolumn{4}{l}{\footnotesize Robust standard errors in parentheses. * p$<$0.1; ** p$<$0.05; *** p$<$0.01} \\
\end{tabular} 
\end{table}

\section*{Discussion}\label{sec:discussion}

Our study is the first to provide evidence that county-level suicide mortality in the US exhibits socio-spatial structuring: namely, the risk of suicide death in a given county varies not only with geographical location but also with the strength of its social ties to other counties and states (\textbf{H1}).
Using two-way fixed effects regression models, we found that a one-standard-deviation increase in the suicide mortality rates of socially connected counties was associated with an increase of \textcolor{teal}{2.469 and 2.75} suicide deaths per 100,000 people in the focal county, with and without controlling for spatial proximity, respectively. 

The directionality and magnitude of this association are consistent with research on suicide contagion in adolescent peer networks. For example, Mueller and Abrutyn analyzed longitudinal friendship data from a national adolescent study and identified a contagion phenomenon at the individual level~\cite{mueller2015suicidal}. If a teenager \emph{knew} that a close friend had attempted suicide, their odds of suicidal ideation were approximately 72\% higher than if they did not have such exposure and their odds of attempting suicide within one year were approximately 65\% higher. \textcolor{teal}{While that study examined micro-level contagion in a youth network, our findings show population-level associations between mortality rates in socially connected counties. These patterns are consistent with, but do not establish, a network-mediated contagion mechanism at the population level. The overall association persisted after adjustment for age and deaths in spatial proximity, although the estimate was not statistically significant among adults aged 65 and older.}  


Beyond the direct effects of social connectedness on suicide mortality, we sought to understand the potential spillover effects of firearm access restriction policies to states that did not implement such policies. We identified a robust, indirect effect of Extreme Risk Protection Orders (ERPOs) among counties located in non-ERPO-implementing states, likely mediated by social ties: a one-standard-deviation increase in social exposure to counties located in ERPO-implementing states was associated with a small but statistically significant reduction of \textcolor{teal}{0.266} suicide deaths per 100,000 people. \textcolor{teal}{Our exposure measure treats ERPO enactment as a binary state--year attribute and therefore does not capture variation in how these laws are written, operationalized or enforced in practice. Statutes differ structurally, including in whether they operate as warrant or order provisions~\cite{doj2021erpo,zeoli2025extreme}, and implementation varies substantially across states: average monthly petition rates range from $0.07$ per $100{,}000$ in California to $1.03$ in Maryland, and the proportion of final orders granted from $50.9\%$ in Connecticut to $94.2\%$ in Florida, with comparable differences in referral source and in the share of petitions citing suicidal behavior~\cite{barnard2025extreme}. Comparable variation is also present among counties within a state, which statewide averages do not reflect~\cite{barnard2025extreme}. Enactment is thus a coarse proxy for the intensity with which a law is used, and because a county's social exposure aggregates ties to states that differ in petition volume and case composition, $\widehat{\delta}_1$ estimates an average association across them rather than the effect of any particular pattern of implementation. It should therefore be read as the association at prevailing levels of ERPO use during our study period, rather than as an estimate of what a more or less actively implemented law would produce. An exposure measure weighted by petition rates would separate these cases, but county-level petition data are not available for all states over our study period.}

Our results show that both elevated suicide mortality and exposure to firearm-restriction policies are patterned along inter-county social ties. Counties socially connected to
higher-mortality counties had higher suicide mortality rates, whereas counties more strongly connected to ERPO-implementing states had lower rates. Both associations persisted after adjustment for geographical proximity. These findings indicate that measured social connectedness is associated with both risk-related and potentially protective exposures across counties.

Current suicide prevention strategies often target counties based on their sociodemographic characteristics or socioeconomic deprivation indicators. Our results suggest that new prioritization schemes could benefit from combining such geography‑based targeting with decentralized, locally adaptive coordination among socially connected counties---a direction already reflected in some local policy agendas. For example, the Suicide Prevention Council of San Diego County has adopted a network-informed multi-stakeholder approach by convening municipalities, behavioral health providers, and faith-based institutions to foster an integrated response system~\cite{sandiego2015spc}. 



Network-informed infrastructures are critical for three operational reasons: (1) targeted  allocation in structurally-central counties can maximize the indirect impact of limited and costly resources such as gatekeeper training, telepsychiatry expansion, and school-based referrals; (2) socially connected populations enable faster diffusion of norms, including help-seeking behaviors and firearm safety practices; and (3) early warning systems built on socially connected surveillance networks can detect emerging suicide clusters before geographical clustering becomes apparent~\cite{zeoli2021use,klimes2021gatekeeper,bernal2022publichealth}.

Going forward, prevention strategies should rely on existing frameworks, such as the Suicide Prevention Resource Center \cite{sprc2020framework} and the CDC’s Community-Based Suicide Prevention programs \cite{cdc2022csp}, to strengthen cross-sectoral coordination, local coalitions, and mental health infrastructure at the county level. 
Indeed, our findings suggest that strategically implementing interventions in counties that serve as ``social hubs'' could amplify the benefits of social policies via network spillovers and be highly effective at mitigating suicide risk.

\subsection*{Limitations and Avenues for Future Research}

A central component of our study is the use of the SCI to quantify indirect exposure to both suicide mortality and firearm restriction policies, through inter-county social ties that go beyond geographical boundaries. The SCI captures the strength of Facebook friendship ties between any two US counties, providing a scalable and validated measure of population-level social connectedness~\cite{bailey2018social,kuchler2022jue}. Research has shown that the SCI accurately predicts a range of real-world outcomes, including migration flows, economic mobility, and infectious disease spread, underscoring its utility as a proxy for underlying social networks~\cite{bailey2018social,kuchler2022jue}. Moreover, the SCI matrix is nearly complete: all US county pairs have non-zero SCI values, enabling comprehensive analysis of spatially distributed social influence. In our setting, this fact allowed estimating the effects of exposure to suicide mortality and to firearm restriction policies through social ties with non-neighboring counties, by leveraging the broad geographical coverage and county-level granularity of the SCI.

However, the SCI's reliance on Facebook user data introduces several limitations. Because the SCI is defined based only on the activity of Facebook users, it does not reflect the interactions of people who are not on the platform. This results in the under-representation of older adults and people with lower socioeconomic status or limited internet access. For instance, 57\% of US adults aged 65 and older report using Facebook, compared with 80\% of those aged 30--49~\cite{gottfried2025americans}. \textcolor{teal}{The index measures the relative probability that two individuals in different counties are friends on Facebook, but not how actively those friendships are used, which may differ across age groups. Social ties on Facebook may therefore represent age groups unevenly. Facebook use also varies with educational attainment, being lower among adults with a high school diploma or less than among those with a four-year degree~\cite{gottfried2025americans}.} Consequently, the SCI-derived exposure metrics used in our models may disproportionately reflect the social interactions of younger, more urban, and digitally connected populations, potentially missing or misrepresenting the extent of social influence in other population subgroups. Although this fact does not compromise the structural completeness of the SCI network, it affects the generalizability of the estimated indirect effects through social diffusion. To alleviate these limitations, future work should consider integrating auxiliary data sources to validate findings in age-group or cohort-specific social networks (e.g., the National Longitudinal Study of Adolescent to Adult Health, commonly known as the ``Add Health'' dataset). 


An additional consideration in interpreting our two-way fixed effects estimates is the reflexive nature of social influence among counties. Although our approach identifies robust associations between direct or indirect exposure to policy interventions and suicide rates in socially connected counties, we cannot ascertain the directionality of social influence, i.e., determine which specific counties affect others. The bidirectional nature of social ties implies that a county can simultaneously influence and be influenced by counties to which it is socially connected. Although the existence of such reflexivity does not undermine our finding of network-like social structuring in suicide mortality, we refrain from formulating causal hypotheses about the nature of underlying influence mechanisms at play.

\textcolor{teal}{A further consideration concerns the interpretation of our ERPO estimates. ERPO adoption is staggered across states, and recent work in the difference-in-differences literature shows that two-way fixed effects estimators need not recover an average of the underlying treatment effects when those effects vary across cohorts or over time~\cite{wang2024advances}. We report the ERPO coefficients as associations throughout and do not interpret them causally. This consideration applies to the direct-effect specification, where $ERPO_{it}$ is a binary indicator that switches on in a state's adoption year and stays on. It does not apply in the same form to the indirect exposure models, where exposure measures the share of each county's social ties that connect to counties in other states where ERPOs have been adopted, varies continuously over time as adoption spreads across states, and enters alongside state-by-year fixed effects that absorb each county's own state's adoption status. Beyond estimator choice, the design itself limits what can be identified here. ERPOs are narrowly targeted, so the expected change in aggregate county-level suicide mortality following adoption is small, and suicide death is itself a rare event at the county-year level. An ERPO must also be petitioned, granted, and result in firearm removal before it can prevent a death; as a result, few adoptions translate into observable changes in mortality. The estimators recommended for staggered adoption compound this difficulty, because they estimate a separate effect for each group of states and each year before and after adoption, dividing the available data into smaller pieces. Establishing the causal effect of ERPO adoption will therefore require dynamic treatment-effect estimation applied to an outcome of higher event frequency and closer to the intervention than suicide mortality, such as crisis-line contact volume, emergency-department presentations for self-harm, or population-level measures of suicidal ideation. Such data are not currently available at county-year resolution across our study period, and we identify this as a priority extension of this work.}

Our current specification defines social proximity weights $w_{ij}$ through a parametric kernel based on the SCI and population scaling. This pre-specified structure imposes a fixed functional form on how social influence decays across counties. Although the current model incorporates a dynamic specification for social exposure by adjusting the strength of social influence based on suicide rates, the overall structure of the social proximity weights remains pre-specified and fixed. Albeit transparent and computationally efficient, this parametric formulation may not reflect the true extent and heterogeneity of influence processes, particularly in settings where socially-central counties serve as ``social hubs," exerting disproportionate or nonlinear effects on others. Going forward, these limitations could be addressed by adopting a fully Bayesian hierarchical framework in which social influence weights are not assumed \textit{a priori} but are instead learned from data. For example, one could introduce latent county interaction structures or flexible prior distributions that allow the strength and topology of social ties to vary across space and time. These statistical frameworks would enable context-sensitive modeling of influence propagation through social networks. In the future, advancing toward data-driven Bayesian inference, either semi-parametric or fully nonparametric, represents a rigorous path for improving the fidelity and generalizability of network-based models of suicide mortality in population health research.

\section*{Methods}
\label{sec:meth}

This study integrates multiple county-level datasets to construct time-varying measures of suicide mortality, socio-spatial influence associated with social ties, social and spatial exposures to firearm restriction policies, sociodemographic and socioeconomic covariates \textcolor{teal}{as well as COVID-19 mortality, drug-overdose mortality, alcohol-related mortality, poor mental health days, and state-level firearm policies (waiting periods, child-access prevention laws, and minimum purchase age of 21).}

\subsection*{Suicide mortality}
\label{sec:mortality_data}
Data on suicide deaths were obtained from the National Vital Statistics System (NVSS), managed by the National Center for Health Statistics (NCHS). We extracted county-level counts of suicide deaths that occurred between 2010 and 2022 using the following International Classification of Diseases, Tenth Revision (ICD-10) codes: \texttt{X60}--\texttt{X84} and \texttt{Y87.0}. These codes are conventionally used to identify deaths due to intentional self-harm. For confidentiality and consistency, all death counts were aggregated by year and standardized per 100{,}000 people using county-level population denominators from the US Census Bureau. We considered both crude and age-adjusted mortality rates. Age adjustment was performed using the 2000 US Standard Population.


\subsection*{Social connectedness index (SCI)}
\label{sec:sci}
The Social Connectedness Index (SCI) was obtained from the Meta Data for Good program (see \hyperref[sec:data-availability]{\it Data Availability}). It quantifies the relative strength of Facebook friendship ties between any two US counties, normalized by population size. Specifically, SCI$_{ij}$ reflects the likelihood that a Facebook user living in county \(i\) is a friend of a Facebook user living in county \(j\). This measure provides a high-resolution empirical proxy of social ties among US counties. Formally, the SCI between two counties $i$ and $j$ is defined as follows ~\cite{bailey2018social}:
\begin{align*}
\mbox{SCI}_{ij} &=\frac{\mbox{Facebook Connections}_{ij}}{\mbox{Facebook Users}_i \times \mbox{Facebook Users}_j}.
\end{align*}
Here, $\mbox{Facebook Users}_i$ represents the number of Facebook users living in county $i$. 

$\mbox{Facebook Connections}_{ij}$ is the total number of Facebook friendship connections between individuals in counties $i$ and $j$. Friendship data was first released by Meta in 2018 and was subsequently updated in 2021. Here, we used the 2021 SCI data, which was the most recent friendship data available at the time of our analysis in November 2025. After our analysis, in February 2026, Meta released an updated version of the SCI. We replicated our analysis using the updated friendship data and obtained substantively identical results (see Supplementary Materials, Section ``Robustness to the Updated Social Connectedness Index").

\subsection*{Socioeconomic covariates}
\label{covariates}


Time-varying, county-level sociodemographic characteristics were drawn from two sources. First, we used data provided by the Agency for Healthcare Research and Quality (AHRQ) to derive yearly community-level indicators, including educational attainment, unemployment, median household income, and racial/ethnic composition, for 2010--2020. Second, for years not covered by AHRQ, we accessed the American Community Survey (ACS) via the \texttt{tidycensus} package in \texttt{R}, using 5-year estimates for all US counties. The final list of harmonized covariates includes population density, age structure, median household income, educational attainment, unemployment rate, and proportion of residents with limited English proficiency.
\textcolor{teal}{
\subsection*{Drug-overdose mortality}
\label{sec:overdose_data}
Suicide, drug overdose, and alcohol-related death share proximate social and economic determinants and are frequently analyzed jointly under the ``deaths of despair'' framework~\cite{case2015rising,case2017mortality}. We therefore constructed county-level drug-overdose mortality rates from the same NVSS Multiple Cause of Death files used for our outcome, using the ICD-10 underlying-cause codes \texttt{X40}--\texttt{X44} (accidental poisoning), \texttt{X85} (assault by poisoning), and \texttt{Y10}--\texttt{Y14} (poisoning of undetermined intent). The conventional CDC drug-poisoning definition additionally includes \texttt{X60}--\texttt{X64} (intentional self-poisoning), which we exclude because those codes fall within the suicide case definition used for our outcome. Counts were aggregated by county and year and expressed per $100{,}000$ people using the same population denominators as the outcome.}
\textcolor{teal}{
\subsection*{Alcohol-related mortality}
\label{sec:alcohol_data}
County-level alcohol-related mortality was likewise derived from the NVSS Multiple Cause of Death files, following the NCHS alcohol-induced case definition: \texttt{E24.4}, \texttt{F10}, \texttt{G31.2}, \texttt{G62.1}, \texttt{G72.1}, \texttt{I42.6}, \texttt{K29.2}, \texttt{K70}, \texttt{K85.2}, \texttt{K86.0}, \texttt{R78.0}, \texttt{X45}, and \texttt{Y15}. We depart from that definition in one respect: we omit \texttt{X65} (intentional self-poisoning by alcohol), which falls within \texttt{X60}--\texttt{X84} and would otherwise double-count deaths already recorded in our outcome. We use realized alcohol-related mortality rather than a policy-based measure of alcohol availability or control because it captures the realized burden of alcohol-related harm through which such policies would plausibly affect suicide risk, and because it varies annually within counties rather than only at the state level. Rates are expressed per $100{,}000$ people.}

\textcolor{teal}{
\subsection*{COVID-19 mortality}
\label{sec:covid_data}
County-level COVID-19 deaths were obtained from the \textit{New York Times} county-level COVID-19 database~\cite{nyt2021covid} and expressed per $100{,}000$ people. Because that source reports cumulative running totals rather than period counts, we recovered annual deaths by differencing consecutive year-end cumulative values within each county, flooring differences at zero to accommodate occasional downward revisions arising from retrospective corrections. Values are zero for $2010$--$2019$. The pandemic imposed unequal burdens across counties and over time, disrupting behavioral health service delivery, employment, and social contact. Because the year fixed effects absorb the common national shock, this covariate is identified from within-year variation across counties.}
\textcolor{teal}{
\subsection*{Population mental health}
\label{sec:mental_health_data}
We adjust for the average number of mentally unhealthy days reported by county residents in the preceding $30$ days, obtained from County Health Rankings \& Roadmaps~\cite{remington2015county}, which derives the measure from the Behavioral Risk Factor Surveillance System using small-area estimation. We selected this measure because it has the most complete coverage of our study period among the County Health Rankings mental health indicators: mental-health-provider density, frequent mental distress, and excessive drinking are each missing for entire years in the early part of the period. Where the retained measure is absent for isolated county--years, values were carried forward or backward within county from adjacent years.}
\textcolor{teal}{
\subsection*{State firearm policies}
\label{sec:firearm_data}
Because firearm access is the most proximate determinant of firearm suicide, and because ERPOs are frequently enacted alongside other firearm legislation, we adjust for three state policies with an established evidence base linking them to suicide: mandatory waiting periods, child-access prevention laws, and a minimum handgun purchase age of $21$~\cite{smart2024science}. Policy status was constructed from the RAND State Firearm Law Database~\cite{hoch2026rand} by tracking implementation and repeal events over $2010$--$2022$ to obtain, for each state and year, a binary indicator of whether the policy was in force. These indicators vary at the state--year level and are therefore identified in the specifications that include county and year fixed effects (Equations~\ref{eq:social_proximity} and~\ref{eq:direct_effect}). These indicators cannot be separately estimated in the ERPO exposure specifications (Equations~\ref{eq:indirect_exposure} and~\ref{eq:social_spatial_exposure}), because the state-by-year fixed effects in those specifications absorb all variation shared by counties within the same state and year. 
}
\subsection*{Political leaning}
\label{sec:political_affiliation}
County-level political leaning was obtained from the MIT Election Data and Science Lab~\cite{MIT_Election_Lab_2018} for the 2008, 2012, 2016, and 2020 presidential elections. Political leaning was operationalized as a binary indicator, coded as Republican-leaning (1) if the Republican candidate received a majority of votes and as Democratic-leaning (0) otherwise. For non-election years, we assigned political leaning values based on the most recent preceding election (by carrying forward the latest value). For Alaska counties with incomplete county-level election reporting ($n=27$ counties), political leaning values were assigned based on district-level returns when available ($n=39$ observations across 3 counties; FIPS codes 02013, 02016, 02020) or imputed based on statewide voting patterns for the remaining counties in the corresponding year ($n=312$ observations across 24 counties).

\subsection*{Socio-spatial influence metrics}
\label{socio-spatial}
To quantify the role of suicide mortality in socio-spatially connected counties, we considered two metrics: \textit{social proximity to suicide deaths} ($s_{-it}$) and \textit{spatial proximity to suicide deaths} ($d_{-it}$). These are formally defined as:

\begin{align}
\label{eq:social_spaital_proximity}
s_{-it} &= \sum_{j \neq i} w_{ij} y_{jt},\ d_{-it} = \sum_{j \neq i} a_{ij} y_{jt},
\end{align}
where $y_{jt}$ denotes the suicide death rate in county $j$ at year $t$. The weights $w_{ij}$ and $a_{ij}$ represent social and spatial proximity, respectively, and are defined as:

\begin{equation}
\label{eq:weights}
w_{ij} = \frac{n_j \text{SCI}_{ij}}{\sum_{k \neq i} n_k \text{SCI}_{ik}}, \quad a_{ij} = \frac{1 / d_{ij}}{\sum_{k \neq i} (1 / d_{ik})},
\end{equation}
where $n_j$ is the population of county $j$, $\text{SCI}_{ij}$ is the social connectedness index between counties $i$ and $j$, and $d_{ij}$ is the geodesic distance between them. Consistent with the literature ~\cite{kuchler2022jue}, we account for the correlation between these two metrics---due to the spatial clustering of social networks---by including both terms jointly in the model specification.

To examine the association between socio-spatial influence metrics and suicide mortality in the focal county, we estimate the following two-way fixed effects regression model:

\begin{align}
\label{eq:social_proximity}
    y_{it} = \zeta_1 s_{-it} + \zeta_2 d_{-it} + \overline{\zeta}_3^T \overline{X}_{it} + \mu_i + \phi_t + \varepsilon_{it},
\end{align}
{where $y_{it}$ denotes the suicide death rate in the focal county $i$ in year $t$; the \textcolor{teal}{$19$}-dimensional vector \( \overline{X}_{it} \) includes \textcolor{teal}{population density, age distribution (percent aged 0--17, 18--44 and 45--64), racial composition (percent Asian, Black, and Other racial subgroups), ethnic composition (percent Hispanic), median household income, percent with limited English proficiency, percent unemployed, and percent with less than high school education, as well as COVID-19 mortality, drug-overdose mortality, alcohol-related mortality, poor mental health days, and state-level firearm policies (waiting periods, child-access prevention laws, and minimum purchase age of 21)}; $\mu_i$ and $\phi_t$ correspond to county and year fixed effects, respectively. Model estimation was conducted using population-weighted ordinary least squares (OLS) and standard errors were clustered at the state level to account for within-state correlation in the error structure.}

\subsection*{Direct and indirect exposures to ERPO policies}
\label{social_spatial_exposure}
To assess the direct association between Extreme Risk Protection Order (ERPO) policies and suicide mortality at the county level, we estimated a two-way fixed effects model. The baseline model specification is as follows:
\begin{equation}
\label{eq:direct_effect}
    y_{it} = \psi_1 ERPO_{it} + \overline{\psi}_2^T \overline{X}_{it} + \phi_i + \gamma_t + \varepsilon_{it}, 
\end{equation}
{where \( y_{it} \) denotes the suicide death rate per 100,000 people in county \( i \) and year \( t \) and \( ERPO_{it} \) is a binary indicator equal to 1 if the ERPO policy has already been enacted by year \( t \) in the state corresponding to county \( i \). 
The \textcolor{teal}{$20$}-dimensional vector \(\overline{X}_{it}\) includes
\textcolor{teal}{population density; age distribution (percent aged 0--17,
18--44, and 45--64); racial composition (percent Asian, Black, and Other
racial subgroups); ethnic composition (percent Hispanic); median household
income; percent with limited English proficiency; percent unemployed; percent
with less than a high school education; political affiliation; COVID-19
mortality; drug-overdose mortality; alcohol-related mortality; poor mental
health days; and three state firearm-policy indicators: waiting periods,
child-access prevention laws, and a minimum purchase age of 21}. 
\( \phi_i \) and \( \gamma_t \) correspond to county and year fixed effects (controlling for unobserved, time-invariant heterogeneity and national temporal shocks), respectively. In this model, we do not include state-by-year fixed effects, since values of \( ERPO_{it} \) vary at the state--year level. Model estimation was performed using population-weighted OLS and standard errors are clustered by state (51 clusters), providing conservative inference that is robust to within-state correlation.}


To quantify the indirect association between ERPO policies and suicide mortality, through social networks, we define the following metric:

\begin{equation}
\label{eq:social_exposure}
    ERPO\ Social\ Exposure_{it} = \sum_{s(i) \neq s(j)} \mathds{1}(\text{ERPO in state } s(j))_t \times \frac{SCI_{ij}}{\sum_h SCI_{ih}},
\end{equation}
where for each county $i$, the state to which it belongs is denoted by $s(i)$, \( SCI_{ij} \) denotes the social connectedness index between counties \( i \) and \( j \) and the indicator function reflects ERPO policy implementation in other states than that corresponding to county \( i \) (i.e., \( s(j) \neq s(i) \)) in year \( t \). This standardized metric, coined ``ERPO social exposure'', captures the share of social ties that county \( i \) has with counties located in ERPO-implementing states.

To estimate the association between ERPO social exposure and suicide mortality at the county level, we estimated the following model:

\begin{equation}
\label{eq:indirect_exposure}
    y_{it} = \delta_1 ERPO\ Social\ Exposure_{it} + \overline{\delta}_2^T \overline{X}_{it} + \phi_i + \gamma_{st} + \varepsilon_{it},
\end{equation}
{where \( \phi_i \) and \( \gamma_{st} \) represent county and state-by-year fixed effects, respectively, and the \textcolor{teal}{$17$}-dimensional vector \( \overline{X}_{it} \) includes \textcolor{teal}{population density, age distribution (percent aged 0--17, 18--44 and 45--64), racial composition (percent Asian, Black, and Other racial subgroups), ethnic composition (percent Hispanic), median household income, percent with limited English proficiency, percent unemployed, percent with less than high school education, political affiliation, COVID-19 mortality, drug-overdose mortality, alcohol-related mortality, and poor mental health days}. \textcolor{teal}{Local ERPO implementation and the three state firearm-policy indicators vary only at the state--year level and are therefore absorbed by the state-by-year fixed effects \(\gamma_{st}\); their coefficients cannot be separately identified in this specification. The coefficient on ERPO social exposure is identified from differences among counties within the same state and year in their out-of-state connections to ERPO-adopting states.} 

As a robustness check, we define a \textit{spatial exposure} metric that captures geographical proximity to states where ERPOs have been implemented:
\begin{equation}
\label{eq:spatial_exposure}
ERPO\ Spatial\ Exposure_{it} = 
\sum_{j: s(j) \neq s(i)}^{} 
\mathds{1}(\text{ERPO in state } s(j))_t 
\times 
\left(
\frac{1 / d_{ij}}{\sum_{k \neq i} (1 / d_{ik})}
\right),
\end{equation} where \( d_{ij} \) denotes the geodesic distance between counties \( i \) and \( j \), excluding counties in the same state \( s(i) \). This additional metric will help determine whether ERPO implementation in geographically close counties is associated with suicide mortality in the focal county, through spatial diffusion.

To evaluate the joint association between indirect social and spatial exposures to ERPOs, through social networks and geographical proximity, and suicide mortality in the focal county, we estimate the following model:
\begin{align}
    y_{it}  = & \; \theta_1 ERPO\ Social\ Exposure_{it} + \theta_2 ERPO\ Spatial\ Exposure_{it} \nonumber \\ & \; + \overline{\theta}_3^T \overline{X}_{it} + \phi_i + \gamma_{st} + \varepsilon_{it}, \label{eq:social_spatial_exposure}
\end{align}
{where \( \phi_i \) and \( \gamma_{st} \) denote county and state-by-year fixed effects, respectively, and \( \overline{X}_{it} \) is \textcolor{teal}{the vector of covariates defined for Equation~\ref{eq:indirect_exposure}}}. This model specification allows for simultaneous estimation of the independent associations between social vs. spatial exposure to ERPOs and suicide mortality in the focal county. To ensure inference is robust to differential population sizes and intra-state dependence in error terms, model estimation was performed using population-weighted OLS and standard errors were clustered at the state level.

\subsection*{Code Availability}\label{sec:code}
The full analysis code is available at \url{https://github.com/kut97/suicide-sci} and can be used to reproduce all the figures and tables in this paper.

\subsection*{Data Availability}\label{sec:data-availability}
Mortality data was obtained from NVSS, managed by the National Center for Health Statistics (NCHS) \cite{NCHS_NVSS_Mortality_RU_2010_2022}. Due to confidentiality concerns, this data set is not publicly accessible, but can be requested from NCHS at \url{https://www.cdc.gov/nchs/nvss/nvss-restricted-data.htm}. Social determinants of health (SDOH) covariates for 2010–2020 were obtained from the Agency for Healthcare Research and Quality (AHRQ) SDOH Database (\url{https://www.ahrq.gov/sdoh/data-analytics/sdoh-data.html}); for 2020–2022, SDOH covariates were constructed from US Census Bureau products (American Community Survey [ACS] 5-year estimates; see \url{https://www.census.gov/programs-surveys/acs/data.html}). \textcolor{teal}{County-level drug-overdose and alcohol-related mortality were derived from the same restricted-use NVSS Multiple Cause of Death files described above. County-level COVID-19 deaths were obtained from the \textit{New York Times} COVID-19 data repository (\url{https://github.com/nytimes/covid-19-data}). County-level mental health measures were obtained from County Health Rankings \& Roadmaps, University of Wisconsin Population Health Institute (\url{https://www.countyhealthrankings.org}). State firearm policy indicators were obtained from the RAND State Firearm Law Database (\url{https://www.rand.org/pubs/tools/TLA243-2-v4.html}).}
County-level political affiliation data were obtained from the MIT Election Data and Science Lab, County Presidential Election Returns 2000-2024, available through Harvard Dataverse at \url{https://doi.org/10.7910/DVN/VOQCHQ} \cite{MIT_Election_Lab_2018}. The social connectedness index (SCI) is available through Facebook (Meta) Data for Good at \url{https://dataforgood.facebook.com/dfg/tools/social-connectedness-index}. For age-adjusted suicide mortality, annual population denominators stratified by 18 five-year age groups were drawn from CDC WONDER Bridged-Race Postcensal Population Estimates (\url{https://wonder.cdc.gov/single-race-population.html}).

\section*{Acknowledgment}

We express our sincere gratitude to Dr. David Brent for his invaluable guidance and inspiration in shaping this study. His insights into the role of Extreme Risk Protection Order (ERPO) policy within the context of social network dynamics were instrumental in refining the direction of our research. His thoughtful feedback significantly strengthened this manuscript. We also extend our appreciation to Professor Jeffrey Shaman, whose work ``Quantifying Suicide Contagion at Population Scale" served as a key motivation for pursuing this line of research. His thoughtful guidance and feedback have been invaluable in strengthening our study.


\bibliography{refs}
\bibliographystyle{Science}

\clearpage
\FloatBarrier
\renewcommand{\theequation}{S\arabic{equation}}
\renewcommand{\thefigure}{S\arabic{figure}}
\renewcommand{\thetable}{S\arabic{table}}
\setcounter{page}{1}
\renewcommand{\thepage}{S\arabic{page}}

\newcommand{\figS}[1]{Fig.~S#1}
\newcommand{\tabS}[1]{Table~S#1}
\newcommand{\eq}[1]{Eq.~(S#1)}
\newcommand{\sectS}[1]{Section~S#1}
\newcommand{\CI}[2]{\,[#1,\ #2]}

\newcommand{\AAR}{\mathrm{AAR}}
\newcommand{\CRUDE}{\mathrm{CRUDE}}
\newcommand{\w}{w}
\newcommand{\D}{D}
\newcommand{\Pop}{P}
\newcommand{\rate}{r}

\begin{center}
{\LARGE \textbf{Supplementary Information for}}\\[4pt]
{\Large \textbf{Socio-Spatial Patterns of Suicide Mortality in the United States}}\\[8pt]
\textbf{Tiwari, Kushagra et al.}\\
Correspondence to: \texttt{rahimian@pitt.edu} and \texttt{mariecharpignon@berkeley.edu}\\[12pt]
\end{center}

\noindent \textbf{This document includes the following:}
\begin{itemize}[leftmargin=1.5em]
  \item Supplementary Text
  \item Supplementary Figures S1 to S8
  \item Supplementary Tables S1 to S10
\end{itemize}

\setcounter{equation}{0}
\setcounter{figure}{0}
\setcounter{table}{0}

\subsection*{Data Description}
All analyses used death counts aggregated at county level for years 2010 to 2022. Suicide mortality data were obtained from the National Vital Statistics System (NVSS) Multiple Cause-of-Death files, restricted to death records with an underlying cause-of-death corresponding to ICD-10 codes X60--X84 or Y87.0. The annual population size of each county was used to calculate mortality rates. Annual population sizes were drawn from the CDC Bridged-Race Postcensal Population Estimates and extracted for a total of 18 five-year age groups (0--4, 5--9, 10--14, 15--19, 20--24, 25--29, 30--34, 35--39, 40--44, 45--49, 50--54, 55--59, 60--64, 65--69, 70--74, 75--79, 80--84, 85+). These estimates are harmonized for intercensal consistency and aligned with official age groupings. Demographic and socioeconomic variables used for adjustment, including median household income, racial/ethnic composition, educational attainment, unemployment rate, and limited English proficiency, were derived from the American Community Survey (ACS) 5-Year Estimates. Political affiliation was obtained from the MIT Election Data and Science Lab. We considered county-level presidential election returns for 2008, 2012, 2016, and 2020, and defined a binary indicator based on plurality vote share. For non-election years, the value was assigned based on the most recent preceding election. Social connectedness metrics for the primary analyses were
calculated using the 2021 release of Meta's Social Connectedness Index (SCI). We repeated the analyses using the updated 2026 release as a robustness check. Geospatial coordinates and land area were sourced from the US Census Bureau's TIGER/Line shapefiles. All datasets were merged using 5-digit FIPS county codes.

\subsection*{Supplementary Analysis Using Age-Adjusted Suicide Mortality Rates}

We evaluated the robustness of our findings to different outcome specifications. Our primary analysis used crude mortality rates as outcomes and included three variables characterizing the age composition of a given county in a given year as regressors (i.e., the share of the county's population aged below 18, 18--44, and 45--64). In a supplementary analysis, we evaluated whether the associations reported in the main manuscript were robust to the use of age-adjusted suicide mortality rates rather than crude suicide mortality rates. To that end, we re-estimated all models using age-adjusted suicide mortality rates, enabling the comparison of county-level outcomes standardized to the same reference population. In these models, we removed the three variables related to a given county's age structure and listed above from the set of regressors.

\subsection*{Age Standardization of Suicide Mortality Rates}

\textcolor{teal}{Age-adjusted suicide mortality rates were derived using direct standardization with the \textit{2000 US Standard Population}~\cite{anderson1998age}. The 2000 US population simply constitutes a reference population. The purpose of age adjustment is to ensure that mortality rates can be compared across different sociodemographic and geographic subgroups. Specifically, death counts and population sizes were derived using NVSS age recodes (27-category), harmonized to the following \textbf{18 age groups}: 0--4, 5--9, 10--14, 15--19, 20--24, 25--29, 30--34, 35--39, 40--44, 45--49, 50--54, 55--59, 60--64, 65--69, 70--74, 75--79, 80--84, and 85 years and older.}

Let \( d_{it}^{(a)} \) and \( p_{it}^{(a)} \) denote the number of suicide deaths and the population size of age group \( a \) in county \( i \) and year \( t \). Let \( w^{(a)} \) denote the population share corresponding to age group \( a \) in the 2000 US Standard Population. The age-adjusted suicide mortality rate \( \tilde{y}_{it} \) is given by:

\begin{equation}
\tilde{y}_{it} = 100{,}000 \times \sum_{a} w^{(a)}\frac{d_{it}^{(a)}}{p_{it}^{(a)}}
\end{equation}

Following CDC guidelines, estimates were suppressed for any county-year pair with fewer than 10 suicide deaths to ensure reliability and confidentiality.

\subsection*{Alternative Outcome Definitions and Model Specifications to Learn Socio-spatial Patterns of Suicide Mortality in the US}

As mentioned above, we conducted a supplementary analysis by re-estimating the models presented in the main manuscript with \textit{age-adjusted suicide mortality rates} ($\tilde{y}_{it}$) as outcomes. This adjustment accounts for the heterogeneity in the age composition of US counties. Age-standardization was performed using direct standardization over 18 age groups: 0--4, 5--9, 10--14, 15--19, 20--24, 25--29, 30--34, 35--39, 40--44, 45--49, 50--54, 55--59, 60--64, 65--69, 70--74, 75--79, 80--84, 85+, as defined in NVSS age recodes (27-category). Age-adjusted suicide mortality rates are expressed as the number of suicide deaths per 100,000 people.

Similarly to the analysis presented in the main manuscript, we estimated two-way fixed effects regression models, incorporating county and year fixed effects to account for unmeasured time-invariant county characteristics and nationwide temporal shocks. The first model specification included ``deaths in social proximity'' ($\tilde{s}_{-it}$) as the main exposure. The second model specification additionally included ``deaths in spatial proximity'' ($\tilde{d}_{-it}$) as a control to disentangle the role of social connectedness from that of geographical proximity. Demographic and socioeconomic covariates included population density, racial composition (shares of Asian, Black, and Other subgroups), ethnic composition (share of Hispanic population), median household income, unemployment rate, limited English proficiency, and educational attainment. The three variables related to a given county's age structure were removed from the set of regressors.

Using age-adjusted suicide mortality rates ($\tilde{y}_{jt}$), we recomputed the two proximity-based exposure metrics:

\begin{equation}
\tilde{s}_{-it} = \sum_{j \neq i} w_{ij} \tilde{y}_{jt}
\end{equation}

\begin{equation}
\tilde{d}_{-it} = \sum_{j \neq i} a_{ij} \tilde{y}_{jt}
\end{equation}

where $w_{ij}$ and $a_{ij}$ represent SCI-based and inverse-distance-based weights, respectively, defined identically to those in the main manuscript.

The model estimating the association between the suicide mortality rate of a given focal county and those in socially connected and spatially proximal counties was specified as follows:

\begin{equation}
\tilde{y}_{it} = \eta_1 \tilde{s}_{-it} + \eta_2 \tilde{d}_{-it} + \boldsymbol{\eta}_3^\top \overline{X}_{it} + \mu_i + \phi_t + \varepsilon_{it}
\end{equation}

Here, $\tilde{y}_{it}$ denotes the age-adjusted suicide mortality rate in county $i$ and year $t$; $\tilde{s}_{-it}$ represents the standardized suicide mortality rate in counties socially connected to county $i$ according to Facebook's SCI; $\tilde{d}_{-it}$ represents the inverse-distance-weighted suicide mortality rate in spatially proximal counties; $\overline{X}_{it}$ is a vector of county- and year-specific demographic and socioeconomic covariates\textcolor{teal}{, together with COVID-19 mortality, drug-overdose mortality, alcohol-related mortality, poor mental health days, and state firearm policies}; $\mu_i$ and $\phi_t$ denote county and year fixed effects, respectively; and $\varepsilon_{it}$ is the error term.

In a first model, we estimated the association between $\tilde{y}_{it}$ and $\tilde{s}_{-it}$ without controlling for spatial exposure. A one-standard-deviation increase in the suicide mortality rate of socially connected counties was associated with an increase of \textcolor{teal}{1.21} deaths per 100,000 people in a focal county (cluster-robust 95\% CI: [\textcolor{teal}{0.82}, \textcolor{teal}{1.59}], $p<0.01$; Fig.~\ref{fig:sensitivity_ci_peer_effects}, red point). The magnitude of this association is sizable, relative to the average suicide mortality rate across US counties over the study period.
In a second model, we included deaths in spatial proximity ($\tilde{d}_{-it}$) as part of the regressors. Under this second specification, the association between the suicide mortality rate of a focal county and that of socially connected counties remained statistically significant, albeit of smaller magnitude than under the first specification (\textcolor{teal}{1.020}; 95\% CI: [\textcolor{teal}{0.61}, \textcolor{teal}{1.43}], $p<0.01$; Fig.~\ref{fig:sensitivity_ci_peer_effects}, blue point). Compared to the results of the main analysis presented in Table 1, where the point estimates for the first and second models were \textcolor{teal}{2.755} and \textcolor{teal}{2.469} respectively, the effect sizes estimated using age-adjusted suicide mortality rates as outcomes are attenuated by approximately 55-60\%. This attenuation likely owes to differences in outcome definitions: while crude suicide mortality rates cannot be compared among counties with differing age composition, age-adjusted suicide mortality rates can be compared since they reflect the burden in a population whose age composition has been aligned with the 2000 US Standard Population.

Despite this attenuation, the direction and statistical significance of the estimated associations are consistent across outcome definitions and model specifications, providing evidence for the robustness of the relation between social exposure and suicide mortality. Full regression results are provided in Table~\ref{tab:sensitivity_socio_spatial}. Overall, our findings suggest that suicide risk is shaped by social connectedness among counties, beyond the role of spatial proximity and age structure.

\begin{table}[!htbp] \centering 
\scriptsize
  \caption{{\bf Estimates of socio-spatial correlates of county-level suicide mortality obtained via two-way fixed effects regressions, using \textit{age-adjusted} death rates}. In Model~(1), county-level age-adjusted suicide mortality rates ($\tilde{y}_{it}$) are regressed on standardized deaths in social proximity ($\tilde{s}_{-it}$). Model~(2) additionally controls for standardized deaths in spatial proximity ($\tilde{d}_{-it}$) to disentangle the role of social ties from that of geographical proximity. Both models include county and year fixed effects and adjust for time-varying county-level characteristics: population density, racial composition (percent Asian, Black, and Other racial subgroups), ethnic composition (percent Hispanic), median household income, percent with limited English proficiency, percent unemployed, and percent with less than high school education\textcolor{teal}{, as well as COVID-19 mortality, drug-overdose mortality, alcohol-related mortality, poor mental health days, and state-level firearm policies (waiting periods, child-access prevention laws, and minimum purchase age of 21)}. Age distribution variables are excluded, as the dependent variable is already adjusted for population structure. Standard errors are clustered at the state level.}
  \label{tab:sensitivity_socio_spatial} 
  \tiny
\begin{tabular}{@{\extracolsep{1pt}}lcc} 
\\[-1.8ex]\hline 
\hline \\[-1.8ex] 
 & \multicolumn{2}{c}{Outcome variable: county-level age-adjusted suicide mortality rate} \\ 
\cline{2-3} 
\\[-1.8ex] & Model 1  & Model 2 \\ 
\hline \\[-1.8ex] 
 Deaths in social proximity $\tilde{s}_{-it}$ & \textcolor{teal}{1.206$^{***}$} & \textcolor{teal}{1.020$^{***}$} \\ 
  & \textcolor{teal}{(0.198)} & \textcolor{teal}{(0.209)} \\ 
 Deaths in spatial proximity $\tilde{d}_{-it}$ &  & \textcolor{teal}{0.521$^{**}$} \\ 
  &  & \textcolor{teal}{(0.226)} \\ 
 Population density & \textcolor{teal}{$-$0.939$^{**}$} & \textcolor{teal}{$-$0.853$^{**}$} \\ 
  & \textcolor{teal}{(0.388)} & \textcolor{teal}{(0.388)} \\ 
  Percent Asian & \textcolor{teal}{$-$0.728$^{***}$} & \textcolor{teal}{$-$0.725$^{***}$} \\ 
  & \textcolor{teal}{(0.188)} & \textcolor{teal}{(0.185)} \\ 
 Percent Black & \textcolor{teal}{$-$1.247$^{**}$} & \textcolor{teal}{$-$1.288$^{**}$} \\ 
  & \textcolor{teal}{(0.594)} & \textcolor{teal}{(0.614)} \\  
 Percent Other & \textcolor{teal}{0.177} & \textcolor{teal}{0.148} \\ 
  & \textcolor{teal}{(0.173)} & \textcolor{teal}{(0.170)} \\ 
 Percent Hispanic & \textcolor{teal}{$-$2.811$^{***}$} & \textcolor{teal}{$-$2.829$^{***}$} \\ 
  & \textcolor{teal}{(0.765)} & \textcolor{teal}{(0.720)} \\ 
 Median household income & \textcolor{teal}{$-$0.657$^{***}$} & \textcolor{teal}{$-$0.634$^{***}$} \\ 
  & \textcolor{teal}{(0.154)} & \textcolor{teal}{(0.156)} \\ 
 Percent with limited English proficiency & \textcolor{teal}{$-$0.051} & \textcolor{teal}{$-$0.038} \\ 
  & \textcolor{teal}{(0.070)} & \textcolor{teal}{(0.069)} \\ 
 Percent unemployed  & \textcolor{teal}{0.128} & \textcolor{teal}{0.118} \\ 
  & \textcolor{teal}{(0.150)} & \textcolor{teal}{(0.143)} \\ 
 Percent with less than high school education & \textcolor{teal}{0.024} & \textcolor{teal}{0.020} \\ 
  & \textcolor{teal}{(0.102)} & \textcolor{teal}{(0.101)} \\ 
 \textcolor{teal}{COVID-19 deaths per 100k} & \textcolor{teal}{0.100$^{*}$} & \textcolor{teal}{0.106$^{*}$} \\ 
  & \textcolor{teal}{(0.057)} & \textcolor{teal}{(0.058)} \\ 
 \textcolor{teal}{Drug-overdose deaths per 100k} & \textcolor{teal}{$-$0.006} & \textcolor{teal}{0.014} \\ 
  & \textcolor{teal}{(0.072)} & \textcolor{teal}{(0.070)} \\ 
 \textcolor{teal}{Alcohol-related deaths per 100k} & \textcolor{teal}{0.598$^{***}$} & \textcolor{teal}{0.585$^{***}$} \\ 
  & \textcolor{teal}{(0.110)} & \textcolor{teal}{(0.111)} \\ 
 \textcolor{teal}{Poor mental health days} & \textcolor{teal}{0.093} & \textcolor{teal}{0.096} \\ 
  & \textcolor{teal}{(0.088)} & \textcolor{teal}{(0.088)} \\ 
 \textcolor{teal}{Firearm: waiting period} & \textcolor{teal}{0.011} & \textcolor{teal}{0.035} \\ 
  & \textcolor{teal}{(0.098)} & \textcolor{teal}{(0.103)} \\ 
 \textcolor{teal}{Firearm: child-access prevention} & \textcolor{teal}{$-$0.040} & \textcolor{teal}{0.032} \\ 
  & \textcolor{teal}{(0.166)} & \textcolor{teal}{(0.129)} \\ 
 \textcolor{teal}{Firearm: minimum purchase age 21} & \textcolor{teal}{$-$0.581$^{***}$} & \textcolor{teal}{$-$0.594$^{***}$} \\ 
  & \textcolor{teal}{(0.177)} & \textcolor{teal}{(0.172)} \\ 
 \hline \\[-1.8ex] 
Observations & 40,794 & 40,794 \\ 
R$^{2}$ & \textcolor{teal}{0.567} & \textcolor{teal}{0.567} \\ 
Adjusted R$^{2}$ & \textcolor{teal}{0.530} & \textcolor{teal}{0.530} \\ 
\hline 
\hline \\[-1.8ex] 
\multicolumn{3}{l}{\footnotesize Robust standard errors in parentheses. * p$<$0.1; ** p$<$0.05; *** p$<$0.01}
\end{tabular} 
\end{table}

\begin{figure}[htbp]
    \centering
    \includegraphics[width=0.8\textwidth]{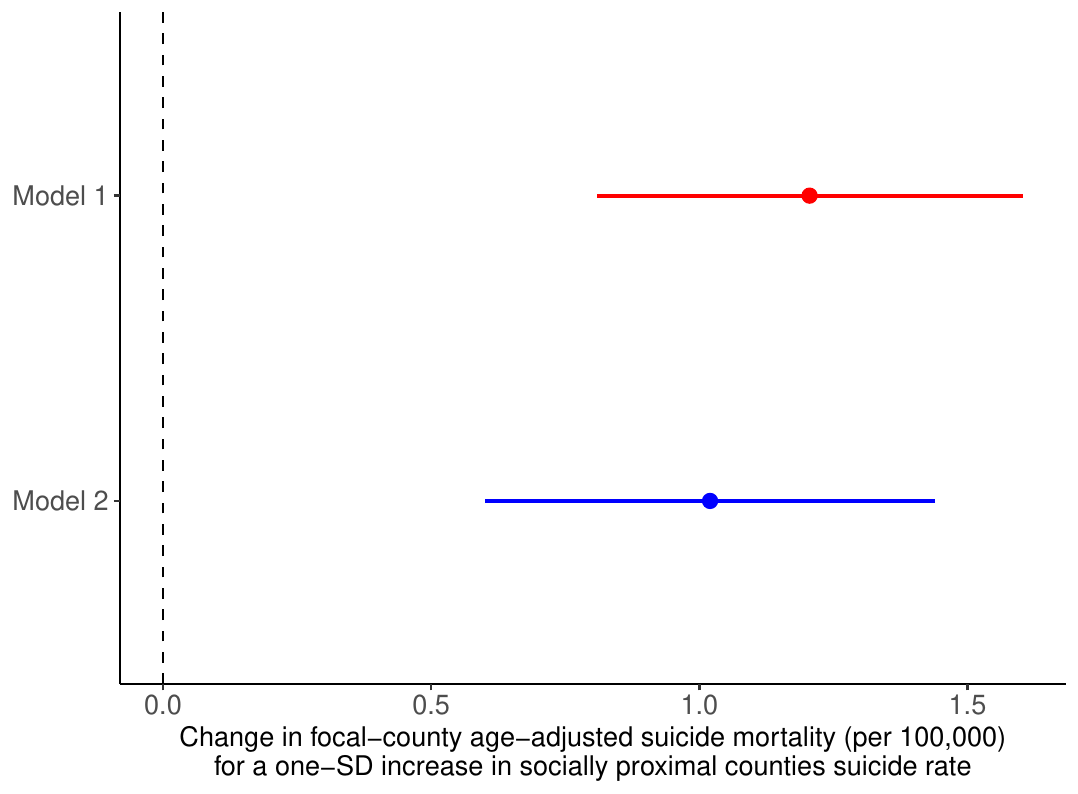}
    \caption{{\bf Role of social ties in county-level suicide mortality, with and without controlling for geographical proximity.} 
    Estimated regression coefficients ($\hat{\eta}_1$) for suicide mortality rates in socially connected counties ($\tilde{s}_{-it}$) in two models, using age standardization. Model 1 (red): without adjustment for deaths in spatial proximity ($\tilde{d}_{-it}$). Model 2 (blue): with adjustment for deaths in spatial proximity ($\tilde{d}_{-it}$). Horizontal lines denote 95\% confidence intervals (CI). The vertical dashed line indicates the null hypothesis ($\eta_1 = 0$). Point estimate in model 1: \textcolor{teal}{1.21} (95\% CI: [\textcolor{teal}{0.82}, \textcolor{teal}{1.59}]); point estimate in model 2: \textcolor{teal}{1.020} (95\% CI: [\textcolor{teal}{0.61}, \textcolor{teal}{1.43}]). All models include county and year fixed effects, time-varying county-level demographic and socioeconomic control variables\textcolor{teal}{, COVID-19 mortality, drug-overdose mortality, alcohol-related mortality, poor mental health days, and state firearm policies} (see Table~\ref{tab:sensitivity_socio_spatial}).
    }
    \label{fig:sensitivity_ci_peer_effects}
\end{figure}

\subsection*{ERPO Exposure: Related Evidence and Sensitivity Analyses}

Prior evaluations of the impact of intensified ERPO enforcement following high-profile violent incidents found significant reductions in firearm-related suicide mortality, from 7.5\% in Indiana to 13.7\% in Connecticut~\cite{kivisto2018effects}. A recent study examining responses from 4,583 ERPO respondents across California, Connecticut, Maryland, and Washington estimated that 17 to 23 ERPOs could prevent one suicide~\cite{swanson2024suicide}. Further, effectiveness was even greater when the petitions explicitly documented suicidal behavior, with only 13 to 18 ERPOs required to prevent one suicide. Additionally, county-level evidence from Oregon highlights the social aspects of ERPO effectiveness, indicating that approximately 73\% of petitions explicitly mentioned concerns about suicidal behavior~\cite{zeoli2021use}.

To assess the robustness of our findings to the outcome specification, we re-estimated the models evaluating the association between ERPO exposure and suicide mortality using \textit{age-adjusted suicide mortality rates} ($\tilde{y}_{it}$) as outcomes. This alternative outcome definition accounts for demographic heterogeneity between counties and helps to disentangle variation in suicide risk from variation in age structure. Age standardization was performed using direct standardization across 18 age groups defined by the NVSS age recode system, as explained above. Age-adjusted suicide mortality rates were expressed in terms of the number of suicide deaths per 100,000 people.

We estimated three fixed effects models. The first model considered the direct association with ERPO implementation. The second model considered the indirect association with social exposure to ERPOs, through social ties. The third model included both social and spatial ERPO exposure as regressors. The first model specification included county $(\phi_i)$ and year $(\gamma_t)$ fixed effects, while the second and third model specifications included county fixed effects and state-by-year fixed effects $(\gamma_{st})$.

The specification for the first model is as follows:
\begin{equation*}
\tilde{y}_{it} = \kappa_1 ERPO_{it} + \boldsymbol{\kappa}_2^\top \overline{X}_{it} + \phi_i + \gamma_t + \varepsilon_{it}.
\end{equation*}

The specification for the second model is as follows:
\begin{equation*}
\tilde{y}_{it} = \tau_1 {ERPO\ Social\ Exposure}_{it} + \boldsymbol{\tau}_2^\top \overline{X}_{it} + \phi_i + \gamma_{st} + \varepsilon_{it}.
\end{equation*}

The specification for the third model is as follows:
\begin{equation*}
\tilde{y}_{it} = \omega_1 {ERPO\ Social\ Exposure}_{it} + \omega_2 {ERPO\ Spatial\ Exposure}_{it} + \boldsymbol{\omega}_3^\top \overline{X}_{it} + \phi_i + \gamma_{st} + \varepsilon_{it}.
\end{equation*} In the above, $\tilde{y}_{it}$ is the age-adjusted suicide mortality rate in county $i$ and year $t$; ${ERPO\ Social\ Exposure}_{it}$ represents the standardized share of social ties to counties located in ERPO-implementing states; ${ERPO\ Spatial\ Exposure}_{it}$ corresponds to the inverse-distance-weighted spatial exposure to counties located in ERPO-implementing states; and $\overline{X}_{it}$ represents county- and year-specific demographic and socioeconomic control variables\textcolor{teal}{, together with COVID-19 mortality, drug-overdose mortality, alcohol-related mortality, and poor mental health days. The first model additionally includes the three state firearm policies; these are constant within each state--year and therefore cannot be estimated in the second and third models, where the state-by-year fixed effects absorb all state-level variation over time}.\\

\textbf{Direct association with ERPO implementation.} Using the first model, we estimated that ERPO implementation in a given state was associated with a reduction of $\widehat{\kappa} = \textcolor{teal}{-0.550}$ (cluster-robust 95\% CI: [\textcolor{teal}{-1.00}, \textcolor{teal}{-0.0964}]) deaths per 100,000 people at the county level. Of note, because the ERPO variable varies at the state--year level, we clustered standard errors by state to ensure conservative inference.

\textbf{Indirect association with ERPO implementation through social ties.} Using the second model, we estimated that a one-standard-deviation increase in social connectedness to counties located in ERPO-implementing states was associated with a reduction of $\widehat{\tau}_1 = \textcolor{teal}{-0.211}$ (cluster-robust 95\% CI: [\textcolor{teal}{-0.306}, \textcolor{teal}{-0.115}]) deaths per 100,000 people in a given focal county. This finding indicates that greater social proximity to counties with active ERPO policies is associated with lower age-adjusted suicide mortality, independently of local legislation.

\textbf{Social and spatial exposures to ERPO implementation.} When additionally controlling for geographical proximity to counties located in ERPO-implementing states in the third model, we estimated a reduction of $\widehat{\omega}_1 = \textcolor{teal}{-0.239}$ (cluster-robust 95\% CI: [\textcolor{teal}{-0.380}, \textcolor{teal}{-0.0972}]) deaths per 100,000 people for a one-standard-deviation increase in social connectedness to counties located in ERPO-implementing states. \textcolor{teal}{The coefficient on ERPO spatial exposure was positive and not statistically distinguishable from zero ($\widehat{\omega}_2 = 0.228$, standard error $0.268$).} The robustness of the estimated indirect association with social exposure to ERPOs indicates that it operates beyond geographical proximity to ERPO-implementing states.

\textbf{Comparison between results from main versus supplementary analyses.} \textcolor{teal}{The magnitude of the estimated indirect effects of ERPO implementation, through social ties, was comparable in our supplementary analyses using age-adjusted rather than crude suicide mortality rates. The directionality and significance of the results remained consistent with the main analysis. The findings were therefore robust to the choice of outcome, for both direct and indirect exposure to ERPO implementation.}

\begin{figure}[htbp]
    \centering
    \includegraphics[width=\linewidth]{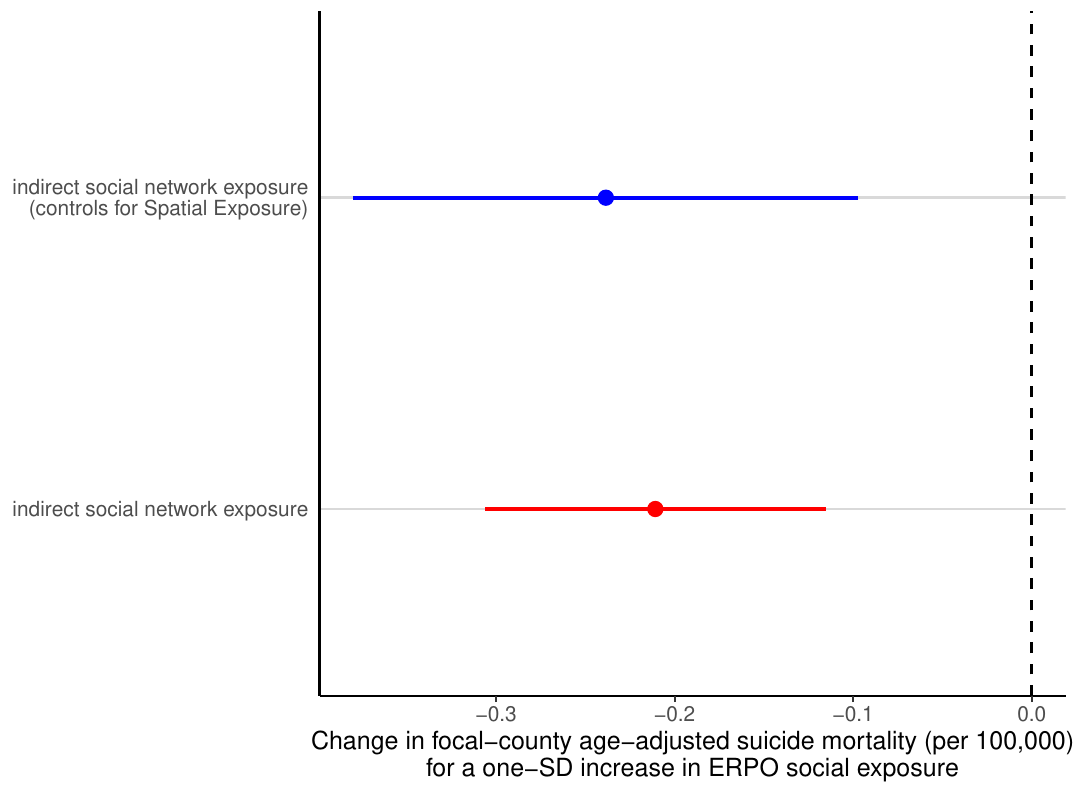}
    \caption{{\bf Indirect effects of ERPO implementation on suicide mortality, through social ties, with and without controlling for geographical proximity.} Estimated regression coefficients ($\hat{\tau}_1$, $\hat{\omega}_1$) for ${ERPO\ Social\ Exposure}$ in two models. Baseline model (red): without controlling for ${ERPO\ Spatial\ Exposure}$. Alternative model (blue): controlling for ${ERPO\ Spatial\ Exposure}$. Point estimate in baseline model: ($\hat{\tau}_1 = \textcolor{teal}{-0.211}$, cluster-robust 95\% CI: [$\textcolor{teal}{-0.306}$, $\textcolor{teal}{-0.115}$]); point estimate in alternative model: ($\hat{\omega}_1 = \textcolor{teal}{-0.239}$, cluster-robust 95\% CI: [$\textcolor{teal}{-0.380}$, $\textcolor{teal}{-0.0972}$]). 
    Horizontal lines denote 95\% confidence intervals (CI). The vertical dashed line indicates the null hypothesis ($\tau_1 = 0$). 
    Both models included county and state--year fixed effects ($\phi_i$, $\gamma_{st}$) as well as demographic and socioeconomic control variables ($\overline{X}_{it}$)\textcolor{teal}{, COVID-19 mortality, drug-overdose mortality, alcohol-related mortality, and poor mental health days (see Supplementary Table~\ref{si_policy_exposure}).}}
    \label{fig:si_erpo_social_robustness}
\end{figure}
\begin{table}[!htbp] \centering 
\scriptsize
  \caption{{\bf Estimated effects of ERPO policy exposure on county-level \textit{age-adjusted} suicide mortality rates (expressed in terms of the number of deaths per 100,000 people).} Column~(1) reports direct effects of local ERPO implementation with county and year fixed effects. Column~(2) reports indirect effects of ERPO social exposure, measured through inter-county social ties, estimated with county and state--year fixed effects. Column~(3) reports results from the indirect social exposure model, with an additional control for ERPO spatial exposure as well as county and state--year fixed effects. All models adjust for population density, racial composition (percent Asian, Black, and Other racial subgroups), ethnic composition (percent Hispanic), median household income, percent with limited English proficiency, percent unemployed, percent with less than high school education, and political affiliation\textcolor{teal}{, COVID-19 mortality, drug-overdose mortality, alcohol-related mortality, and poor mental health days. The state firearm policies (waiting periods, child-access prevention laws, and a minimum purchase age of 21) are constant within each state--year and therefore perfectly collinear with the state--year fixed effects; they can be estimated only in Column~(1)}. Standard errors are clustered at the state level.} 
  \label{si_policy_exposure} 
  \renewcommand{\arraystretch}{0.92}
  \setlength{\tabcolsep}{4pt}
  \tiny
\begin{tabular}{@{\extracolsep{1pt}}lccc} 
\hline 
 & \multicolumn{3}{c}{Outcome variable: county-level age-adjusted suicide mortality rate} \\ 
\cline{2-4} 
 & (1) & (2) & (3) \\ 
\hline \\[-1.8ex]
 ERPO & \textcolor{teal}{$-$0.550$^{**}$} &  &  \\ 
  & \textcolor{teal}{(0.226)} &  &  \\ 
  ERPO social exposure &  & \textcolor{teal}{$-$0.211$^{***}$} & \textcolor{teal}{$-$0.239$^{***}$} \\ 
  &  & \textcolor{teal}{(0.048)} & \textcolor{teal}{(0.070)} \\ 
  ERPO spatial exposure &  &  & \textcolor{teal}{0.228} \\ 
  &  &  & \textcolor{teal}{(0.268)} \\ 
  Population density & \textcolor{teal}{$-$1.315$^{***}$} & \textcolor{teal}{$-$0.532} & \textcolor{teal}{$-$0.560} \\ 
  & \textcolor{teal}{(0.391)} & \textcolor{teal}{(0.522)} & \textcolor{teal}{(0.506)} \\ 
  Percent Asian & \textcolor{teal}{$-$1.382$^{**}$} & \textcolor{teal}{$-$2.238$^{***}$} & \textcolor{teal}{$-$2.177$^{***}$} \\ 
  & \textcolor{teal}{(0.535)} & \textcolor{teal}{(0.617)} & \textcolor{teal}{(0.603)} \\ 
  Percent Black & \textcolor{teal}{$-$0.872$^{***}$} & \textcolor{teal}{$-$0.454$^{**}$} & \textcolor{teal}{$-$0.441$^{**}$} \\ 
  & \textcolor{teal}{(0.217)} & \textcolor{teal}{(0.180)} & \textcolor{teal}{(0.177)} \\ 
  Percent Other & \textcolor{teal}{0.228} & \textcolor{teal}{0.014} & \textcolor{teal}{0.002} \\ 
  & \textcolor{teal}{(0.189)} & \textcolor{teal}{(0.173)} & \textcolor{teal}{(0.175)} \\ 
  Percent Hispanic & \textcolor{teal}{$-$3.175$^{***}$} & \textcolor{teal}{$-$1.454$^{***}$} & \textcolor{teal}{$-$1.488$^{***}$} \\ 
  & \textcolor{teal}{(0.938)} & \textcolor{teal}{(0.542)} & \textcolor{teal}{(0.550)} \\ 
  Median household income  & \textcolor{teal}{$-$0.660$^{***}$} & \textcolor{teal}{$-$0.520$^{**}$} & \textcolor{teal}{$-$0.498$^{**}$} \\ 
  & \textcolor{teal}{(0.177)} & \textcolor{teal}{(0.220)} & \textcolor{teal}{(0.216)} \\ 
  Percent with limited English proficiency & \textcolor{teal}{$-$0.076} & \textcolor{teal}{$-$0.155$^{**}$} & \textcolor{teal}{$-$0.160$^{**}$} \\ 
  & \textcolor{teal}{(0.071)} & \textcolor{teal}{(0.073)} & \textcolor{teal}{(0.070)} \\ 
  Percent unemployed & \textcolor{teal}{0.162} & \textcolor{teal}{$-$0.059} & \textcolor{teal}{$-$0.056} \\ 
  & \textcolor{teal}{(0.164)} & \textcolor{teal}{(0.142)} & \textcolor{teal}{(0.143)} \\ 
  Percent with less than high school education & \textcolor{teal}{0.035} & \textcolor{teal}{$-$0.104} & \textcolor{teal}{$-$0.117} \\ 
  & \textcolor{teal}{(0.124)} & \textcolor{teal}{(0.116)} & \textcolor{teal}{(0.116)} \\ 
  Political affiliation & \textcolor{teal}{0.022} & \textcolor{teal}{0.116} & \textcolor{teal}{0.102} \\ 
  & \textcolor{teal}{(0.159)} & \textcolor{teal}{(0.146)} & \textcolor{teal}{(0.149)} \\ 
  \textcolor{teal}{COVID-19 deaths per 100k} & \textcolor{teal}{0.117$^{*}$} & \textcolor{teal}{0.189$^{*}$} & \textcolor{teal}{0.186$^{*}$} \\ 
  & \textcolor{teal}{(0.063)} & \textcolor{teal}{(0.097)} & \textcolor{teal}{(0.098)} \\ 
  \textcolor{teal}{Drug-overdose deaths per 100k} & \textcolor{teal}{$-$0.057} & \textcolor{teal}{0.071} & \textcolor{teal}{0.076} \\ 
  & \textcolor{teal}{(0.075)} & \textcolor{teal}{(0.070)} & \textcolor{teal}{(0.070)} \\ 
  \textcolor{teal}{Alcohol-related deaths per 100k} & \textcolor{teal}{0.679$^{***}$} & \textcolor{teal}{0.456$^{***}$} & \textcolor{teal}{0.454$^{***}$} \\ 
  & \textcolor{teal}{(0.125)} & \textcolor{teal}{(0.113)} & \textcolor{teal}{(0.113)} \\ 
  \textcolor{teal}{Poor mental health days} & \textcolor{teal}{0.064} & \textcolor{teal}{0.028} & \textcolor{teal}{0.024} \\ 
  & \textcolor{teal}{(0.089)} & \textcolor{teal}{(0.103)} & \textcolor{teal}{(0.104)} \\ 
  \textcolor{teal}{Firearm: waiting period} & \textcolor{teal}{0.106} &  &  \\ 
  & \textcolor{teal}{(0.131)} &  &  \\ 
  \textcolor{teal}{Firearm: child-access prevention} & \textcolor{teal}{0.117} &  &  \\ 
  & \textcolor{teal}{(0.190)} &  &  \\ 
  \textcolor{teal}{Firearm: minimum purchase age 21} & \textcolor{teal}{$-$0.315} &  &  \\ 
  & \textcolor{teal}{(0.217)} &  &  \\ 
 \hline \\[-1.8ex] 
Observations & 40,794 & 40,794 & 40,794 \\ 
R$^{2}$ & \textcolor{teal}{0.565} & \textcolor{teal}{0.576} & \textcolor{teal}{0.577} \\ 
Adjusted R$^{2}$ & \textcolor{teal}{0.529} & \textcolor{teal}{0.533} & \textcolor{teal}{0.534} \\ 
\hline 
\multicolumn{4}{l}{\tiny Robust standard errors in parentheses. * p$<$0.1; ** p$<$0.05; *** p$<$0.01} \\
\end{tabular} 
\end{table}

\subsection*{Robustness Test: Controlling for Deaths in Social Proximity}

To further validate our findings, we conducted a robustness test by building a model that additionally controls for deaths in socially connected counties. This alternative specification helps address potential confounding that arises from the correlated suicide mortality rates in the social network. Specifically, social connectedness may serve as a proxy for shared environmental characteristics prone to elevated suicide risk or common cultural and structural determinants of suicide that operate independently of ERPO policies. Without accounting for these baseline correlations in suicide mortality, the estimated effects of ERPO social exposure could reflect pre-existing social network-level patterns rather than true policy impacts.

We estimated two fixed effects models. The first model considered crude suicide mortality rates as outcomes, while the second model considered age-adjusted mortality rates as outcomes. Both models included standardized social and spatial ERPO exposure variables. Additionally, both models included county fixed effects $(\phi_i)$ and state-by-year fixed effects $(\gamma_{st})$, with standard errors clustered at the state level. The control variables for deaths in social proximity ($s_{-it}$ in the first model and $\tilde{s}_{-it}$ in the second model) correspond to the SCI-weighted average of contemporaneous suicide mortality in socially connected counties, excluding the focal county itself. By adjusting for these network-level suicide mortality patterns, we can isolate the effect of ERPO social exposure.

The regression equation for the first model is as follows:
\begin{equation}
\label{eq:crude_spill}
\begin{aligned}
y_{it}
= \alpha_{1}\, ERPO\ Social\ Exposure_{it}
+ \alpha_{2}\, ERPO\ Spatial\ Exposure_{it} \\
+ \alpha_{3}\, s_{-it} + \boldsymbol{\alpha}_{4}^{\top}\, \overline{X}_{it}
+ \phi_{i}
+ \gamma_{st}
+ \varepsilon_{it}.  
\end{aligned}
\end{equation}

The regression equation for the second model is as follows:
\begin{equation}
\label{eq:adj_spill}
\begin{aligned}
\tilde{y}_{it} &= \vartheta_{1}\, {ERPO\ Social\ Exposure}_{it}
               + \vartheta_{2}\, {ERPO\ Spatial\ Exposure}_{it} \\
               &\quad + \vartheta_{3}\, \tilde{s}_{-it}
               + \boldsymbol{\vartheta}_{4}^{\top}\, \overline{X}_{it}
               + \phi_{i} + \gamma_{st} + \varepsilon_{it},
\end{aligned}
\end{equation} where $y_{it}$ is the crude suicide mortality rate in county $i$ and year $t$ (expressed in terms of the number of deaths per 100{,}000 people); $\tilde{y}_{it}$ is the corresponding age-adjusted suicide mortality rate; $ERPO\ Social\ Exposure_{it}$ and $ERPO\ Spatial\ Exposure_{it}$ are social and spatial ERPO exposure variables given by Equations (5) and (7) in \textit{Methods}; $s_{-it}$ is the SCI–weighted average of \emph{crude} suicide deaths in socially connected counties ($j\neq i$); $\tilde{s}_{-it}$ is the corresponding \emph{age-adjusted} variable; and $\overline{X}_{it}$ denotes time-varying county-level demographic and socioeconomic control variables\textcolor{teal}{, together with COVID-19 mortality, drug-overdose mortality, alcohol-related mortality, and poor mental health days}. All specifications include county fixed effects $(\phi_i)$ and state-by-year fixed effects $(\gamma_{st})$; inference uses robust standard errors clustered by state.

The negative association between suicide mortality in a given focal county and ERPO social exposure remained robust after controlling for contemporaneous suicide mortality in socially connected counties. As shown in Figure~\ref{fig:si_erpo_social_spill} and Tables~\ref{tab:si_erpo_social_control} and~\ref{tab:si_erpo_social_control_age_adjusted}, estimated coefficients for the ERPO social exposure variable were consistently negative and statistically significant, irrespective of whether crude suicide mortality rates ($\hat{\alpha}_1 = \textcolor{teal}{-0.280}$, 95\% CI: [$\textcolor{teal}{-0.440}$, $\textcolor{teal}{-0.119}$]) or age-adjusted suicide mortality rates ($\hat{\vartheta}_1 = \textcolor{teal}{-0.256}$, 95\% CI: [$\textcolor{teal}{-0.405}$, $\textcolor{teal}{-0.106}$]) were used as outcomes. Notably, the association between suicide mortality in a given focal county and deaths in social proximity was statistically significant in the model using age-adjusted suicide mortality rates ($\hat{\vartheta}_3 = \textcolor{teal}{-0.455}$, $p < 0.05$) but not in the model using crude suicide mortality rates. This divergence likely reflects the role of the age standardization procedure, which removes heterogeneity in county-level age composition and can thus help better reveal underlying correlations in suicide mortality across socially connected counties that may otherwise be masked by differing age structures. Critically, the inclusion of this control variable did not materially alter the magnitude or significance of the association between suicide mortality and ERPO social exposure under any model specification. The robustness of our findings suggests that indirect exposure to ERPOs through social ties independently explains reductions in suicide mortality in counties located in non-implementing states.

\begin{figure}[htbp]
    \centering
    \includegraphics[width=\linewidth]{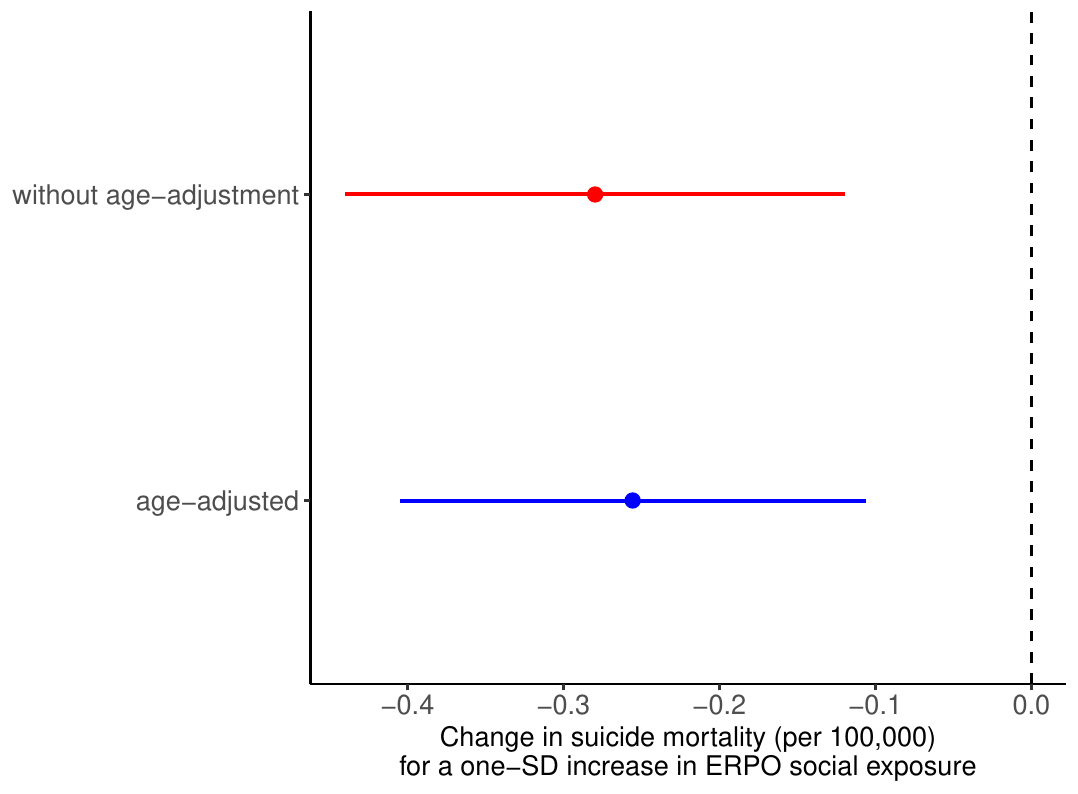}
    \caption{{\bf Role of social ties in county-level suicide mortality, with and without age standardization.} Estimated regression coefficients ($\hat{\alpha}_1$, $\hat{\vartheta}_1$) for ${ERPO\ Social\ Exposure}$ in two models, both controlling for deaths in social proximity. 
Baseline model (red): without age standardization. Alternative model (blue): with age standardization. Point estimate in baseline model: ($\hat{\alpha}_1 = \textcolor{teal}{-0.280}$, cluster-robust 95\% CI: [$\textcolor{teal}{-0.440}$, $\textcolor{teal}{-0.119}$]). Point estimate in alternative model: ($\hat{\vartheta}_1 = \textcolor{teal}{-0.256}$, cluster-robust 95\% CI: [$\textcolor{teal}{-0.405}$, $\textcolor{teal}{-0.106}$]). Horizontal lines denote 95\% confidence intervals (CI). The vertical dashed line indicates the null hypothesis. 
Both models include county and state-by-year fixed effects ($\phi_i$, $\gamma_{st}$), demographic and socioeconomic control variables ($\overline{X}_{it}$)\textcolor{teal}{, COVID-19 mortality, drug-overdose mortality, alcohol-related mortality, and poor mental health days}, and a variable characterizing deaths in social proximity ($s_{-it}$ or $\tilde{s}_{-it}$){; see Tables~\ref{tab:si_erpo_social_control}  and~\ref{tab:si_erpo_social_control_age_adjusted}}.
}
\label{fig:si_erpo_social_spill}
\end{figure}

\begin{table}[!htbp] \centering 
  \caption{\textbf{Estimated effects of ERPO policy exposure on county-level \emph{crude} suicide mortality rates, additionally controlling for suicide deaths in social proximity.} \textcolor{teal}{This specification extends Column~(3) of Table~\ref{policy_exposure}  in the main manuscript by including deaths in social proximity $s_{-it}$ as a regressor, to assess whether the indirect ERPO association is confounded by correlated suicide mortality across the social network.} The model includes county and state--year fixed effects and adjusts for population density, age distribution (percent aged 0--17, 18--44 and 45--64), racial composition (percent Asian, Black, and Other racial subgroups), ethnic composition (percent Hispanic), median household income, percent with limited English proficiency, percent unemployed, percent with less than high school education, and political affiliation\textcolor{teal}{, COVID-19 mortality, drug-overdose mortality, alcohol-related mortality, and poor mental health days}. Standard errors are clustered at the state level.}
  \label{tab:si_erpo_social_control} 
  \tiny
\begin{tabular}{@{\extracolsep{1pt}}lc} 
\\[-1.8ex]\hline 
\hline \\[-1.8ex] 
  \multicolumn{2}{c}{\textbf{Outcome variable:} county-level crude suicide mortality rate (per 100,000 people)} \\ 
\cline{2-2} 
\hline \\[-1.8ex] 
 ERPO social exposure & \textcolor{teal}{$-$0.280$^{***}$} \\ 
  & \textcolor{teal}{(0.080)} \\ 
  ERPO spatial exposure & \textcolor{teal}{0.444} \\ 
  & \textcolor{teal}{(0.311)} \\ 
  Deaths in social proximity $s_{-it}$ & \textcolor{teal}{$-$0.530} \\ 
  & \textcolor{teal}{(1.145)} \\ 
  Population density & \textcolor{teal}{$-$0.568} \\ 
  & \textcolor{teal}{(0.567)} \\ 
  Percent aged 0-17 & \textcolor{teal}{$-$0.210} \\ 
  & \textcolor{teal}{(0.358)} \\ 
  Percent aged 18-44 & \textcolor{teal}{$-$0.176} \\ 
  & \textcolor{teal}{(0.547)} \\ 
  Percent aged 45-64 & \textcolor{teal}{$-$0.540} \\ 
  & \textcolor{teal}{(0.387)} \\ 
  Percent Asian & \textcolor{teal}{$-$0.618$^{**}$} \\ 
  & \textcolor{teal}{(0.260)} \\ 
  Percent Black & \textcolor{teal}{$-$2.757$^{***}$} \\ 
  & \textcolor{teal}{(0.681)} \\ 
  Percent Other & \textcolor{teal}{$-$0.096} \\ 
  & \textcolor{teal}{(0.279)} \\ 
  Percent Hispanic & \textcolor{teal}{$-$2.765$^{***}$} \\ 
  & \textcolor{teal}{(0.810)} \\ 
  Median household income  & \textcolor{teal}{$-$0.568$^{***}$} \\ 
  & \textcolor{teal}{(0.180)} \\ 
  Percent with limited English proficiency & \textcolor{teal}{$-$0.056} \\ 
  & \textcolor{teal}{(0.128)} \\ 
  Percent unemployed & \textcolor{teal}{$-$0.093} \\ 
  & \textcolor{teal}{(0.125)} \\ 
  Percent with less than high school education & \textcolor{teal}{$-$0.092} \\ 
  & \textcolor{teal}{(0.128)} \\ 
  Political affiliation & \textcolor{teal}{0.130} \\ 
  & \textcolor{teal}{(0.155)} \\ 
  \textcolor{teal}{COVID-19 deaths per 100k} & \textcolor{teal}{0.134} \\ 
  & \textcolor{teal}{(0.108)} \\ 
  \textcolor{teal}{Drug-overdose deaths per 100k} & \textcolor{teal}{$-$0.045} \\ 
  & \textcolor{teal}{(0.087)} \\ 
  \textcolor{teal}{Alcohol-related deaths per 100k} & \textcolor{teal}{0.227$^{*}$} \\ 
  & \textcolor{teal}{(0.119)} \\ 
  \textcolor{teal}{Poor mental health days} & \textcolor{teal}{0.113} \\ 
  & \textcolor{teal}{(0.101)} \\ 
 \hline \\[-1.8ex] 
Observations & 40,794 \\ 
R$^{2}$ & 0.947 \\ 
Adjusted R$^{2}$ & 0.941 \\ 
\hline 
\hline \\[-1.8ex] 
\multicolumn{1}{l}{\footnotesize Robust standard errors in parentheses. * p$<$0.1; ** p$<$0.05; *** p$<$0.01}
\end{tabular} 
\end{table}

\begin{table}[!htbp] \centering 
  \caption{\textbf{Estimated effects of ERPO policy exposure on county-level \emph{age-adjusted} suicide mortality (expressed in terms of the number of deaths per 100,000 people).} The model includes county and state--year fixed effects. The model adjusts for deaths in social proximity $\widetilde{s}_{-it}$ population density, racial composition (percent Asian, Black, and Other racial subgroups), ethnic composition (percent Hispanic), median household income, percent with limited English proficiency, percent unemployed, percent with less than high school education, and political affiliation\textcolor{teal}{, as well as COVID-19 mortality, drug-overdose mortality, alcohol-related mortality, and poor mental health days}. Standard errors are clustered at the state level.}
  \label{tab:si_erpo_social_control_age_adjusted} 
  \tiny
\begin{tabular}{@{\extracolsep{1pt}}lc} 
\\[-1.8ex]\hline 
\hline \\[-1.8ex] 
  \multicolumn{2}{c}{\textbf{Outcome variable:} county-level age-adjusted suicide mortality rate (per 100,000 people)} \\
\cline{2-2} 
\hline \\[-1.8ex] 
 ERPO Social Exposure & \textcolor{teal}{$-$0.256$^{***}$} \\ 
  & \textcolor{teal}{(0.074)} \\ 
  ERPO Spatial Exposure & \textcolor{teal}{0.225} \\ 
  & \textcolor{teal}{(0.282)} \\ 
  Deaths in social proximity $\tilde{s}_{-it}$ & \textcolor{teal}{$-$0.455$^{**}$} \\ 
  & \textcolor{teal}{(0.204)} \\ 
  Population density & \textcolor{teal}{$-$0.589} \\ 
  & \textcolor{teal}{(0.523)} \\ 
  Percent Asian & \textcolor{teal}{$-$2.195$^{***}$} \\ 
  & \textcolor{teal}{(0.610)} \\ 
  Percent Black & \textcolor{teal}{$-$0.479$^{**}$} \\ 
  & \textcolor{teal}{(0.180)} \\ 
  Percent Other & \textcolor{teal}{$-$0.010} \\ 
  & \textcolor{teal}{(0.179)} \\ 
  Percent Hispanic & \textcolor{teal}{$-$1.570$^{***}$} \\ 
  & \textcolor{teal}{(0.585)} \\ 
  Median household income  & \textcolor{teal}{$-$0.508$^{**}$} \\ 
  & \textcolor{teal}{(0.217)} \\ 
  Percent with limited English proficiency & \textcolor{teal}{$-$0.163$^{**}$} \\ 
  & \textcolor{teal}{(0.071)} \\ 
  Percent unemployed & \textcolor{teal}{$-$0.051} \\ 
  & \textcolor{teal}{(0.144)} \\ 
  Percent with less than high school education & \textcolor{teal}{$-$0.119} \\ 
  & \textcolor{teal}{(0.118)} \\ 
  Political Affiliation & \textcolor{teal}{0.108} \\ 
  & \textcolor{teal}{(0.149)} \\ 
  \textcolor{teal}{COVID-19 deaths per 100k} & \textcolor{teal}{0.196$^{*}$} \\ 
  & \textcolor{teal}{(0.100)} \\ 
  \textcolor{teal}{Drug-overdose deaths per 100k} & \textcolor{teal}{0.069} \\ 
  & \textcolor{teal}{(0.070)} \\ 
  \textcolor{teal}{Alcohol-related deaths per 100k} & \textcolor{teal}{0.467$^{***}$} \\ 
  & \textcolor{teal}{(0.115)} \\ 
  \textcolor{teal}{Poor mental health days} & \textcolor{teal}{0.018} \\ 
  & \textcolor{teal}{(0.104)} \\ 
 \hline \\[-1.8ex] 
Observations & 40,794 \\ 
R$^{2}$ & 0.577 \\ 
Adjusted R$^{2}$ & 0.534 \\ 
\hline 
\hline \\[-1.8ex] 
\multicolumn{1}{l}{\footnotesize Robust standard errors in parentheses. * p$<$0.1; ** p$<$0.05; *** p$<$0.01}
\end{tabular} 
\end{table}

\subsection*{Robustness to the Updated Social Connectedness Index}
\label{sci_2025}

To assess the sensitivity of our findings to the measurement of social ties, we repeated our main analyses using the updated 2026 release of the SCI. The results are reported in Table~\ref{tab:socio_spatial_model_sci_2025} and Table~\ref{policy_exposure_sci_2025}, with corresponding coefficient plots shown in Figures~\ref{fig:ci_peer_effects_2025} and~\ref{fig:erpo_social_robustness_sci_2025}.

The estimated association between suicide mortality and deaths in socially connected counties remains positive and statistically significant, both without adjustment for spatial proximity (Model~1R: $\hat{\zeta}_1 = \textcolor{teal}{2.75}$, 95\% CI: [\textcolor{teal}{1.36}, \textcolor{teal}{4.14}]) and after controlling for deaths in spatially proximate counties (Model~2R: $\hat{\zeta}_1 = \textcolor{teal}{2.49}$, 95\% CI: [\textcolor{teal}{0.97}, \textcolor{teal}{4.01}]). These estimates are closely aligned with those obtained using the 2021 SCI data, indicating that the role of social ties in county-level suicide mortality is not an artifact of a particular vintage of the connectedness measure.

The estimated associations of indirect ERPO policy exposure through social networks are likewise robust to the updated SCI. The baseline social exposure estimate ($\hat{\delta}_1 = \textcolor{teal}{-0.181}$) and the estimate controlling for spatial exposure ($\hat{\theta}_1 = \textcolor{teal}{-0.259}$) both remain negative and statistically significant, while spatial exposure itself is not significant. The direction, magnitude, and statistical significance of all key coefficients are substantively unchanged relative to the main specification.

These results indicate that our core findings, concerning the association between social connectedness and county-level suicide mortality and the negative association of indirect ERPO exposure through social ties, are robust to measurement updates in the underlying social network data.

\begin{figure}[htbp]
    \centering
    \includegraphics[width=0.8\textwidth]{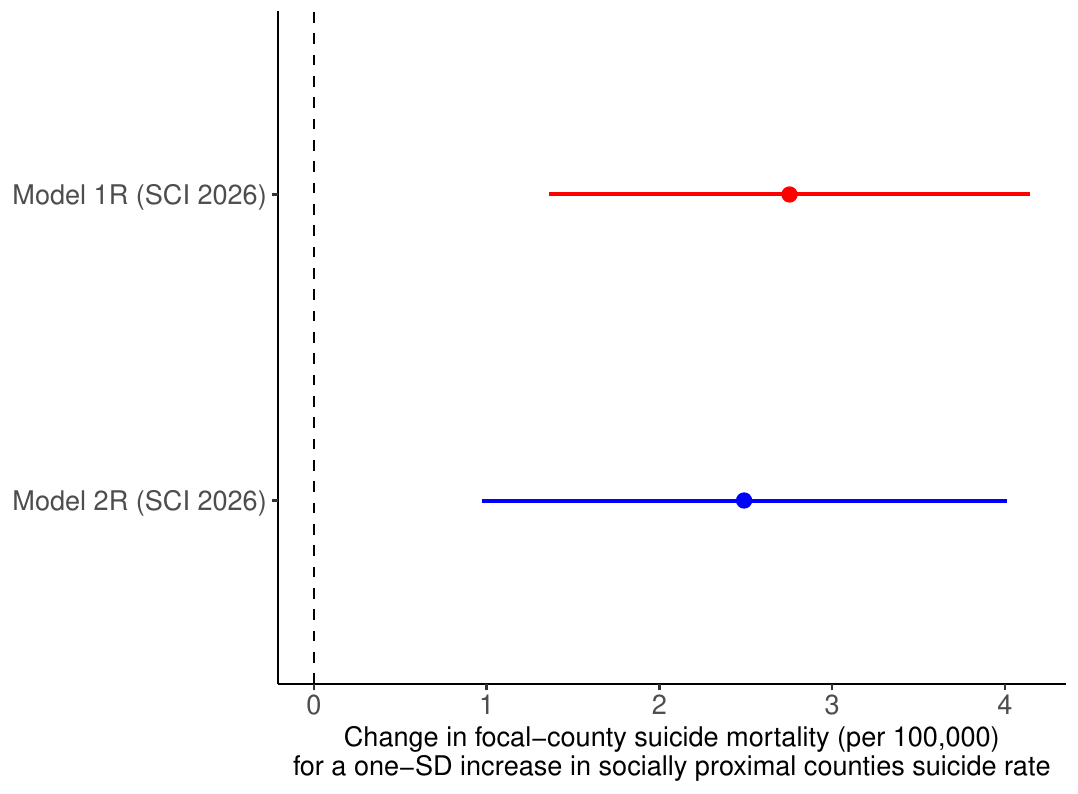}
    \caption{{\bf Role of social ties in county-level suicide mortality.}
    Estimated regression coefficients ($\hat\zeta_1$) for suicide mortality rates in socially connected counties ($s_{-it}$) in two models. Model 1R (SCI 2026) (red): without adjustment for deaths in spatial proximity ($d_{-it}$). Model 2R (SCI 2026) (blue): with adjustment for deaths in spatial proximity ($d_{-it}$). Horizontal lines denote 95\% confidence intervals (CI). The vertical dashed line indicates the null hypothesis ($\zeta_1 = 0$). Point estimate in model 1: \textcolor{teal}{2.75} (cluster-robust 95\% CI: [\textcolor{teal}{1.36}, \textcolor{teal}{4.14}]); point estimate in model 2: \textcolor{teal}{2.49} (cluster-robust 95\% CI: [\textcolor{teal}{0.97}, \textcolor{teal}{4.01}]). Both models include county and year fixed effects and sociodemographic control variables (see Table \ref{tab:socio_spatial_model_sci_2025}).
    }
    \label{fig:ci_peer_effects_2025}
\end{figure}

\begin{table}[!htbp] \centering 
\scriptsize
  \caption{{\bf Estimates of socio-spatial correlates of county-level suicide mortality obtained via two-way fixed effects regressions.} In Model~1R (SCI 2026), county-level suicide mortality rates are regressed on standardized deaths in social proximity ($s_{-it}$). Model~ 2R (SCI 2026) additionally controls for standardized deaths in spatial proximity ($d_{-it}$) to disentangle the role of social ties from that of geographical proximity. Both models include county and year fixed effects and adjust for time-varying county-level characteristics: population density, age distribution (percent aged 0--17, 18--44 and 45--64), racial composition (percent Asian, Black, and Other racial subgroups), ethnic composition (percent Hispanic), median household income, percent with limited English proficiency, percent unemployed, and percent with less than high school education\textcolor{teal}{, as well as COVID-19 mortality, drug-overdose mortality, alcohol-related mortality, poor mental health days, and state-level firearm policies (waiting periods, child-access prevention laws, and minimum purchase age of 21)}. Standard errors are clustered at the state level.} 
  \label{tab:socio_spatial_model_sci_2025} 
  \tiny
\begin{tabular}{@{\extracolsep{1pt}}lcc} 
\\[-1.8ex]\hline 
\hline \\[-1.8ex] 
 & \multicolumn{2}{c}{Outcome variable: county-level crude suicide mortality rate} \\ 
\cline{2-3} 
 & Model 1R (SCI 2026) & Model 2R (SCI 2026) \\ 
\hline \\[-1.8ex] 
 Deaths in social proximity $s_{-it}$ & \textcolor{teal}{2.753$^{***}$} & \textcolor{teal}{2.490$^{***}$} \\ 
  & \textcolor{teal}{(0.692)} & \textcolor{teal}{(0.756)} \\ 
  Deaths in spatial proximity $d_{-it}$ &  & \textcolor{teal}{0.384} \\ 
  &  & \textcolor{teal}{(0.331)} \\ 
  Population density & \textcolor{teal}{$-$0.638$^{*}$} & \textcolor{teal}{$-$0.559} \\ 
  & \textcolor{teal}{(0.375)} & \textcolor{teal}{(0.360)} \\ 
  Percent aged below 18 & \textcolor{teal}{0.003} & \textcolor{teal}{0.009} \\ 
  & \textcolor{teal}{(0.369)} & \textcolor{teal}{(0.365)} \\ 
  Percent aged 18-44 & \textcolor{teal}{0.036} & \textcolor{teal}{$-$0.057} \\ 
  & \textcolor{teal}{(0.490)} & \textcolor{teal}{(0.498)} \\ 
  Percent aged 45-64 & \textcolor{teal}{$-$0.420} & \textcolor{teal}{$-$0.434} \\ 
  & \textcolor{teal}{(0.359)} & \textcolor{teal}{(0.360)} \\ 
  Percent Asian & \textcolor{teal}{$-$0.646$^{***}$} & \textcolor{teal}{$-$0.637$^{***}$} \\ 
  & \textcolor{teal}{(0.233)} & \textcolor{teal}{(0.233)} \\ 
  Percent Black & \textcolor{teal}{$-$1.919$^{**}$} & \textcolor{teal}{$-$1.932$^{**}$} \\ 
  & \textcolor{teal}{(0.878)} & \textcolor{teal}{(0.884)} \\ 
  Percent Other & \textcolor{teal}{0.387$^{*}$} & \textcolor{teal}{0.360$^{*}$} \\ 
  & \textcolor{teal}{(0.203)} & \textcolor{teal}{(0.190)} \\ 
  Percent Hispanic & \textcolor{teal}{$-$2.960$^{***}$} & \textcolor{teal}{$-$2.916$^{***}$} \\ 
  & \textcolor{teal}{(0.837)} & \textcolor{teal}{(0.857)} \\ 
  Median household income  & \textcolor{teal}{$-$0.480$^{***}$} & \textcolor{teal}{$-$0.479$^{**}$} \\ 
  & \textcolor{teal}{(0.177)} & \textcolor{teal}{(0.180)} \\ 
  Percent with limited English proficiency & \textcolor{teal}{$-$0.021} & \textcolor{teal}{$-$0.007} \\ 
  & \textcolor{teal}{(0.072)} & \textcolor{teal}{(0.068)} \\ 
  Percent unemployed & \textcolor{teal}{0.047} & \textcolor{teal}{0.051} \\ 
  & \textcolor{teal}{(0.158)} & \textcolor{teal}{(0.157)} \\ 
  Percent with less than high school education & \textcolor{teal}{$-$0.0001} & \textcolor{teal}{$-$0.003} \\ 
  & \textcolor{teal}{(0.121)} & \textcolor{teal}{(0.121)} \\ 
  \textcolor{teal}{COVID-19 deaths per 100k} & \textcolor{teal}{0.067} & \textcolor{teal}{0.065} \\ 
  & \textcolor{teal}{(0.066)} & \textcolor{teal}{(0.065)} \\ 
  \textcolor{teal}{Drug-overdose deaths per 100k} & \textcolor{teal}{$-$0.939} & \textcolor{teal}{$-$0.915} \\ 
  & \textcolor{teal}{(1.011)} & \textcolor{teal}{(1.027)} \\ 
  \textcolor{teal}{Alcohol-related deaths per 100k} & \textcolor{teal}{9.529$^{***}$} & \textcolor{teal}{9.496$^{***}$} \\ 
  & \textcolor{teal}{(3.246)} & \textcolor{teal}{(3.271)} \\ 
  \textcolor{teal}{Poor mental health days} & \textcolor{teal}{0.127$^{*}$} & \textcolor{teal}{0.132$^{*}$} \\ 
  & \textcolor{teal}{(0.073)} & \textcolor{teal}{(0.073)} \\ 
  \textcolor{teal}{Firearm: waiting period} & \textcolor{teal}{0.338$^{*}$} & \textcolor{teal}{0.376$^{**}$} \\ 
  & \textcolor{teal}{(0.174)} & \textcolor{teal}{(0.156)} \\ 
  \textcolor{teal}{Firearm: child-access prevention} & \textcolor{teal}{$-$0.067} & \textcolor{teal}{$-$0.021} \\ 
  & \textcolor{teal}{(0.158)} & \textcolor{teal}{(0.160)} \\ 
  \textcolor{teal}{Firearm: minimum purchase age 21} & \textcolor{teal}{$-$0.303} & \textcolor{teal}{$-$0.289} \\ 
  & \textcolor{teal}{(0.193)} & \textcolor{teal}{(0.189)} \\ 
 \hline \\[-1.8ex] 
Observations & 40,794 & 40,794 \\ 
R$^{2}$ & \textcolor{teal}{0.949} & \textcolor{teal}{0.949} \\ 
Adjusted R$^{2}$ & \textcolor{teal}{0.945} & \textcolor{teal}{0.945} \\ 
\hline 
\hline \\[-1.8ex] 
\end{tabular} 
\end{table}

\begin{figure}[htbp]
    \centering
    \includegraphics[width=\linewidth]{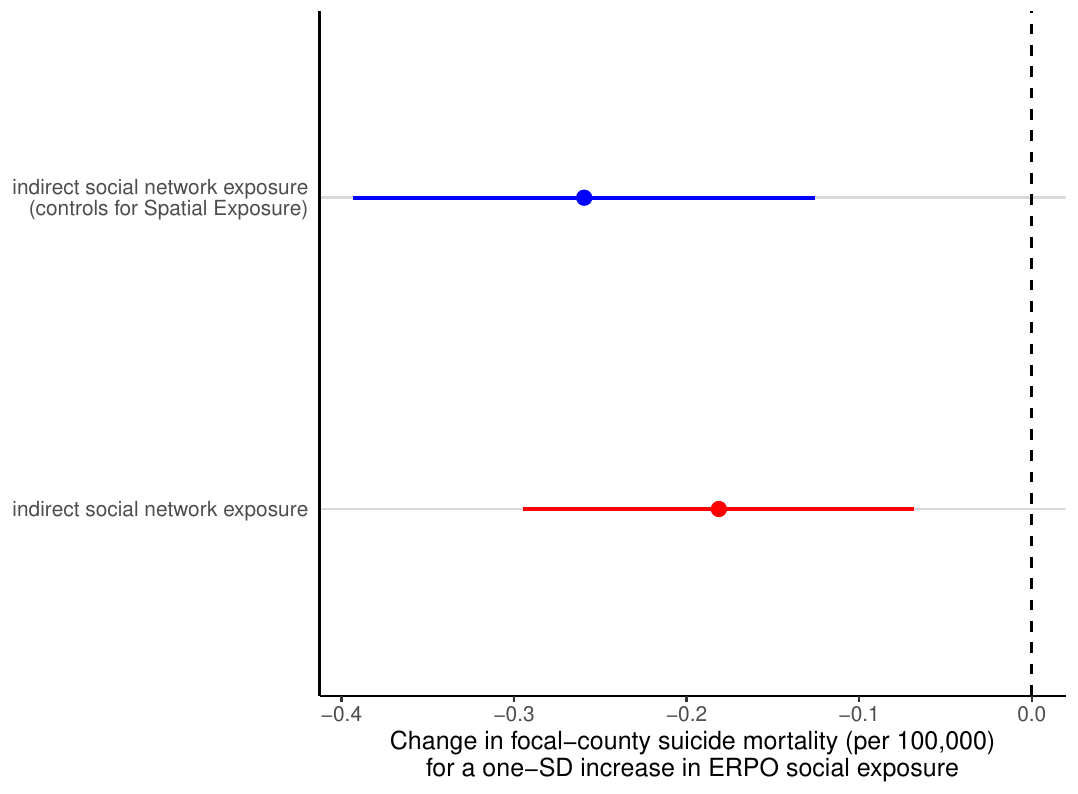}
    \caption{\textbf{Estimated coefficients ($\hat{\delta}_1$, $\hat{\theta}_1$) for ERPO social exposure in two specifications.} 
    Red point indicates estimate from the baseline model without spatial exposure ($\hat{\delta}_1 = \textcolor{teal}{-0.181}$, cluster-robust 95\% CI: [$\textcolor{teal}{-0.295}$, $\textcolor{teal}{-0.0680}$]); 
    blue point indicates  estimate from the specification controlling for $\textit{ERPO Spatial Exposure}_{it}$ ($\hat{\theta}_1 = \textcolor{teal}{-0.259}$, cluster-robust 95\% CI: [$\textcolor{teal}{-0.393}$, $\textcolor{teal}{-0.126}$]). 
    Horizontal lines denote 95\% confidence intervals; vertical dashed line denotes the null hypothesis ($\delta_1 = 0$). 
    Both models include county and state--year fixed effects ($\phi_i$, $\gamma_{st}$) and sociodemographic controls ($\overline{X}_{it}$). Consistent negative and statistically significant estimates indicate the association between suicide mortality and indirect social exposure to ERPO policies is robust to spatial confounding.}
    \label{fig:erpo_social_robustness_sci_2025}
\end{figure}

\begin{table}[p] 
\thisfloatpagestyle{empty}
\centering 
\scriptsize
\caption{\textbf{Estimated effects of ERPO policy exposure using 2026 SCI data on county-level crude suicide mortality rates (expressed in terms of the number of deaths per 100,000 people).} Column~(1) reports direct effects of local ERPO implementation with county and year fixed effects. Column~(2) reports indirect effects of ERPO social exposure, measured through inter-county social ties, estimated with county and state--year fixed effects. Column~(3) reports results from the indirect social exposure model, with an additional control for ERPO spatial exposure as well as county and state--year fixed effects. All models adjust for population density, age distribution (percent aged 0--17, 18--44 and 45--64), racial composition (percent Asian, Black, and Other racial subgroups), ethnic composition (percent Hispanic), median household income, percent with limited English proficiency, percent unemployed, percent with less than high school education, and political affiliation\textcolor{teal}{, COVID-19 mortality, drug-overdose mortality, alcohol-related mortality, and poor mental health days. The state firearm policies (waiting periods, child-access prevention laws, and a minimum purchase age of 21) are constant within each state--year and therefore perfectly collinear with the state--year fixed effects; they can be estimated only in Column~(1)}. Standard errors are clustered at the state level.}
\label{policy_exposure_sci_2025}
\setlength{\tabcolsep}{3pt}
\renewcommand{\arraystretch}{0.85}
\tiny
\begin{tabular}{@{\extracolsep{1pt}}lccc} 
\hline 
 & \multicolumn{3}{c}{Outcome variable: county-level crude suicide mortality rate} \\ 
\cline{2-4} 
 & (1) & (2) & (3) \\ 
\hline \\[-1.8ex]
 ERPO & \textcolor{teal}{$-$0.481$^{**}$} &  &  \\ 
  & \textcolor{teal}{(0.217)} &  &  \\ 
 ERPO social exposure &  & \textcolor{teal}{$-$0.181$^{***}$} & \textcolor{teal}{$-$0.259$^{***}$} \\ 
  &  & \textcolor{teal}{(0.056)} & \textcolor{teal}{(0.067)} \\ 
 ERPO spatial exposure &  &  & \textcolor{teal}{0.484} \\ 
  &  &  & \textcolor{teal}{(0.312)} \\ 
 Population density & \textcolor{teal}{$-$0.897$^{**}$} & \textcolor{teal}{$-$0.253} & \textcolor{teal}{$-$0.277} \\ 
  & \textcolor{teal}{(0.360)} & \textcolor{teal}{(0.550)} & \textcolor{teal}{(0.535)} \\ 
 Percent aged below 18 & \textcolor{teal}{0.089} & \textcolor{teal}{$-$0.263} & \textcolor{teal}{$-$0.245} \\ 
  & \textcolor{teal}{(0.378)} & \textcolor{teal}{(0.344)} & \textcolor{teal}{(0.345)} \\ 
 Percent aged 18-44 & \textcolor{teal}{0.162} & \textcolor{teal}{$-$0.482} & \textcolor{teal}{$-$0.400} \\ 
  & \textcolor{teal}{(0.509)} & \textcolor{teal}{(0.497)} & \textcolor{teal}{(0.487)} \\ 
 Percent aged 45-64 & \textcolor{teal}{$-$0.518} & \textcolor{teal}{$-$0.545} & \textcolor{teal}{$-$0.486} \\ 
  & \textcolor{teal}{(0.384)} & \textcolor{teal}{(0.366)} & \textcolor{teal}{(0.376)} \\ 
 Percent Asian & \textcolor{teal}{$-$0.780$^{***}$} & \textcolor{teal}{$-$0.489$^{*}$} & \textcolor{teal}{$-$0.471$^{**}$} \\ 
  & \textcolor{teal}{(0.263)} & \textcolor{teal}{(0.246)} & \textcolor{teal}{(0.229)} \\ 
 Percent Black & \textcolor{teal}{$-$1.904$^{**}$} & \textcolor{teal}{$-$2.848$^{***}$} & \textcolor{teal}{$-$2.775$^{***}$} \\ 
  & \textcolor{teal}{(0.847)} & \textcolor{teal}{(0.692)} & \textcolor{teal}{(0.695)} \\ 
 Percent Other & \textcolor{teal}{0.475$^{**}$} & \textcolor{teal}{0.081} & \textcolor{teal}{0.074} \\ 
  & \textcolor{teal}{(0.218)} & \textcolor{teal}{(0.203)} & \textcolor{teal}{(0.205)} \\ 
 Percent Hispanic & \textcolor{teal}{$-$3.315$^{***}$} & \textcolor{teal}{$-$2.154$^{***}$} & \textcolor{teal}{$-$2.277$^{***}$} \\ 
  & \textcolor{teal}{(0.928)} & \textcolor{teal}{(0.683)} & \textcolor{teal}{(0.708)} \\ 
 Median household income  & \textcolor{teal}{$-$0.440$^{**}$} & \textcolor{teal}{$-$0.400$^{**}$} & \textcolor{teal}{$-$0.372$^{*}$} \\ 
  & \textcolor{teal}{(0.183)} & \textcolor{teal}{(0.183)} & \textcolor{teal}{(0.186)} \\ 
 Percent with limited English proficiency & \textcolor{teal}{$-$0.045} & \textcolor{teal}{$-$0.052} & \textcolor{teal}{$-$0.069} \\ 
  & \textcolor{teal}{(0.072)} & \textcolor{teal}{(0.116)} & \textcolor{teal}{(0.114)} \\ 
 Percent unemployed & \textcolor{teal}{0.056} & \textcolor{teal}{$-$0.053} & \textcolor{teal}{$-$0.040} \\ 
  & \textcolor{teal}{(0.165)} & \textcolor{teal}{(0.129)} & \textcolor{teal}{(0.130)} \\ 
 Percent with less than high school education & \textcolor{teal}{0.011} & \textcolor{teal}{$-$0.085} & \textcolor{teal}{$-$0.090} \\ 
  & \textcolor{teal}{(0.139)} & \textcolor{teal}{(0.109)} & \textcolor{teal}{(0.111)} \\ 
 Political affiliation & \textcolor{teal}{$-$0.057} & \textcolor{teal}{0.159} & \textcolor{teal}{0.139} \\ 
  & \textcolor{teal}{(0.168)} & \textcolor{teal}{(0.142)} & \textcolor{teal}{(0.146)} \\ 
 \textcolor{teal}{COVID-19 deaths per 100k} & \textcolor{teal}{0.088} & \textcolor{teal}{0.098} & \textcolor{teal}{0.082} \\ 
  & \textcolor{teal}{(0.068)} & \textcolor{teal}{(0.095)} & \textcolor{teal}{(0.098)} \\ 
 \textcolor{teal}{Drug-overdose deaths per 100k} & \textcolor{teal}{$-$1.020} & \textcolor{teal}{$-$0.927} & \textcolor{teal}{$-$0.923} \\ 
  & \textcolor{teal}{(0.983)} & \textcolor{teal}{(1.123)} & \textcolor{teal}{(1.125)} \\ 
 \textcolor{teal}{Alcohol-related deaths per 100k} & \textcolor{teal}{9.671$^{***}$} & \textcolor{teal}{9.671$^{***}$} & \textcolor{teal}{9.667$^{***}$} \\ 
  & \textcolor{teal}{(3.243)} & \textcolor{teal}{(3.442)} & \textcolor{teal}{(3.440)} \\ 
 \textcolor{teal}{Poor mental health days} & \textcolor{teal}{0.104} & \textcolor{teal}{0.091} & \textcolor{teal}{0.081} \\ 
  & \textcolor{teal}{(0.078)} & \textcolor{teal}{(0.090)} & \textcolor{teal}{(0.091)} \\ 
 \textcolor{teal}{Firearm: waiting period} & \textcolor{teal}{0.399$^{*}$} &  &  \\ 
  & \textcolor{teal}{(0.208)} &  &  \\ 
 \textcolor{teal}{Firearm: child-access prevention} & \textcolor{teal}{0.101} &  &  \\ 
  & \textcolor{teal}{(0.196)} &  &  \\ 
 \textcolor{teal}{Firearm: minimum purchase age 21} & \textcolor{teal}{$-$0.016} &  &  \\ 
  & \textcolor{teal}{(0.184)} &  &  \\ 
\hline \\[-1.8ex] 
Observations & 40,794 & 40,794 & 40,794 \\ 
R$^{2}$ & \textcolor{teal}{0.949} & \textcolor{teal}{0.950} & \textcolor{teal}{0.950} \\ 
Adjusted R$^{2}$ & \textcolor{teal}{0.945} & \textcolor{teal}{0.945} & \textcolor{teal}{0.945} \\ 
\hline 
\multicolumn{4}{l}{\tiny Robust standard errors in parentheses. * p$<$0.1; ** p$<$0.05; *** p$<$0.01} \\
\end{tabular} 
\end{table}

 
{\color{teal}
 
\subsection*{Heterogeneity of Socio-Spatial Diffusion Across Age, Sex, and Manner of Death}
\label{sec:subgroup}
 
Our primary analysis established a population-level association between the suicide mortality rate of a focal county and that of its socially proximal counties, using all suicide deaths in the period. An important question is whether this association is uniform across the population or concentrated within particular subgroups. Both theory and prior empirical work on suicide contagion suggest that susceptibility to social influence is age-graded, being more pronounced among younger than older adults, who are embedded in denser peer networks and more exposed to suicide-related content~\cite{mueller2015suicidal, niederkrotenthaler2020role}. A separate consideration concerns manner of death. Because firearm suicides are concentrated in regions with greater firearm availability, an association detectable only for firearm deaths could reflect the spatial clustering of lethal means rather than diffusion through social ties; showing the association for both firearm and non-firearm deaths is therefore an important validity check. We addressed these questions by re-estimating the social-proximity model within subgroups defined by age, sex, the interaction of age and sex, and manner of death.
 
Suicide deaths were grouped into three age bands, $10$--$24$ (adolescents and young adults, consistent with the contemporary definition of adolescence as ages $10$--$24$, inclusive~\cite{sawyer2018age}), $25$--$64$ (working-age adults), and $65{+}$ (older adults), two sex strata (male and female), the six age-by-sex cells, and two manner-of-death strata (firearm suicide, ICD-10 \texttt{X72--X74}; and non-firearm suicide, the complement of the firearm codes within \texttt{X60--X84} and \texttt{Y87.0}). Ages $0$--$9$, in which suicide is effectively absent, were excluded from the age and age-by-sex subgroups. For each subgroup $g$, both the outcome and the denominator are subgroup-specific: death counts were obtained from the NVSS records and population denominators from the American Community Survey ($5$-year estimates, table \texttt{B01001}).
 
Let $y^{(g)}_{jt}$ denote the suicide mortality rate (deaths per $100{,}000$) of subgroup $g$ in county $j$ and year $t$. We defined the subgroup-specific social and spatial exposures of a focal county $i$ exactly as in the main analysis, weighting the subgroup-specific rates of all other counties by the same connectedness and proximity weights:
\begin{equation}
\color{teal}
s^{(g)}_{-it} \;=\; \sum_{j \neq i} w_{ij}\, y^{(g)}_{jt},
\qquad
d^{(g)}_{-it} \;=\; \sum_{j \neq i} a_{ij}\, y^{(g)}_{jt},
\label{eq:subgroup-exposure}
\end{equation}
where the social weights $w_{ij}$ and spatial weights $a_{ij}$ are identical to those defined in the main analysis. Because the Social Connectedness Index is not resolved by age or sex, the weights were held fixed across all subgroups, so that only the diffusing quantity $y^{(g)}$ varies. Differences across subgroups are therefore attributable to the outcome rather than to the construction of the exposure.

For each subgroup $g$, we estimated the two-way fixed effects model
\begin{equation}
\color{teal}
y^{(g)}_{it}
\;=\;
\lambda^{(g)}_{1}\, s^{(g)}_{-it}
\;+\; \lambda^{(g)}_{2}\, d^{(g)}_{-it}
\;+\; \bigl(\boldsymbol{\lambda}^{(g)}_{3}\bigr)^{\!\top} X_{it}
\;+\; \mu_i \;+\; \phi_t \;+\; \varepsilon_{it},
\label{eq:subgroup-model}
\end{equation}
where $\mu_i$ and $\phi_t$ denote county and year fixed effects and $X_{it}$ is the vector of time-varying county-level controls. The vector $X_{it}$ comprises population density; the population shares aged below~$18$, $18$--$44$, and $45$--$64$; the population shares identifying as Asian, Black, and Other racial subgroups; the Hispanic population share; median household income; the share of residents with limited English proficiency; the unemployment rate; the share of residents with less than a high-school education; the COVID-19 mortality rate (per $100{,}000$); the drug-overdose mortality rate (per $100{,}000$); the alcohol-related mortality rate (per $100{,}000$); and the average number of poor mental health days. This is the same set of controls used in the main specification. For the age and age-by-sex subgroups the three age-composition shares were omitted, since the outcome is already restricted to a single age band, following the same reasoning as in our age-adjusted analysis. The sex and manner-of-death subgroups retained the full control vector. The parameter of interest is $\lambda^{(g)}_{1}$, the association between subgroup-specific mortality in the focal county and the subgroup-specific suicide rate of socially proximal counties, controlling for spatial proximity and the covariates in $X_{it}$. As in the main text, all continuous regressors, including $s^{(g)}_{-it}$ and $d^{(g)}_{-it}$, were standardized to zero mean and unit variance within each subgroup, so that $\lambda^{(g)}_{1}$ is expressed per one-standard-deviation increase in social exposure. Estimation was by population-weighted ordinary least squares, with weights equal to the subgroup population $n^{(g)}_{it}$ and standard errors clustered by state.

Expressed in deaths per $100{,}000$, the coefficient $\lambda^{(g)}_{1}$ is directly interpretable but not comparable across subgroups, because subgroups differ in both the level and the cross-county dispersion of suicide mortality. A larger coefficient in a high-rate cohort such as working-age adults may therefore reflect the scale of mortality rather than a greater sensitivity to social influence. To obtain a measure of effect that is comparable across cohorts, we additionally standardized the outcome within each cohort. Letting $n^{(g)}_{it}$ denote the cohort population, which is also the estimation weight, we defined the population-weighted mean and standard deviation of the cohort outcome as
\begin{equation}
\color{teal}
\bar{y}^{(g)}
= \frac{\sum_{i,t} n^{(g)}_{it}\, y^{(g)}_{it}}{\sum_{i,t} n^{(g)}_{it}},
\qquad
\tilde{\sigma}^{(g)}
= \left[
\frac{\sum_{i,t} n^{(g)}_{it}\bigl(y^{(g)}_{it} - \bar{y}^{(g)}\bigr)^{2}}
     {\sum_{i,t} n^{(g)}_{it}}
\right]^{1/2},
\label{eq:wmeansd}
\end{equation}
and the standardized outcome $\tilde{y}^{(g)}_{it} = \bigl(y^{(g)}_{it} - \bar{y}^{(g)}\bigr)/\tilde{\sigma}^{(g)}$. The standardized model retains the right-hand side of Eq.~(\ref{eq:subgroup-model}), with the same control vector $X_{it}$, county fixed effects $\mu_i$, and year fixed effects $\phi_t$, and reads
\begin{equation}
\color{teal}
\tilde{y}^{(g)}_{it}
\;=\;
\tilde{\lambda}^{(g)}_{1}\, s^{(g)}_{-it}
\;+\; \tilde{\lambda}^{(g)}_{2}\, d^{(g)}_{-it}
\;+\; \bigl(\tilde{\boldsymbol{\lambda}}^{(g)}_{3}\bigr)^{\!\top} X_{it}
\;+\; \mu_i \;+\; \phi_t \;+\; \varepsilon_{it},
\label{eq:subgroup-model-std}
\end{equation}
so that $\tilde{\lambda}^{(g)}_{1}$ measures the change in the cohort suicide rate, in within-cohort standard-deviation units, per one-standard-deviation increase in social exposure. We adopted the population-weighted standard deviation in Eq.~(\ref{eq:wmeansd}) rather than its unweighted counterpart for two reasons. First, it is coherent with the population-weighted estimator. Second, the unweighted dispersion of subgroup rates is inflated by small-number variation in sparsely populated counties, where a single death produces an extreme rate; dividing by an unweighted standard deviation would understate the standardized effect for small cohorts, and most of all for adolescents. Because the transformation is a linear rescaling, statistical significance is identical to that of Eq.~(\ref{eq:subgroup-model}), and the two parameterizations are related by $\tilde{\lambda}^{(g)}_{1} = \lambda^{(g)}_{1}/\tilde{\sigma}^{(g)}$.

Estimates of $\lambda^{(g)}_{1}$ and $\tilde{\lambda}^{(g)}_{1}$ for all cohorts are reported in Table~\ref{tab:subgroup} and displayed in Fig.~\ref{fig:subgroup}. The social-proximity association is positive and statistically significant in every non-elderly cohort and for both manners of death, and is statistically indistinguishable from zero in every cohort aged $65{+}$. On the per-$100{,}000$ scale, the point estimate is largest among working-age adults ($\lambda^{(g)}_{1}=2.89$, cluster-robust $95\%$ CI $[1.11, 4.68]$) and smaller among adolescents and young adults ($\lambda^{(g)}_{1}=1.27$, $[0.50, 2.03]$). This ordering reflects the higher baseline suicide rate of working-age adults rather than a stronger sensitivity to social influence: on the standardized scale the two cohorts are close ($\tilde{\lambda}^{(g)}_{1}=0.100$, $[0.038, 0.161]$ for ages $25$--$64$ against $0.075$, $[0.030, 0.121]$ for ages $10$--$24$), with overlapping confidence intervals. The older-adult association is null on both scales ($\tilde{\lambda}^{(g)}_{1}=-0.016$, $[-0.060, 0.028]$). The same pattern holds within sex: the association is present only in the non-elderly strata, and among women the adolescent and working-age estimates are of similar magnitude ($\tilde{\lambda}^{(g)}_{1}=0.050$ and $0.057$). The association is larger for men than for women on both scales ($\tilde{\lambda}^{(g)}_{1}=0.121$ against $0.066$), consistent with the higher baseline suicide rate among men. It is also positive and significant for both firearm ($\tilde{\lambda}^{(g)}_{1}=0.081$, $[0.017, 0.145]$) and non-firearm suicides ($\tilde{\lambda}^{(g)}_{1}=0.089$, $[0.037, 0.141]$), of comparable magnitude. That the association holds for non-firearm as well as firearm deaths, controlling for spatial proximity and the full set of county-level covariates, indicates that it does not reflect only the geographic distribution of lethal means.

The standardized coefficients range from $0.05$ to $0.12$ across the cohorts in which the association is significant, a scale consistent with the associations reported in our main analysis and with what would be expected of a population-level process estimated net of county and year fixed effects. The null association among adults aged $65{+}$ accords with the age-graded susceptibility described above, but we do not read it as evidence of no diffusion in this group: $57\%$ of US adults aged $65{+}$ use Facebook, compared with $80\%$ of those aged $30$--$49$~\cite{gottfried2025americans}, so the SCI captures their social ties less completely than those of younger cohorts, which would attenuate any measured association toward zero. Across subgroups, the association is present among adolescents, young adults, and working-age adults, of comparable strength once differences in baseline mortality are accounted for, null among older adults, and present for both firearm and non-firearm deaths. The data are therefore consistent with a step change at older ages rather than a monotone gradient across the life course.
 
\begin{figure}[!htbp]
\color{teal}
\centering
\includegraphics[width=\linewidth]{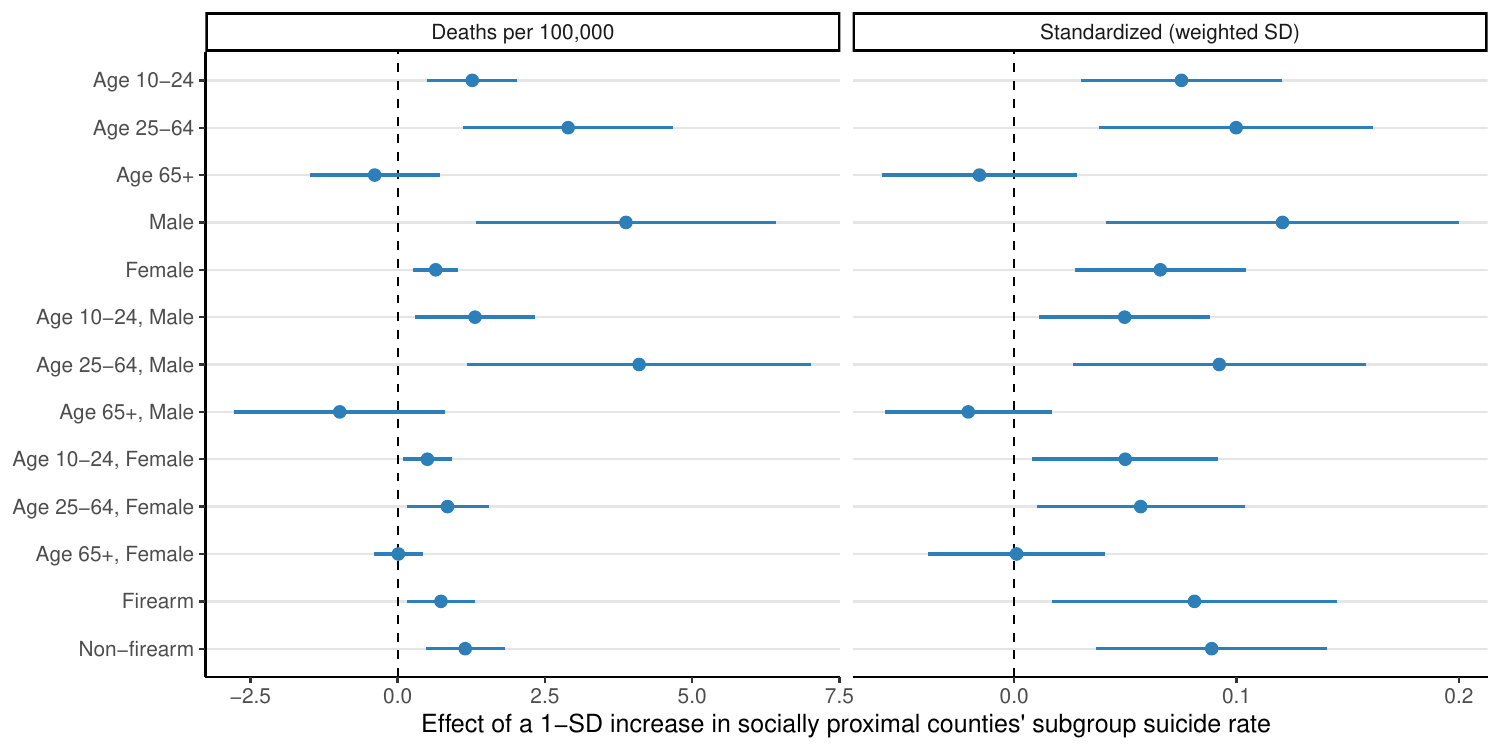}
\caption{\textbf{Heterogeneity of the social-proximity association across demographic and clinical cohorts.} Estimated social-proximity coefficients for each cohort, from separate two-way fixed effects models (Eqs.~(\ref{eq:subgroup-model}) and (\ref{eq:subgroup-model-std})), each including county and year fixed effects, the full set of sociodemographic and health-related controls, and the spatial-proximity term $d^{(g)}_{-it}$. \emph{Left:} $\hat{\lambda}^{(g)}_{1}$, the change in focal-county cohort suicide mortality (per $100{,}000$) for a one-standard-deviation increase in cohort-specific social exposure. \emph{Right:} $\hat{\tilde{\lambda}}^{(g)}_{1}$, the same association with the outcome standardized using the population-weighted standard deviation, expressed in within-cohort standard-deviation units and comparable across cohorts. Points are point estimates; horizontal lines are cluster-robust $95\%$ confidence intervals (standard errors clustered by state); the dashed vertical line marks the null.}
\label{fig:subgroup}
\end{figure}
 
\begin{table}[!htbp]
\color{teal}
\centering
\caption{\textbf{Social proximity and county-level suicide mortality across demographic and clinical cohorts.} Each row reports a separate two-way fixed effects regression of the cohort-specific suicide mortality rate on the cohort-specific social-proximity exposure $s^{(g)}_{-it}$, controlling for spatial proximity $d^{(g)}_{-it}$, the time-varying covariates of the main analysis (including the COVID-19, drug-overdose, and alcohol-related mortality rates and poor mental health days), county fixed effects, and year fixed effects, estimated by population-weighted OLS with standard errors clustered by state. Column~(1) reports $\hat{\lambda}^{(g)}_{1}$ in deaths per $100{,}000$ (Eq.~(\ref{eq:subgroup-model})); column~(2) reports the comparable standardized coefficient $\hat{\tilde{\lambda}}^{(g)}_{1}$, obtained by standardizing the outcome with the population-weighted standard deviation (Eq.~(\ref{eq:subgroup-model-std})). For the age and age-by-sex cohorts the age-composition covariates are omitted, as the outcome is age-homogeneous. Cluster-robust standard errors in parentheses.}
\label{tab:subgroup}
\tiny
\begin{tabular}{@{\extracolsep{4pt}}lccc}
\\[-1.8ex]\hline
\hline \\[-1.8ex]
 & \multicolumn{2}{c}{\textit{Social proximity} $s^{(g)}_{-it}$} & \\
\cline{2-3}
\\[-1.8ex]
Cohort & $\hat{\lambda}^{(g)}_{1}$ (per 100{,}000) & $\hat{\tilde{\lambda}}^{(g)}_{1}$ (standardized) & Observations \\
\hline \\[-1.8ex]
\multicolumn{4}{l}{\textit{Panel A. By age}} \\
\quad Age 10--24 & $1.266^{***}$ & $0.075^{***}$ & 40{,}784 \\
                 & $(0.380)$     & $(0.023)$     & \\
\quad Age 25--64 & $2.894^{***}$ & $0.100^{***}$ & 40{,}786 \\
                 & $(0.888)$     & $(0.031)$     & \\
\quad Age 65+    & $-0.391$      & $-0.016$      & 40{,}784 \\
                 & $(0.549)$     & $(0.022)$     & \\
\hline \\[-1.8ex]
\multicolumn{4}{l}{\textit{Panel B. By sex}} \\
\quad Male       & $3.876^{***}$ & $0.121^{***}$ & 40{,}786 \\
                 & $(1.266)$     & $(0.039)$     & \\
\quad Female     & $0.646^{***}$ & $0.066^{***}$ & 40{,}786 \\
                 & $(0.188)$     & $(0.019)$     & \\
\hline \\[-1.8ex]
\multicolumn{4}{l}{\textit{Panel C. By age and sex}} \\
\quad Age 10--24, Male   & $1.312^{**}$  & $0.050^{**}$  & 40{,}784 \\
                         & $(0.506)$     & $(0.019)$     & \\
\quad Age 25--64, Male   & $4.099^{***}$ & $0.092^{***}$ & 40{,}786 \\
                         & $(1.456)$     & $(0.033)$     & \\
\quad Age 65+, Male      & $-0.984$      & $-0.021$      & 40{,}781 \\
                         & $(0.892)$     & $(0.019)$     & \\
\quad Age 10--24, Female & $0.505^{**}$  & $0.050^{**}$  & 40{,}775 \\
                         & $(0.210)$     & $(0.021)$     & \\
\quad Age 25--64, Female & $0.846^{**}$  & $0.057^{**}$  & 40{,}786 \\
                         & $(0.347)$     & $(0.023)$     & \\
\quad Age 65+, Female    & $0.011$       & $0.001$       & 40{,}784 \\
                         & $(0.207)$     & $(0.020)$     & \\
\hline \\[-1.8ex]
\multicolumn{4}{l}{\textit{Panel D. By manner of death}} \\
\quad Firearm      & $0.735^{**}$  & $0.081^{**}$  & 40{,}786 \\
                   & $(0.289)$     & $(0.032)$     & \\
\quad Non-firearm  & $1.146^{***}$ & $0.089^{***}$ & 40{,}786 \\
                   & $(0.334)$     & $(0.026)$     & \\
\hline
\hline \\[-1.8ex]
\multicolumn{4}{l}{\footnotesize County and year fixed effects, full (extended) controls,} \\
\multicolumn{4}{l}{\footnotesize and $d^{(g)}_{-it}$ included in every model.} \\
\multicolumn{4}{l}{\footnotesize $^{*}p<0.1$; $^{**}p<0.05$; $^{***}p<0.01$.} \\
\end{tabular}
\end{table}
 
 
\subsection*{Robustness: Distinguishing Social from Geographical Proximity}

Social connectedness and geographical proximity are correlated, since people are on average more likely to befriend those who live nearby. An important question is therefore whether the social-proximity exposure $s_{-it}$ defined in Equation~(1) re-encodes geographical adjacency rather than capturing social ties that extend beyond physical space. If it did, the association reported in Table~1 would reflect spatial structure already absorbed by the spatial-proximity exposure $d_{-it}$. We address this in two ways: by quantifying the empirical overlap between social connectedness and geographical distance, and by re-estimating the association after restricting the set of counties over which the connectedness-weighted average is computed.

We compared, across all $4{,}921{,}953$ county pairs, the social connectedness between two counties, $\mathrm{SCI}_{ij}$, with the inverse of the geodesic distance between their centroids, $1/d_{ij}$, which is the quantity underlying the spatial weights $a_{ij}$ in Equation~(2). The two are positively but only moderately associated (Pearson correlation of $0.53$ on the logarithmic scale; Spearman $\rho = 0.52$), so geographical distance accounts for approximately $28\%$ of the variation in social connectedness (Supplementary Fig.~\ref{fig:sci_distance}). Social connectedness therefore overlaps with spatial proximity without being reducible to it, and a substantial share of inter-county friendship ties links counties that are not geographically close.

To assess whether the association between focal-county suicide mortality and $s_{-it}$ persists once geographically close counties are removed, we re-estimated the model under alternative definitions of the set of counties entering the weighted average. For a focal county $i$ and an admissible set $\mathcal{R}_i \subseteq \{\,j : j \neq i\,\}$, we define the restricted social weights and the corresponding exposure as
\begin{equation}
  \color{teal}
  w^{\mathcal{R}}_{ij} \;=\; \frac{n_j\,\mathrm{SCI}_{ij}\,\mathbf{1}\{\,j \in \mathcal{R}_i\,\}}
       {\sum_{k \neq i} n_k\,\mathrm{SCI}_{ik}\,\mathbf{1}\{\,k \in \mathcal{R}_i\,\}},
  \qquad
  s^{\mathcal{R}}_{-it} \;=\; \sum_{j \neq i} w^{\mathcal{R}}_{ij}\, y_{jt},
  \label{eq:restricted_weights}
\end{equation}
where $n_j$ is the population of county $j$ and $y_{jt}$ is its suicide death rate in year $t$, as in Equations~(1) and~(2). The unrestricted set $\mathcal{R}_i = \{\,j : j \neq i\,\}$ recovers the baseline weights $w_{ij}$ and exposure $s_{-it}$ of the main analysis. We considered two restrictions:
\begin{itemize}
  \item \textbf{Distance-based restriction.} $\mathcal{R}_i = \mathcal{D}_i = \{\,j \neq i : d_{ij} > 100\text{ km}\,\}$, which excludes all contiguous counties by construction, since no two contiguous US counties have centroids more than $100$~km apart.
  \item \textbf{Connection-strength restriction.} Let $Q_i(0.99)$ denote the $99$th percentile of the empirical distribution of $\{\mathrm{SCI}_{ij}\}_{j \neq i}$. The admissible set is $\mathcal{R}_i = \mathcal{T}_i = \{\,j \neq i : \mathrm{SCI}_{ij} \geq Q_i(0.99)\,\}$, comprising the one percent of counties with which the focal county has the highest connectedness.
\end{itemize}
For each definition we estimated the two-way fixed effects model
\begin{equation}
  \color{teal}
  y_{it} \;=\; \lambda_1\, s^{\mathcal{R}}_{-it} \;+\; \lambda_2\, d_{-it}
       \;+\; \lambda_3^{\top} X_{it} \;+\; \mu_i \;+\; \phi_t \;+\; \varepsilon_{it},
  \label{eq:robustness_model}
\end{equation}
holding the spatial-proximity exposure $d_{-it}$ at its baseline inverse-distance definition (Equation~(2)), so that only the social weights vary across specifications. Here $\mu_i$ and $\phi_t$ denote county and year fixed effects and $X_{it}$ is the same vector of time-varying county-level covariates used throughout. We denote the social- and spatial-proximity coefficients in these regressions by $\lambda_1$ and $\lambda_2$ to distinguish them from the coefficients of the main models.

The social-proximity association is positive and statistically significant under all three definitions (Supplementary Table~\ref{tab:sci_robustness}). The baseline specification reproduces Model~2 of Table~1: a one-standard-deviation increase in $s_{-it}$ is associated with an increase of $\hat{\lambda}_1 = 2.47$ suicide deaths per $100{,}000$ in the focal county (cluster-robust $95\%$ CI: $[0.97, 4.00]$, $p<0.01$). Under the distance-based restriction, the estimate is of comparable magnitude and more precisely estimated ($\hat{\lambda}_1 = 2.23$; $95\%$ CI: $[1.13, 3.34]$, $p<0.001$). Under the connection-strength restriction, it is $\hat{\lambda}_1 = 1.57$ ($95\%$ CI: $[0.40, 2.73]$, $p<0.01$); the point estimate is smaller, as would be expected when the exposure is constructed from a narrower subset of ties. Since the distance-based estimate uses no county pair separated by less than $100$~km, it is not driven by contiguity, and any explanation resting on short-range mobility would have to operate over distances exceeding that threshold. The spatial-proximity coefficient $\hat{\lambda}_2$ is not statistically distinguishable from zero under the baseline or distance-based definitions, and reaches only $p<0.1$ under the connection-strength restriction, so the social channel is the one that holds consistently across specifications.

These results also clarify how the coefficients should be read. The exposure $s_{-it}$ is a population- and connectedness-weighted average of the suicide mortality rates prevailing among a county's social connections (Equations~(1) and~(2)), and is standardized to unit variance. The coefficient $\hat{\zeta}_1 = 2.47$ reported in Model~2 of Table~1 therefore indicates that a one-standard-deviation increase in the suicide mortality rate experienced across a county's social network is associated with $2.47$ additional suicide deaths per $100{,}000$ residents of the focal county, holding geographical proximity and all county-level covariates fixed. Because the county fixed effects absorb every time-invariant feature of a county, including its overall level of social connectedness, the coefficient is identified from year-to-year variation in the outcomes prevailing among a county's connections rather than from how connected a county is. This distinction separates a county's level of connectedness from the mortality experienced by the counties to which it is connected, and it is the latter that $s_{-it}$ is designed to isolate. The same reading applies to the coefficients in the restricted specifications above and to the other regressors in the model.

These analyses indicate that the social-proximity measure captures social structure rather than a relabeling of geographical proximity. The association persists when geographically close counties are excluded, it persists when the average is computed over only the most strongly connected counties, and social connectedness shares approximately a quarter of its variation with geographical distance. This supports interpreting $s_{-it}$ as exposure to suicide mortality diffusing through inter-county social ties, and indicates that the social channel operates beyond the geographical channel captured by $d_{-it}$. We note that the SCI is constructed from Facebook friendship links aggregated to the county level. The index is also distributed at ZIP-code-tabulation-area level, but annual population denominators are not available at that resolution, and the ZIP-code-level index excludes areas with few active users, which would remove network nodes from sparsely populated areas where suicide mortality is highest. County is therefore the level at which the exposure and the outcome can be jointly measured, and it is the standard unit of analysis in the social-connectedness literature. The aggregation may nonetheless understate finer-grained heterogeneity in network structure, a limitation we revisit in the Discussion.
\begin{figure}[!htbp]
  \color{teal}
  \centering
  \includegraphics[width=0.72\textwidth]{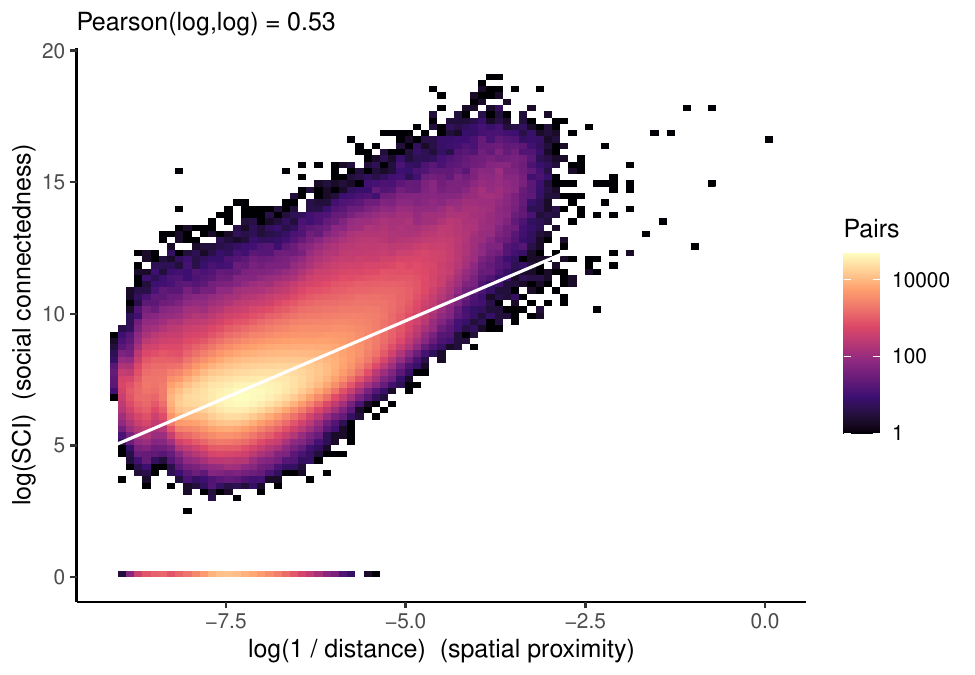}
  \caption{\textbf{Social connectedness is related to but distinct from geographical proximity.}
  Joint distribution of the Facebook Social Connectedness Index ($\mathrm{SCI}_{ij}$) and inverse geodesic distance ($1/d_{ij}$) across all $4{,}921{,}953$ county pairs, on logarithmic axes; color denotes the number of pairs in each hexagonal bin and the white line is the ordinary least squares fit. The two quantities are positively correlated (Pearson $r = 0.53$ on the log scale; Spearman $\rho = 0.52$), yet geographical distance accounts for only about $28\%$ of the variation in social connectedness, indicating that $\mathrm{SCI}_{ij}$ encodes substantial social structure beyond spatial adjacency.}
  \label{fig:sci_distance}
\end{figure}
 
\begin{table}[!htbp] \centering
  \color{teal}
  \caption{\textbf{Robustness of the social-proximity association to the definition of the social neighborhood.} Each column reports a separate population-weighted two-way fixed effects regression (Equation~\ref{eq:robustness_model}) of the county-level crude suicide mortality rate on the connectedness-weighted suicide mortality rate of socially connected counties, $s^{\mathcal{R}}_{-it}$. The admissible set $\mathcal{R}$ differs across columns: all counties (baseline, Equation~(1)); only counties whose centroids lie more than $100$~km from the focal county, which excludes all contiguous counties; and only the one percent of counties with which the focal county has the highest connectedness. All specifications hold the spatial-proximity exposure $d_{-it}$ at its baseline inverse-distance definition (Equation~(2)) and include county and year fixed effects together with the full vector of time-varying controls $X_{it}$: population density; age distribution (percent aged $0$--$17$, $18$--$44$, $45$--$64$); racial composition (percent Asian, Black, and Other); ethnic composition (percent Hispanic); median household income; percent with limited English proficiency; percent unemployed; percent with less than a high school education; the COVID-19, drug-overdose, and alcohol-related death rates; poor mental-health days; and state firearm policies. Column~(1) is estimated as in Table~1, Model~2. Standard errors are clustered at the state level.}
  \label{tab:sci_robustness}
  \begingroup
  \small
  \setlength{\tabcolsep}{4pt}

  \begin{adjustbox}{max width=\textwidth}
  \begin{tabular}{@{\extracolsep{1pt}}lccc}
  \\[-1.8ex]\hline \hline \\[-1.8ex]
   & \multicolumn{3}{c}{\textit{Dependent variable:} crude suicide mortality rate (per 100{,}000)} \\
  \cline{2-4}
  \\[-1.8ex]
   & Baseline & Distant counties & Strongest ties \\
   & (all counties) & ($d_{ij} > 100$ km) & (top $1\%$) \\
  \\[-1.8ex] & (1) & (2) & (3) \\
  \hline \\[-1.8ex]
  Deaths in social proximity $s^{\mathcal{R}}_{-it}$ & $2.469^{***}$ & $2.235^{***}$ & $1.569^{***}$ \\
   & (0.763) & (0.548) & (0.580) \\
  Deaths in spatial proximity $d_{-it}$ & $0.409$ & $0.271$ & $0.624^{*}$ \\
   & (0.324) & (0.267) & (0.319) \\
  \hline \\[-1.8ex]
  Time-varying controls $X_{it}$ & Yes & Yes & Yes \\
  County fixed effects & Yes & Yes & Yes \\
  Year fixed effects & Yes & Yes & Yes \\
  Observations & 40{,}794 & 40{,}794 & 40{,}794 \\
  \hline \hline \\[-1.8ex]
  \end{tabular}
  \end{adjustbox}
  \endgroup
  \\[0.6ex]
  {\footnotesize Robust standard errors, clustered at the state level, in parentheses. $^{*}p<0.1$; $^{**}p<0.05$; $^{***}p<0.01$.}
\end{table}

\subsection*{Robustness to Residual Spatial Autocorrelation: A Panel Spatial-Error Model}
\label{sec:sem}
 
A remaining concern is whether the association we attribute to social proximity could instead reflect residual spatial autocorrelation. In our main specification, geographical structure is addressed in two ways: by including the spatial-proximity exposure $d_{-it}$ as a regressor, and by clustering standard errors at the state level. Neither device, however, fully models the dependence between geographically adjacent counties. Clustering by state permits arbitrary within-state correlation but treats states as independent blocks; it does not encode the fact that two counties sharing a border are more alike than two counties at opposite ends of the same state, and it does nothing for dependence that crosses state lines. Including $d_{-it}$ adjusts the conditional mean for the mortality of nearby counties but leaves the covariance structure of the disturbances unmodeled. If the regression errors are themselves spatially correlated, the independence assumption underlying the reported standard errors is violated, the errors may be anti-conservative, and a portion of what we ascribe to social connectedness could in principle be residual spatial structure. To address this directly, we re-estimated the socio-spatial association with a model that places the spatial dependence in the error process itself, rather than only among the regressors.
 
We adopted a two-way fixed effects spatial-error model (SEM) estimated on the full county-year panel. Let $y_{it}$ denote the crude suicide mortality rate in county $i$ and year $t$, let $s_{-it}$ and $d_{-it}$ denote the standardized social- and spatial-proximity exposures defined in the main text, and let $Z_{it}$ collect the same time-varying county-level controls used throughout. The model is
\begin{equation}
\color{teal}
y_{it} \;=\; \psi\, s_{-it} \;+\; \chi\, d_{-it} \;+\; \boldsymbol{\pi}^{\top} Z_{it} \;+\; c_i \;+\; \tau_t \;+\; u_{it},
\label{eq:sem-mean}
\end{equation}
where $c_i$ and $\tau_t$ are county and year fixed effects and $u_{it}$ is a disturbance that, unlike in an ordinary fixed effects regression, is permitted to be spatially correlated. The spatial dependence is modeled as a first-order spatial-autoregressive process in the errors,
\begin{equation}
\color{teal}
u_{it} \;=\; \rho \sum_{j \neq i} M_{ij}\, u_{jt} \;+\; \nu_{it},
\qquad
\nu_{it} \sim \mathrm{iid}(0,\ \sigma^{2}_{\nu}),
\label{eq:sem-error}
\end{equation}
in which $M$ is a row-standardized spatial-weights matrix, $\rho$ is the scalar spatial-error parameter, and $\nu_{it}$ is a well-behaved innovation that does satisfy the independence assumption. Equation~(\ref{eq:sem-error}) lets each county's disturbance depend on the contemporaneous disturbances of its geographical neighbors; estimating $\rho$ jointly with the mean-equation coefficients removes that dependence from the residuals, so that the reported standard errors on $\psi$ and $\chi$ are corrected for spatial autocorrelation rather than merely being robust to within-state clustering. The parameter $\rho$ is itself informative: a statistically significant $\rho$ constitutes direct evidence that spatial autocorrelation is present in the outcome, and its sign and magnitude summarize the structure that the model absorbs.
 
The neighbor structure in $M$ is defined by $k$-nearest-neighbor (kNN) adjacency among county centroids. For each county we identified its $k$ geographically closest counties by geodesic distance between centroids and assigned them equal, row-standardized weights, so that $\sum_{j} M_{ij} = 1$ for every $i$ and $M_{ii}=0$. We chose a kNN graph rather than a contiguity (shared-border) graph for two reasons. First, it guarantees that every county has exactly $k$ neighbors, avoiding the islands and irregular-degree problems that arise for coastal counties, water boundaries, and the geographically isolated counties of Alaska and Hawaii under a contiguity rule. Second, it yields a symmetric, connected graph whose density is controlled by the single interpretable parameter $k$, which we vary below to confirm that the conclusions do not depend on any particular choice. Our primary specification uses $k=5$; the model is a two-way within (fixed effects) estimator with the spatial-error process of Equation~(\ref{eq:sem-error}), fit by maximum likelihood.
 
The results confirm that the social-proximity finding is robust to explicit correction for spatial autocorrelation (Table~\ref{tab:sem_fe}). The spatial-error parameter is negative and highly significant ($\hat{\rho} = -0.057$, $95\%$ CI $[-0.073, -0.041]$, $p<0.001$), which is itself the key diagnostic: it constitutes direct evidence that spatial autocorrelation is present in the outcome and is now being explicitly modeled through the error covariance rather than left in the residuals. This directly addresses the concern that the reported associations might rest on an unwarranted independence assumption. With that dependence absorbed into the error process---so that the standard errors on the proximity coefficients are corrected for spatial autocorrelation rather than merely made robust to within-state clustering---the social-proximity association remains positive, sizable, and highly significant: $\hat{\psi} = 3.93$ additional suicide deaths per $100{,}000$ per one-standard-deviation increase in social exposure ($95\%$ CI $[2.91, 4.96]$, $p<0.001$). The spatial-proximity term is also positive ($\hat{\chi} = 1.26$, $95\%$ CI $[0.12, 2.39]$), consistent with a genuine but secondary geographical component operating alongside the social channel. That the social-proximity association retains its significance under a model that explicitly accounts for spatially correlated errors indicates that it does not reflect residual spatial structure: the effect cannot be explained away as unmodeled autocorrelation.
 
To ensure that these conclusions are not an artifact of the particular neighbor count, we re-estimated the model across a range of kNN graphs, $k \in \{6, 8, 10\}$ (Table~\ref{tab:sem_fe_knn}). The social-proximity coefficient is stable and highly significant throughout ($\hat{\psi}$ ranging from $2.90$ to $3.14$, all $p<0.001$), and the spatial-error parameter remains negative and significant at every $k$ ($\hat{\rho}$ between $-0.038$ and $-0.048$, all $p<0.001$), confirming both that spatial autocorrelation is a robust feature of the data and that its correction leaves the social-proximity association intact. Taken together, the spatial-error model directly addresses the concern that our findings might reflect unmodeled spatial dependence: spatial autocorrelation is present, it is now explicitly estimated in the error covariance rather than assumed away, and the social-proximity association survives---indeed sharpens---once it is accounted for.
 
\begin{table}[!htbp]
\color{teal}
\centering
\caption{\textbf{Panel spatial-error model of socio-spatial diffusion, correcting for residual spatial autocorrelation.} The model estimates the mean specification of Equation~(\ref{eq:sem-mean})---county-level crude suicide mortality (deaths per $100{,}000$) regressed on the standardized social-proximity exposure $s_{-it}$ and spatial-proximity exposure $d_{-it}$, with county and year fixed effects and the full vector of time-varying controls $Z_{it}$ (population density; age distribution; racial and ethnic composition; median household income; limited English proficiency; unemployment; educational attainment; COVID-19, drug-overdose, and alcohol-related mortality; poor mental health days; state firearm policies; and political affiliation) together with the first-order spatial-autoregressive error process of Equation~(\ref{eq:sem-error}), using a $k=5$ nearest-neighbor, row-standardized spatial-weights matrix $M$, estimated by maximum likelihood. The spatial-error parameter $\hat{\rho}$ is reported below the proximity coefficients; a significant $\hat{\rho}$ indicates spatial autocorrelation in the disturbances that the model absorbs, so that the standard errors on the proximity coefficients are corrected for that dependence.}
\label{tab:sem_fe}
\begin{tabular}{@{\extracolsep{4pt}}lc}
\\[-1.8ex]\hline \hline \\[-1.8ex]
 & Panel SEM-FE (kNN error, $k=5$) \\
\\[-1.8ex] & Outcome: crude suicide mortality rate (per $100{,}000$) \\
\hline \\[-1.8ex]
Deaths in social proximity $s_{-it}$  & $3.932^{***}$ \\
                                      & $(0.523)$ \\
Deaths in spatial proximity $d_{-it}$ & $1.255^{*}$ \\
                                      & $(0.577)$ \\
Spatial-error parameter $\rho$        & $-0.057^{***}$ \\
                                      & $(0.008)$ \\
\hline \\[-1.8ex]
Time-varying controls $Z_{it}$ & Yes \\
County fixed effects           & Yes \\
Year fixed effects             & Yes \\
Spatial-error correction       & Yes \\
Observations                   & 40{,}794 \\
\hline \hline \\[-1.8ex]
\multicolumn{2}{l}{\footnotesize Maximum-likelihood standard errors from the spatial-error model, in parentheses.} \\
\multicolumn{2}{l}{\footnotesize $^{*}p<0.1$; $^{**}p<0.05$; $^{***}p<0.01$.} \\
\end{tabular}
\end{table}
 
\begin{table}[!htbp]
\color{teal}
\centering
\caption{\textbf{Robustness of the spatial-error model to the neighbor count $k$.} Each column re-estimates the panel spatial-error model of Equations~(\ref{eq:sem-mean})--(\ref{eq:sem-error}) with a different number of nearest neighbors $k$ defining the row-standardized spatial-weights matrix $M$, holding the mean specification, fixed effects, and controls fixed. Reported are the social- and spatial-proximity coefficients ($\hat{\psi}$, $\hat{\chi}$) and the spatial-error parameter $\hat{\rho}$, with maximum-likelihood standard errors in parentheses. The social-proximity association and the significance of the spatial-error parameter are stable across all neighbor counts.}
\label{tab:sem_fe_knn}
\begin{tabular}{@{\extracolsep{4pt}}lccc}
\\[-1.8ex]\hline \hline \\[-1.8ex]
 & \multicolumn{3}{c}{Nearest-neighbor count} \\
\cline{2-4}
\\[-1.8ex] & $k=6$ & $k=8$ & $k=10$ \\
\hline \\[-1.8ex]
Deaths in social proximity $s_{-it}$  & $3.144^{***}$ & $2.920^{***}$ & $2.896^{***}$ \\
                                      & $(0.527)$     & $(0.526)$     & $(0.525)$ \\
Deaths in spatial proximity $d_{-it}$ & $0.873$       & $0.915$       & $1.095^{*}$ \\
                                      & $(0.586)$     & $(0.586)$     & $(0.582)$ \\
Spatial-error parameter $\rho$        & $-0.038^{***}$ & $-0.039^{***}$ & $-0.048^{***}$ \\
\hline \\[-1.8ex]
County and year fixed effects & Yes & Yes & Yes \\
Time-varying controls $Z_{it}$ & Yes & Yes & Yes \\
Observations & 40{,}794 & 40{,}794 & 40{,}794 \\
\hline \hline \\[-1.8ex]
\multicolumn{4}{l}{\footnotesize Maximum-likelihood standard errors in parentheses. $^{*}p<0.1$; $^{**}p<0.05$; $^{***}p<0.01$.} \\
\end{tabular}
\end{table}

\begin{figure}[!htbp]
\color{teal}
\centering
\includegraphics[width=\linewidth]{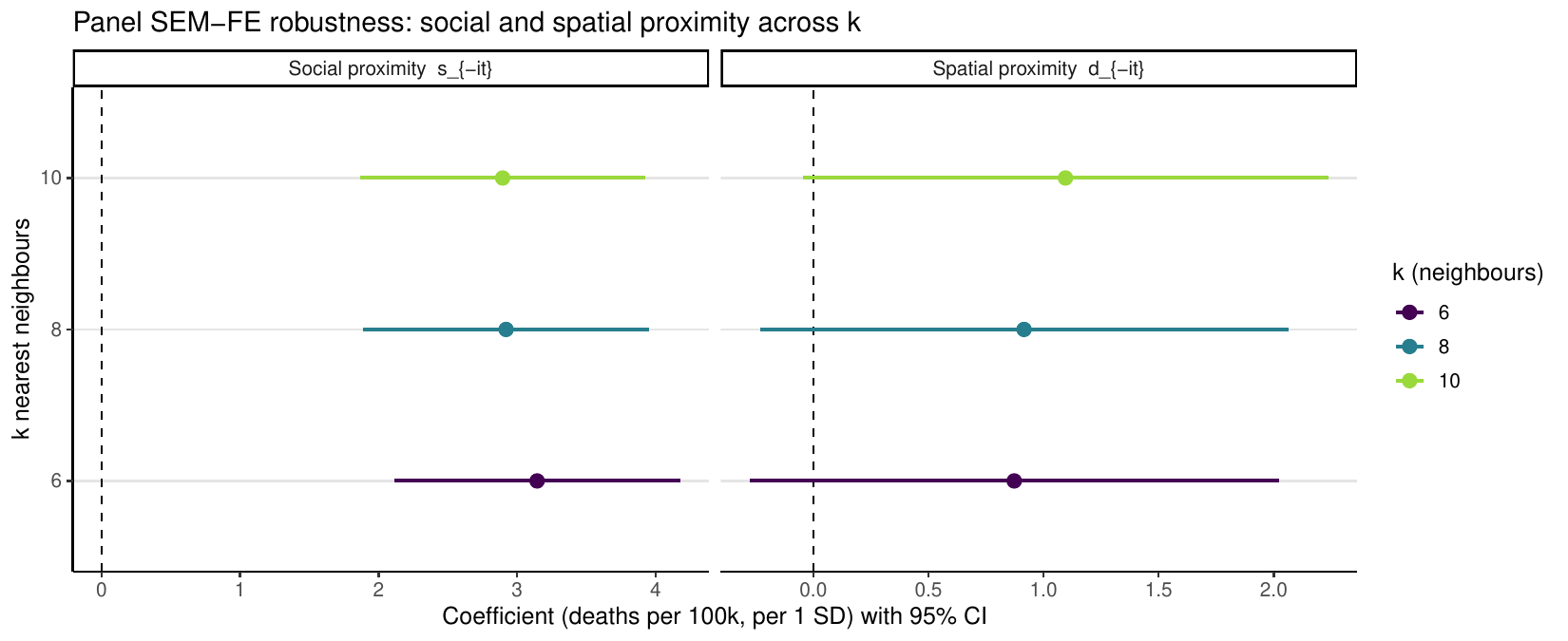}
\caption{\textbf{Stability of the spatial-error model across the number of nearest neighbors.} Estimated social-proximity ($\hat{\psi}$, left panel) and spatial-proximity ($\hat{\chi}$, right panel) coefficients from the panel spatial-error model of Equations~(\ref{eq:sem-mean})--(\ref{eq:sem-error}), re-estimated with $k \in \{6, 8, 10\}$ nearest neighbors defining the row-standardized spatial-weights matrix $M$. Points are point estimates and horizontal lines are $95\%$ confidence intervals; coefficients are expressed in deaths per $100{,}000$ per one-standard-deviation increase in the exposure, and the vertical dashed line marks the null. The social-proximity association is positive, of stable magnitude, and clears the null at every neighbor count, confirming that the correction for spatial autocorrelation and the resulting social-proximity estimate do not depend on the particular choice of $k$.}
\label{fig:sem_fe_knn}
\end{figure}

}

\restoregeometry

\end{document}